\documentclass[fleqn,usenatbib]{mnras}
\usepackage{amssymb}	
\usepackage{newtxtext,newtxmath}
\usepackage[T1]{fontenc}
\DeclareRobustCommand{\VAN}[3]{#2}
\let\VANthebibliography\thebibliography
\def\thebibliography{\DeclareRobustCommand{\VAN}[3]{##3}\VANthebibliography}

\usepackage{graphicx}	
\usepackage{amsmath}	

\usepackage{float}
\usepackage{pifont}
\usepackage{tabularx}
\usepackage{threeparttable}
\usepackage{xspace}
\usepackage{soul,xcolor}

\newcommand{\orcid}[1]{\href{https://orcid.org/#1}{\includegraphics[scale=0.08]{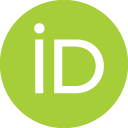}}}

\newcommand{\beq}{\begin{eqnarray}}
\newcommand{\eeq}{\end{eqnarray}}
\newcommand{\mr}{\mathrm}
\newcommand{\Lbol}{L^{\mathrm{AGN}}_{\mathrm{bol}}}
\newcommand{\Mtwohcrit}{M_{\mr{200,crit}}}
\newcommand{\msun}{\,M$_{\odot}$\xspace}

\newcommand{\illustris}{\texttt{Illustris}\xspace}
\newcommand{\illustrisTNG}{\texttt{IllustrisTNG}\xspace}
\newcommand{\TNGcluster}{\texttt{TNG-Cluster}\xspace}
\newcommand{\TNGthreeh}{\texttt{TNG300}\xspace}
\newcommand{\TNG}{\texttt{TNG300}+\texttt{TNG-Cluster}\xspace}
\newcommand{\TNGoneh}{\texttt{TNG100}\xspace}
\newcommand{\TNGfifty}{\texttt{TNG50}\xspace}
\newcommand{\flamingo}{\texttt{FLAMINGO}\xspace}
\newcommand{\flone}{\texttt{L1\_m8}\xspace}
\newcommand{\fltwo}{\texttt{L2p8\_m9}\xspace}

\newcommand{\simba}{\texttt{Simba}\xspace}
\newcommand{\eagle}{\texttt{EAGLE}\xspace}

\title[The Halo Masses of AGNs and quasars at $z=3-7$]{The host halo masses of AGNs and quasars at $z \sim 3-7$ with TNG-Cluster, FLAMINGO and other cosmological galaxy simulations}

\author[Akanksha Kapahtia et. al.]{Akanksha Kapahtia\orcid{0000-0003-0348-0065},$^{1}$\thanks{E-mail:akanksha.kapahtia@gmail.com}
Annalisa Pillepich\orcid{0000-0003-1065-9274},$^{1}$, Joey Braspenning\orcid{0009-0003-3956-4890},$^{1}$, Dylan Nelson\orcid{0000-0001-8421-5890},$^{2,3}$
Joop Schaye\orcid{0000-0002-0668-5560},$^{4}$ 
\newauthor
Eduardo Ba\~nados\orcid{0000-0002-2931-7824},$^{1}$
Silvia Belladitta \orcid{0000-0003-4747-4484},$^{1,5}$
and Frederick Davies$^{1}$\\
$^{1}$Max-Planck-Institut für Astronomie, Königstuhl 17, D-69117 Heidelberg, Germany \\
$^{2}$Universit\"at Heidelberg, Zentrum f\"ur Astronomie, ITA, Albert-Ueberle-Str. 2, D-69120 Heidelberg, Germany\\
$^{3}$Universität Heidelberg, Interdisziplinäres Zentrum für Wissenschaftliches Rechnen, INF 205, 69120 Heidelberg\\
$^{4}$Leiden Observatory, Leiden University, PO Box 9513, 2300 RA Leiden, the Netherlands\\
$^{5}$INAF — Osservatorio di Astrofisica e Scienza dello Spazio, via Gobetti 93/3, I-40129, Bologna, Italy\\
}
\date{Accepted XXX. Received YYY; in original form ZZZ}

\pubyear{2026}

\begin{document}

\label{firstpage}
\pagerange{\pageref{firstpage}--\pageref{lastpage}}
\maketitle
\setstcolor{red}
\begin{abstract}
Most observations and clustering analyses suggest quasars inhabit a narrow range of dark-matter halo masses ($10^{12-13}$\msun) across cosmic time ($z \lesssim 7$). Thanks to recent cosmological hydrodynamical simulations in gigaparsec-scale comoving volumes, it is now possible to directly compare this observational picture with self-consistent galaxy-formation models. In this work, we quantify the relation between AGN bolometric luminosity and host halo mass before Cosmic Noon in the \TNGthreeh, \TNGcluster, \flamingo \flone and \fltwo, and smaller-volume simulations (\illustris, \eagle, \TNGoneh, and \simba).
We focus on AGNs with bolometric luminosities $\Lbol \geq 10^{42}$ erg~s$^{-1}$. Across all models, more massive haloes tend, on average, to host more luminous AGNs, although this trend persists only up to a certain halo mass.
The median relation is highly non-linear, non strictly monotonic, with large scatter, and flattens (\flamingo) or even turns over (\TNG) at a halo mass of $\Mtwohcrit \gtrsim 10^{12}$ \msun, at least at $z<5-6$. This high-mass AGN quenching manifests also as a characteristic host halo mass for quasars: in \TNG, quasars ($\Lbol \sim 10^{45-47}$ erg~s$^{-1}$) typically reside in haloes of $10^{12-12.5}$\msun at $z=3-6$ and \flamingo's quasars extend to median masses of $\sim 10^{12.8}$\msun at $z \sim 3-4$.
However, object-to-object variations are large: all simulations predict much larger scatter in AGN bolometric luminosity at fixed host halo mass than in halo mass at fixed luminosity (up to 3 dex versus $\lesssim 1$ dex for 5th–95th percentiles). This implies weak coupling between halo growth and instantaneous SMBH accretion. Also thanks to this diversity, the host halo masses of simulated quasars broadly match observational estimates. On average, the most luminous AGNs reside in increasingly rarer haloes at earlier epochs but, typically, do not inhabit the most massive haloes at any given time up to $z \approx 7$.

\end{abstract}

\section{Introduction} 
\label{sec:intro}

In hierarchical structure formation, the most massive dark-matter (DM) haloes at any epoch correspond to the rarest peaks of the primordial density field. As baryons collapse into these haloes, they cool and form galaxies at their centers \citep[e.g.][]{Peebles}. Observations indicate that nearly all massive galaxies (stellar mass $\gtrsim 10^{10}$~\msun) host a central supermassive black hole (SMBH) with mass $\gtrsim10^6$~\msun, whose accretion may be limited by the balance between gravity and radiation pressure, the so-called Eddington limit \citep{Eddington}. However, most SMBHs accrete at rates far below this limit, and their masses correlate strongly with the properties of their host galaxies \citep[e.g.][]{KormendyRichstone95,Magorrian98,Ferrarese2000,Gebhardt2000}. These correlations suggest that SMBH growth is regulated on scales extending well beyond their immediate gravitational environment, potentially involving the entire host galaxy \citep[e.g.][]{Heckman2014,Alexander2025} and possibly the surrounding halo \citep[e.g.][]{Booth2010,Shankar2025}. Understanding how this connection extends to halo scales, and how it evolves with cosmic time, is therefore central to studies of galaxy--SMBH co-evolution.

The accretion of gas onto SMBHs releases large amounts of energy, producing active galactic nuclei (AGNs) whose luminosity traces the instantaneous accretion rate onto the black hole. AGNs and their luminosities therefore provide a powerful observational probe of SMBH growth. The most luminous AGNs (i.e., quasars) can reach bolometric luminosities of $\sim10^{48-49}$ erg~s$^{-1}$, allowing them to be detected across cosmic time and at very large distances \citep[e.g.][]{LyndenBell69,Rees1984, Fan2023,Banados2026} including well into the Epoch of Reionization \citep[e.g.][ for $z=7.642$, the current redshift record]{Wang2021}. It is also believed that most galaxies experience at least one phase of {\it observable} AGN activity during their lifetime \citep[e.g.][]{Soltan1982,Hopkins2006,Conroy2013}. Observations have revealed quasars hosting SMBHs with masses $\gtrsim10^9$~\msun at redshifts $z\gtrsim6$ \citep[e.g.][]{Mazzuccheli2017,Farina2022,Banados2023,Yang2020,Yang2021,Wang2021,Belladitta2022,Belladitta2025}, posing a challenge to simple models of SMBH growth through gas accretion within the first billion years after the Big Bang. Recent observations with the James Webb Space Telescope (JWST) extend these studies to lower SMBH masses ($10^{6-8}$~\msun) at similarly high redshifts \citep[e.g.][]{Ubler2024,Larson2023,Bogdan,Kokorev2023,Miaolino2024,Matsuoka2025}. Identifying the hosts of these early AGNs, including both galaxy and halo properties, may therefore provide key insights into aspects of galaxy--SMBH co-evolution prior to Cosmic Noon ($z \gtrsim 3$), i.e. before feedback processes significantly shape the broader galaxy wide statistics \citep[e.g.][]{Fabian2012,Somerville2015}.

The properties of DM haloes might influence SMBH growth, directly or indirectly, beginning with their total mass \citep{JahnkeMaccio2011,BogdanGoulding2015}.
For both practical and conceptual reasons, observational studies predominantly examine the properties of host galaxies rather than haloes, such as bulge mass and stellar velocity dispersion \citep[e.g.][]{KormendyHo2013}, which  in principle can be used as proxy for halo mass. At high redshift, however, these measurements become increasingly challenging due to the faintness and compactness of galaxies and the requirement for high signal-to-noise, high-resolution spectroscopy to resolve their internal kinematics.

Observational constraints on the total halo masses of AGN hosts are often derived from clustering measurements of their spatial position, especially for the more luminous ones ($\gtrsim10^{44}$ erg~s$^{-1}$). Measurements from large surveys such as the Sloan Digital Sky Survey \citep[SDSS,][]{sdss2000} and the 2dF Quasar Survey \citep[2QZ,][]{2df2004} indicate that quasars reside in haloes of characteristic mass $\sim10^{12}$~\msun up to $z\simeq2$ \citep{Croom2005,Ross2009}. 
Barring the inferences by \cite{Shen2007} at $z\sim4$ and \cite{Arita2023} at $z\sim6$, all clustering measurements at $z=0-6$ are broadly consistent with luminous AGNs and quasars residing in hosts with total mass $10^{12-13}$~\msun \citep[e.g.][and references in the rest of the paper]{White2012,Sarah2015,Timlin2018,He2018,Mascarell2025}. This range is also broadly consistent with the host halo masses inferred from density field measurements \citep[e.g.][]{Chen2022} and gas kinematics \citep[][]{Qinyue2025} around quasars at high redshifts as well as inferences on host halo masses of AGNs and quasars selected in X-ray at $z\lesssim2.5$ redshifts \citep[e.g.][]{Leauthaud2015, Viitanen2019, Comparat2023, Mountrichas2026}. 
At higher redshift, clustering measurements become increasingly difficult, due to the decreasing abundance of bright quasars \citep{Schindler2023}, therefore an alternative way is to use cross-correlation of quasars with their galaxy environment \citep[e.g.][]{Garcia2022}. Using JWST observations from the EIGER survey, \cite{Eilers} used this technique during the Epoch of Reionization. This quasar-galaxy cross-correlation technique has since been applied to two quasars at $z\sim7$ by \cite{Schindler2026} and two dozens quasars at $z=6.5-6.8$ from the ASPIRE survey \citep{Wang2026, Huang2026}.

The halo mass function grows with time: therefore, the roughly constant typical host halo mass derived from clustering measurements over the last 12–13 billion years of cosmic history has been interpreted as indicating that luminous quasars trace progressively rarer, higher-density peaks at earlier epochs. Conversely, for the same reasons, at lower redshift, the most luminous quasars do not necessarily reside in the rarest and most massive systems. These notable broad-picture conclusions remain, however, based on relatively indirect inferences. 

An important limitation of clustering-based estimates is that they necessarily rely on flux-limited samples, implicitly linking the minimum observed AGN luminosity to a {\it minimum} host halo mass \citep[e.g.][]{White2012,Sarah2015}. Such approaches hence provide only limited constraints on the distribution of AGN and quasar luminosities across halo masses, as they cannot account for the full populations below the observable threshold and for the effects that these undetected systems may have on the inferred trends.

Moreover, interpreting these measurements sometimes requires (implicit or explicit) assumptions about the relation between AGN luminosity and host halo mass.  Halo-occupation and conditional-luminosity models often assume a monotonic relation, with empirical constraints providing limits on its scatter and deviations from linearity. For example, models motivated by the strong clustering of $z\sim4$ quasars \citep{Shen2007} suggest that a roughly linear AGN luminosity -- host halo mass relation with modest scatter ($\sim0.3$ dex) can reproduce the observed clustering signal \citep{White2008}. However, other semi-empirical models indicate that the quasar–halo connection may involve substantially larger scatter \citep[e.g.][]{Zhang2023}. Models that incorporate mergers and more complex accretion histories indicate that a scatter approaching $\sim1$ dex reconciles clustering measurements with the rapidly evolving quasar luminosity function \citep{Wyithe2009}, while other similar models reveal apparent tensions between clustering and luminosity statistics \citep{Shankar2010}. These important subtleties have been discussed also in the context of more recent halo-occupation models and are crystallized by conditional luminosity-function models informed by cosmological DM-only large-volume simulations of latest generation: reconciling clustering measurements with the observed quasar luminosity function at $z\sim2.5, 4,$ and 6 requires varying levels of scatter (typically $\lesssim 1$ dex) in the AGN luminosity -- host halo mass relation, and assumptions or inferences on the scatter are often related to the implied host halo mass estimates at the high-mass end \citep{Pizzati2024,Pizzati2024b}.

Cosmological simulations of galaxy formation and evolution in representative large-scale volumes offer a complementary avenue to explore AGN demographics and the physical processes that regulate SMBH accretion, particularly given the level of realism achieved over the last decade \citep{Vogelsberger_review}. Spearheaded by projects such as \illustris \citep{Vogelsberger2014-illustris, Genel-Illustris,Nelson-illustris-data} and \eagle \citep{Eagle-Schaye, Eagle-Crain, Eagle-data-release}, these simulations aim to capture the unfolding and effects of astrophysical processes spanning a wide range of scales, from $\gg \mathrm{Mpc}$ scales where gravitational collapse begins, down to sub-$\mathrm{kpc}$ scales where star formation, SMBH physics, and feedback originate. Resolving this enormous dynamic range is (at the very least) computationally prohibitive, so simulations employ sub-grid models for processes below their resolution limits, with free parameters and choices typically calibrated against selected observational constraints such as galaxy scaling relations in the local Universe or galaxy population statistics while retaining predictive power for other properties.

Previous simulation studies have investigated the statistical properties of SMBH and AGN populations and how their luminosities are shaped by gas supply, feedback processes, and SMBH fueling efficiencies \citep{Habouzit2021,Habouzit2022a,Habouzit2022b, Astrid,Hirschmann2014, Steinborn2015,Schulze2015}. However, simulations with moderate box sizes (e.g. $\lesssim100-150$ comoving Mpc) cannot capture the rare and bright quasars probed by modern wide-field surveys, while larger-volume simulations often compromise resolution and may not fully resolve the small-scale gas dynamics  that drive AGN luminosities. Alternatively, some models focus on the very high-redshift Universe only. For example, the BlueTides simulation \citep{Feng2016}, analyzed down to $z\sim7$ \citep{Tenneti2019, Marshall2021}, indicate that the SMBHs powering the most luminous AGNs do not necessarily evolve into the most massive haloes at later times \citep[e.g.,][]{Tenneti2018}. A recent study based on zoom-in simulations compared velocities of galaxies in quasar environments around $z>6$ quasars to those in observations and inferred that the halo masses of quasars is $\gtrsim5\times10^{12}$\msun \citep{Costa2024}. Recently, exciting avenues for the demographics of AGNs at $z>3$ \citep{Shen2026} are being uncovered by combining radiation-hydrodynamics with galaxy-formation physics in $(\sim500\ \mathrm{Mpc})^3$ comoving volumes \citep[see the {\tt LUMINA} project,][]{Zier2026}.

Extending existing studies to even larger volumes and statistics and to higher redshifts is particularly important, given that observations at these epochs are now more robust and statistically well constrained. In this work, we investigate the relation between AGN bolometric luminosity and host halo mass at $z\gtrsim3$ using two new simulation suites that enable such improvements. We use the \flamingo simulations \citep{Flamingo-Schaye2023, Flamingo-Roi2023, Helly2026}, encompassing volumes of 1–3 comoving Gpc per side, and the combined \TNG simulations, one of the flagship runs and an extension \citep{Nelson-TNG-Cluster}, respectively, of the \illustrisTNG project \citep[][and references therein]{Nelson-TNG-data}. These encompass volumes of 300–1000 comoving Mpc per side at roughly two orders of magnitude better resolution and with a different underlying galaxy-formation model than \flamingo. Together with the outcomes of other galaxy-formation simulations of $(\sim100\ \mathrm{Mpc})^3$ volumes of the previous generation (e.g., \illustris, \eagle, \TNGoneh, and \simba), these simulations allow us to explore how modeling assumptions, as well as volume and resolution, influence the predicted relationship between AGN bolometric luminosity and host halo mass.

Recently, \cite{flamingo_quasar_clustering} showed that the large-volume $2.8^3$~Gpc$^3$ box of the \flamingo suite reproduces the observed quasar clustering measurements at $0 \lesssim z \lesssim 3$. Bolstered by this agreement with observations, and to focus on cosmic epochs when feedback processes have not yet left a significant imprint on galaxy and SMBH population statistics, in this paper we deliberately concentrate on $z=3-7$. Our goal is to map the AGN bolometric luminosity -- host halo mass relation and its scatter across contemporary cosmological galaxy formation simulations during the first couple of billion years of cosmic evolution, within different modeling frameworks, and to place these predictions in the context of observationally-inferred host halo masses. 
We first provide a quantitative overview of how AGNs populate the plane of AGN bolometric luminosity vs. host halo mass across the population of observationally plausible systems. We then examine high-luminosity objects (bolometric luminosity $\geq 10^{45}$ erg~s$^{-1}$), i.e. quasars, and assess whether and how their host halo masses vary across time.

In Section~\ref{sec:methods} we describe the simulations and AGN selection criteria. Section~\ref{sec:results_plane} presents the AGN bolometric luminosity vs. host halo mass plane for all simulated AGNs with bolometric luminosity $\geq 10^{42}$ erg~s$^{-1}$ at $z=3-7$, whereas Section~\ref{sec:results_quasars} focuses on the redshift evolution of the host halo masses of quasars. We discuss implications and limitations of our findings and of the comparisons to observations in Section~\ref{sec:disc} and summarize our conclusions in Section~\ref{sec:conclusions}. 

\begin{table*}
\caption{\textbf{Key characteristics and SMBH physics of the cosmological (magneto-)hydrodynamical galaxy simulations used in this work.} The columns read: 1) simulation name;  2) size of the cubic periodic boundary-condition box; 3–4) mass of the baryonic (gas and stars) and DM components; 5–8) SMBH seeding mass, halo mass threshold for SMBH seeding, accretion and feedback models. The simulations are listed in approximate chronological order, starting with the \eagle and \illustris simulations of 2014, and finishing, in the bottom two sections, with the primary simulation suites used in this study, \TNG and \flamingo, which encompass much larger volumes and hence greater statistics at the high halo mass end than the earlier-generation, $\sim100$ cMpc simulation projects. The \texttt{TNG100}, \TNGthreeh and \TNGcluster runs are based on the same underlying \illustrisTNG galaxy-formation model. The \flamingo $\flone$ and $\fltwo$ runs have different box size and resolution but are based on the same fiducial \flamingo, with model-parameter values adjusted for resolution. In all simulations studied in this paper, the SMBH growth via Bondi gas accretion is capped at the Eddington limit. Importantly, the ensuing accretion rates of the SMBHs in all simulation models depend on a host of physical conditions and mechanisms acting in concert, including e.g. the effects of stellar feedback on the availability of gas.  We determine the AGN bolometric luminosity in simulations uniformly in post-processing based on the SMBH accretion rates, as per Eq.~\ref{eq:luminosities}.}
\begin{center}
\begin{tabular}{l c c c c c c c}   
\hline
Simulation &
Box size &
Baryon mass& 
DM mass&
SMBH& 
SMBH&
SMBH&
SMBH\\

& & & &
Seed mass & 
Seed halo mass&
Gas accretion law&
Feedback mode(s) \\

&(cMpc) & [$10^6$\msun] & [$10^6 $\msun] &$[10^5$\msun]& [\msun]& &\\

\hline
\illustris &106.5&1.26&6.26&1.4 & $7.1 \times 10^{10}$ & Bondi with boost factor & Quasar + Radio$^{(\rm d)}$\\
\eagle &100&1.8&9.7&1.5 &$1.47 \times 10^{10}$& Bondi, modified & Thermal\\
\simba &147&18&96&0.14 & $3\times10^9$$^{(\rm b)}$& Torque-limited + Bondi & Kinetic + Jet\\
\TNGoneh &100.7&1.4 & 7.5 &12 &$7.4 \times 10^{10}$& Bondi& Thermal + Kinetic\\
& & & & &&\\
\TNGthreeh &302.6&11& 59& 12 & $7.4 \times 10^{10}$ & Bondi& Thermal + Kinetic\\
\TNGcluster &1004$^{(\rm a)}$&11& 59& 12 & $7.4 \times 10^{10}$ & Bondi& Thermal + Kinetic\\
& & & & &&\\
\flamingo \flone &1000&134&706&1 &$3.5 \times 10^{10}$& Bondi with boost factor$^{(\rm c)}$ & Thermal$^{(\rm e)}$\\
\flamingo \fltwo &2800&1070&5650&1 &$2.8 \times 10^{11}$& Bondi with boost factor$^{(\rm c)}$ & Thermal$^{(\rm e)}$\\
\hline
\end{tabular}
\end{center}
\begin{tablenotes}[para,flushleft]
\footnotesize
\item[a] Size of the parent box: \TNGcluster consists of zoom-in simulations of 352 haloes selected from a 1~comoving Gpc box at the \TNGthreeh resolution.\\
\item[b] This is a galaxy stellar mass: in \simba, SMBHs are seeded based on a stellar-mass rather than a halo-mass threshold, as typical instead in other models.\\
\item[c] The boost factor in \flamingo is density dependent whereas, for example, in \illustris this is a constant number \citep[100, see][]{Vogelsberger-illustris-model}.\\
\item[d] Strictly speaking, both the \illustris and \illustrisTNG models include a third SMBH feedback mode, dubbed ``radiative'' and acting irrespective of the SMBH accretion rate, whereby the cooling of the gas around AGN is modulated to account for their radiation field \citep{Vogelsberger-illustris-model}. Similarly, the \simba model also includes an ``X-ray feedback'' to mimic the effect of heating from X-ray off the accretion disk. Both are implemented as modifications to the cooling/heating tables of the gas and we hence prefer to mention them separately. \\
\item[e] The precise implementations of the thermal mode feedback differ across simulation projects. For example, in \eagle and \flamingo, SMBHs do not inject thermal energy at every time-step but rather store energy for release until certain conditions are met. This episodic nature of SMBH feedback is similar in essence to the one implemented for the radio mode of \illustris and the kinetic mode of the \illustrisTNG model (i.e. \TNGoneh, \TNGthreeh and \TNGcluster) but is different than thermal mode feedback of \illustris and \illustrisTNG, which is continuous.
\end{tablenotes}
\label{Tab:simulations}
\end{table*}

\section{Methodology: Simulations, SMBH models and AGN populations} 
\label{sec:methods}

This paper is based on the outcome of a series of large-volume cosmological (magneto-)hydrodynamical simulations of galaxies and, consequently, of SMBHs and AGNs. Therein, within a given cosmological framework and given specific initial conditions, galaxy evolution is modeled by following the coupled evolution of collisionless gravitational dynamics (for DM, gas, stellar particles and SMBHs) and hydrodynamics of gas in expanding and representative portions of synthetic universes \citep[e.g.][]{Vogelsberger_review}. 

In particular, our analysis is primarily focused on the outcome of two sets of simulations. On the one hand, we use the \TNGthreeh simulation of the \illustrisTNG project \citep{ Pillepich-TNG,Nelson-TNG-data,Springel-TNG-clustering,Naiman-TNG, Nelson-TNG-color,Marinacci-TNG} combined with the more recent \TNGcluster \citep{Nelson-TNG-Cluster}: these are based on the same underlying IllustrisTNG galaxy-formation model \citep{Pillepich-TNG, Weinberger-TNG} and we will refer to them as \TNG. On the other hand, we analyze two main runs of the \flamingo simulation suite \citep{Flamingo-Schaye2023, Flamingo-Roi2023, Helly2026}, particularly the \flone and \fltwo runs based on the fiducial \flamingo galaxy-formation model\footnote{For the \flamingo \fltwo we carry out our analysis only at $z= 3, 4$ and 5, since these are the only high-redshift outputs available for this run.}. These together provide unprecedented access to the high-mass end of the halo mass distribution (up to $\sim 10^{13-14}$\msun haloes also before Cosmic Noon, $z \gtrsim 3$) while simultaneously allowing us to bracket systematic differences due to different modeling choices.

For benchmarking and to connect to past results in the literature, we contextualize our results within those from relatively smaller-volume $(\sim100\ \mathrm{Mpc})^3$ simulations of previous generation: \illustris \citep{Vogelsberger2014-illustris, Genel-Illustris,Nelson-illustris-data}, \eagle \citep{Eagle-Schaye, Eagle-Crain, Eagle-data-release}, the \TNGoneh run of the \illustrisTNG project \citep[][and references therein and above]{Nelson-TNG-data}, and \simba \citep{Simba-Dave,Simba-Wu}. These allow comparisons for halo masses $\lesssim 10^{12.5}$\msun at $z \ge 3$ across an even wider range of models. 

Together, all the runs considered in this work span three orders of magnitudes in mass resolution, a factor of 30,000 in volume, and a few different parameterizations of the standard $\Lambda$CDM cosmological model: a summary is given in Table~\ref{Tab:simulations}.
Below, we briefly provide more details about the simulations used in our analysis by highlighting similarities and differences in their SMBH seeding, accretion, and feedback models. We refer the reader to the respective literature for a detailed description of the astrophysical modeling employed in these simulations. Cosmological parameters are adopted as defined in each respective simulation, all consistent within the final nine-year WMAP and the Planck 2016 constraints.

\subsection{Large-volume simulations: \TNGcluster and \flamingo}
\label{sec:methods_sims_big}

The \TNG and \flamingo simulations extend AGN modeling into large-volume regimes that capture structures up to the galaxy-cluster scale, enabling a systematic exploration of quasars and of SMBH growth and feedback in the most massive environments.

The \illustrisTNG project\footnote{\url{https://www.tng-project.org/}} is a suite of cosmological magneto-hydrodynamical simulations of galaxies that builds and improves upon the original \illustris model (see Section~\ref{sec:methods_sims_small}). \TNGthreeh is one of its flagship runs, together with \TNGoneh (Section~\ref{sec:methods_sims_small}) and \TNGfifty \citep{Pillepich-TNG50,Nelson-TNG50}, from poorer to better resolution, respectively (baryonic target mass from $1.1 \times 10^7$ to $8.5 \times 10^4$\msun) and from larger to smaller volume, respectively (box size from 303 to 52 comoving Mpc). 

The \TNGcluster project\footnote{\url{https://www.tng-project.org/cluster/}}\citep{Nelson-TNG-Cluster} extends the \illustrisTNG framework to a suite of 352 zoom-in simulations of massive haloes: these have been selected at $z=0$ from a DM-only parent simulation with side of about 1 Gpc and evolved with the \illustrisTNG galaxy-formation model and at the same resolution as \TNGthreeh, i.e. target baryonic mass of $1.1 \times 10^7$\msun. Effectively, the \TNGcluster haloes are sampled from the parent volume so as to mitigate the drop-off in the \TNGthreeh halo mass function beyond halo masses of $\sim 10^{14.5}$\msun, producing a roughly flat halo mass distribution in the halo mass range $10^{14.2-15}$\msun at $z=0$. Above this mass, all haloes identified in the 1-Gpc parent simulation are included, yielding a complete volume-limited cluster sample, at least at the current epoch. This approach leads to 636 haloes more massive than $10^{14}$\msun in the combined \TNG simulations at $z=0$, compared to 280 haloes in \TNGthreeh alone. We use all the \TNGcluster zoom haloes (as identified at $z=0$) along with all their progenitors (main and secondary) at higher redshifts with no low-resolution contamination, allowing us to work with hundreds of high-mass systems also before Cosmic Noon\footnote{We use haloes having \texttt{GroupPrimaryZoomTarget}$>0$ and subsequently exclude haloes which have non-0 contamination. This leads to more than 352 haloes from \TNGcluster alone at higher redshifts.}. 

The SMBHs in the \illustrisTNG model are seeded with a mass of $1.2 \times 10^6$\msun within Friends-of-Friends \citep[FoF,][]{Press_and_davis_1982,Davis_fof} haloes with total mass larger than $7.4 \times 10^{10}$\msun. This SMBH seed mass then grows either via gas accretion, at the Eddington-limited Bondi accretion rate, or via mergers with other SMBHs. For the feedback, at high accretion rates, the surrounding gas is heated thermally, continuously and isotropically;  at low accretion rates, a kinetic feedback model is applied, which transfers momentum to nearby gas cells, in an episodic manner and in different random directions for different energy injection events, effectively removing gas from the SMBH’s immediate environment. The transition between these two feedback modes is governed by a SMBH-mass-dependent threshold on the Eddington ratio $\chi$, such that lower-mass SMBHs are less likely to enter the kinetic mode \citep[see][for details]{Weinberger-TNG}. 

The \textbf{F}ull-hydro \textbf{L}arge-scale structure simulations with \textbf{A}ll-sky \textbf{M}apping for the \textbf{I}nterpretation of \textbf{N}ext \textbf{G}eneration \textbf{O}bservations (\flamingo) simulation suite\footnote{\url{https://flamingo.strw.leidenuniv.nl/}} constitutes even larger-volume, cosmological galaxy simulations, with flagship runs spanning box sizes of 1 comoving Gpc (e.g. the \flone used here) and 2.8 comoving Gpc (e.g. \fltwo) and achieving mass resolutions that are approximately one and two orders of magnitude lower than \TNGthreeh, respectively (see Table~\ref{Tab:simulations}). The vast simulated volumes of \flamingo enable robust sampling of the high-mass end of the halo mass function, providing superior statistics for cluster-scale systems and rare, massive haloes that are sparsely represented in smaller-volume simulations. For example at $z=0$, the \fltwo run has $4600$ haloes with mean halo mass $\gtrsim 10^{15}$\msun. 

In the fiducial \flamingo runs analysed in this work, SMBHs are seeded in FoF haloes with mass greater than $2.8 \times 10^{11}$ and $3.5 \times 10^{10}$\msun in the \fltwo and \flone simulations, respectively, at the location of the densest gas particle, and assigned a seed mass of $10^5$\msun. The mass of the SMBH grows according to an Eddington-limited, Bondi-Hoyle accretion rate modified by a density dependent boost factor. The AGN feedback model employed in the fiducial \flamingo runs used throughout this work is the same as the thermal feedback model of \cite{Eagle-feedback2009} employed in the \eagle simulations described below. The \flamingo simulations have shown to reproduce the observed quasar luminosity function at $z \sim 1$ and for faint quasars $\Lbol \lesssim 10^{45}$ erg~s$^{-1}$ \citep{flamingo_quasar_clustering}. 

Therefore, the better resolution of \TNG and the large number of massive haloes captured by \flamingo offer two complementary frameworks, one sufficient to resolve well the internal structure of haloes and galaxies and the other optimized for statistical sampling of the high-mass population, particularly at $z \gtrsim 3$. 

\subsection{Smaller-volume simulations: \illustris, \eagle, \texttt{TNG100}, and \simba} \label{sec:methods_sims_small}

To assess how sensitive the properties of AGN hosts are to additional variations in galaxy-formation physics and numerical resolution, we include in our analysis a set of smaller-volume cosmological simulations: \illustris, \eagle, and \texttt{TNG100} (with improved resolution compared to \TNG), and \simba (with similar resolution as \TNG), based on yet alternative galaxy-formation models. These simulations cover comoving volumes of order $100-150$ comoving Mpc a side, and achieve baryonic mass resolutions of $\sim 10^{6}$\msun (except \simba, with $1.8\times10^{7}$\msun). 

Despite differences in implementation, these models share a broadly similar conceptual framework for SMBH formation and growth and, in fact, similar to those of \TNG and \flamingo (see Table~\ref{Tab:simulations}). As a reminder, \TNGoneh, \TNGthreeh and \TNGcluster share the same \illustrisTNG galaxy-formation model, with subgrid parameter values kept unchanged across resolutions \citep[see more details and strong resolution convergence checks in the Appendices of e.g.][]{Pillepich-TNG, Pillepich2018, Pillepich-TNG50}. 

Also in the smaller-volume simulations, SMBHs are seeded with initial masses in the range $10^{4}\text{--}10^{6}~\mathrm{M_\odot}$ once haloes exceed masses of about $10^{10-10.5}~\mathrm{M_\odot}$ (but for \simba, whereby the host mass threshold for seeding a SMBH is based on galaxy stellar rather than total halo mass). SMBHs grow via gas accretion, in all cases modeled through an Eddington-limited, Bondi-like prescription (possibly in addition to other gas accretion mechanisms), and through SMBH mergers. In all models, the dynamics of SMBHs should be interpreted with care, as repositioning schemes of the SMBHs to e.g. the potential local minima are commonly adopted.

All simulation models include feedback from SMBHs, either in a single fashion only (e.g. the thermal mode feedback of \eagle and \flamingo) or by invoking a two-mode feedback mechanism depending on the SMBH accretion rates (\illustris, \illustrisTNG, and \simba). In fact, in the absence of radiation-hydrodynamics, none of the models used in this work include truly self-consistent treatments of radiative feedback from SMBHs, but rather attempt to mimic its effects via dedicated ``radiative'' feedback channels (\illustris, \illustrisTNG, and \simba). The parameterization and the nature of the energy injections vary from model to model too. All models but \simba invoke a thermal mode, with internal energy donated isotropically to the gas in the immediate surrounding of the SMBHs, in a continuous (\illustris at high accretion rates, \TNGoneh, \TNGthreeh, and \TNGcluster) or episodic manner (\eagle and \flamingo). The ``radio'' or ``bubble'' mode feedback of \illustris also entails a thermal (non-continuous) dump, but in gas in the middle of the gaseous haloes around galaxies, to mimic the long-range effects of jets. In the \illustrisTNG and \simba models, SMBHs also inject kinetic energy or momentum: as described for \TNG, also in \TNGoneh low-accreting SMBHs undergo episodic \emph{kinetic} energy injections in random directions (but isotropic once averaged in time); in \simba, similar episodic kinetic winds transition into a jet-dominated (and so bipolar) regime once SMBHs become massive and accretion rates decline.   

Although these simulations successfully reproduce, to varying degrees, a number of  observed galaxy properties and statistics, their comparatively limited cosmological volumes restrict the sampling of  massive haloes ($\gtrsim 10^{13-14}~\mathrm{M_\odot}$) and the luminous AGN populations, particularly at $z \gtrsim 3$. Nevertheless, examining the distribution of AGN and quasar host halo masses in these models provides valuable qualitative insight into how differences in subgrid physics, feedback coupling, and SMBH fueling channels may influence the environments in which AGN activity is triggered and sustained. 

Finally, in this comparative summary we have emphasized the SMBH physics. However, the growth of SMBHs in cosmological simulations -- and thus the properties and luminosities of AGNs and quasars therein -- depends on a wide range of coupled physical processes. These include not only seeding prescriptions, gas accretion models, and SMBH feedback, but also hierarchical structure growth \citep[e.g.][for IllustrisTNG]{Weinberger-feedback, Truong2021} and stellar feedback \citep[e.g.][]{Dubois2015, Bower2017, Pillepich-TNG, Habouzit2019, Truong2021}, which regulate the gas supply available for SMBH growth, particularly during the early, low-mass, high-redshift phases.

\subsection{Definition and properties of haloes, galaxies, and SMBHs}
\label{sec:methods_defs}

As alluded to above, all simulations identify haloes of DM, gas and stars using the FoF algorithm. Gravitationally-bound structures (subhaloes) are then identified by the \texttt{subfind} algorithm \citep{Springel2001} in all simulations except in \flamingo, where they are identified using the \texttt{HBT-HERONS} structure finder \citep{Moreno2025}, which in turn is an updated version of the Hierarchical Bound Tracing algorithm \citep{HBT}. Throughout this paper, haloes refer specifically to those identified by the FoF method and galaxies refer to {\it central} \texttt{subfind} or \texttt{HBT-HERONS} objects with non-zero stellar mass.

We quantify the mass of haloes using $\Mtwohcrit$, defined as the total mass within a spherical region of radius $R_{200,\mr{crit}}$ such that the average density of the enclosed mass is 200 times the critical density of the Universe at that redshift. In the following, halo mass will refer to $\Mtwohcrit$ of the halo that hosts the AGN.

For all simulations except \flamingo, by galaxy stellar masses we mean all the stellar mass that is gravitationally-bound and within twice the stellar half-mass radius from the center of any given galaxy. In \flamingo we use the gravitationally-bound stellar mass within an aperture of $50~\mr{pkpc}$ also used for calibrating the simulations to the observations. 

It should be noted that not all simulated galaxies are stored in the simulations catalogs irrespective of their mass i.e. irrespective of their number of particles and cells. For example, in the \illustrisTNG runs, only \texttt{subfind} objects with at least 32 among particles and cells are stored whereas, in \flamingo, the instantaneous accretion rate of SMBHs is stored only for subhaloes comprising of at least 100 total particles. This sets a limit on the smallest halo mass (and hence galaxy stellar mass and, possibly, AGN luminosity) that can be studied -- we expand on this below.

Depending upon the mergers and (re)positioning prescriptions, a simulated galaxy might occasionally host multiple SMBHs. Throughout this work, we define the SMBH powering the AGN under inspection as the most massive SMBH that is gravitationally bound to any given galaxy i.e. halo. For each SMBH, the masses and instantaneous accretion rates are self-consistently determined by the respective simulations. AGN luminosity, on the other hand, is not a direct outcome of the simulations adopted here. We calculate the AGN luminosity from the instantaneous accretion rates of SMBHs in post-processing, based on the model of \cite{Lbol-Churazov}:

\begin{equation}
L^{\mr{AGN}}_{\mr{bol}}= 
\begin{cases}
\epsilon_r \, \dot{M} \, c^2, & \dot{M}\ge 0.1 \, \dot{M}_{\text{Edd}} \\
(10 \chi )^2 \, \epsilon_r \, \dot{M}_{\text{SMBH}} \, c^2, & \dot{M}_{\text{SMBH}} < 0.1 \, \dot{M}_{\text{Edd}}
\end{cases}
\label{eq:luminosities}
\end{equation}

where $\dot{M}_{\text{SMBH}}$ is the \textit{instantaneous} accretion rate of the SMBH (which is a direct outcome and prediction of the simulations), $\epsilon_r$ is a radiative efficiency that describes the fraction of SMBH's accreted rest mass energy that is converted into radiation, and $\chi$ is the Eddington ratio. We adopt throughout $\epsilon_r \equiv 0.1$.

This ``bimodal'' definition of AGN luminosity is similar to that used in other works \citep[e.g.][]{Habouzit2022a,Habouzit2022b, Weinberger25}, but not all. For example, \cite{flamingo_quasar_clustering} use a single-formula conversion between SMBH accretion rate and bolometric luminosity similar to the high-accretion case of Eq.~\ref{eq:luminosities}, as they primarily focus on the (more luminous) quasars. As we aim to explore SMBHs also toward low levels of luminosities, we prefer to distinguish between radiatively-efficient and -inefficient SMBHs. 

We also elect to adopt a common value for the radiative efficiency parameter, even though within the galaxy-formation models used throughout different choices are in place (e.g. \simba, \eagle and \flamingo use 0.1, whereas all the \illustris and \illustrisTNG runs and \TNGcluster use 0.2). In this way, possible differences among simulations can be traced more directly to the underlying SMBH accretion rates predicted by the simulations (which account for {\it all} modeling differences), rather than also reflecting different post-processing choices. Moreover, $\epsilon_r \sim 0.1$ is consistent with values inferred from applications of the Soltan argument \citep{Soltan1982}, based on  comparing the integrated AGN luminosity density to the observed $z=0$ SMBH mass density \citep[e.g.][]{Yu2002, Shankar2010}, even though some studies indicate that this efficiency can be much lower at the high redshifts explored here \citep{Davies2019, Knud2025}.

Finally, throughout this paper and unless differently stated, we use the word AGN loosely, to denote luminous SMBHs irrespective of their bolometric luminosity. On the other hand, when emphasis and focus is placed on the most luminous ones, we use the word \textit{quasars} to specifically refer to AGN with bolometric luminosities $\Lbol \geq 10^{45}$~ erg~s$^{-1}$, and distinguish between \textit{faint} and \textit{bright quasars}, with the latter exceeding $10^{46}$ erg~s$^{-1}$.

\subsection{Selecting simulated AGN populations}
\label{sec:methods_selection}
The simulations introduced in the previous sections naturally return AGNs across wide ranges of mass accretion rates and hence bolometric luminosities. As an example, within the \illustrisTNG model, bolometric luminosities of SMBHs as low as $10^{35-40}$ erg~s$^{-1}$ are typical of the accretion rates corresponding to the SMBHs at the centers of Milky Way-like galaxies at $z=0$, yet capable of producing vast bipolar feedback features similar to the bubbles of X-ray emission in our Galaxy \citep{Pillepich2021}. 

Throughout this paper, we restrict our analysis to simulated AGNs with $\Lbol \geq 10^{42}$ erg~s$^{-1}$. Furthermore, as motivated in Section~\ref{sec:intro}, we deliberately focus on the high-redshift Universe, before Cosmic Noon, i.e. at redshifts $z=3-7$. 

Such a selection is agnostic, a priori, of halo, galaxy or SMBH mass. However, AGN bolometric luminosities, halo masses, galaxy stellar masses, and SMBH masses are all correlated with each other, to varying and different degrees across different simulations and across different mass and redshift regimes within the same simulation. An overview of this and of the populations of SMBHs realized by the \TNG and \flamingo simulations are given in Appendix~\ref{app:SMBHpopulations}. As shown there, restricting the analysis to AGNs more luminous than $10^{42}$ erg~s$^{-1}$ reduces our sensitivity to the regimes that are most dependent -- by construction and unavoidably -- on specific choices of the galaxy-formation models, in particular the mass with which SMBHs are seeded (typically $10^{5-6}$~\msun) and the halo mass at which seeding occurs (typically total FoF halo masses of $10^{10-10.8}$~\msun). 

Although AGNs at luminosities as low as $10^{42-43}$ erg~s$^{-1}$ are observationally difficult to distinguish from normal galaxies \citep[e.g.][]{Banados2026}, such a threshold allows us to probe the full range of SMBH accretion activity, from low-luminosity AGNs to luminous quasars, and within the aforementioned limits of the simulation models. In fact, for SMBH masses spanning $10^{6-9}$ ($10^{5-9}$)\msun in \TNG (\flamingo), this luminosity cut corresponds to Eddington ratios in the ranges of $10^{-3}$ to $10^{-1.5}$ ($10^{-3}$ to $10^{-1}$). Moreover, the median Eddington ratio of SMBHs with $\Lbol=10^{42}$ erg~s$^{-1}$ is even higher, reaching about $10^{-2}$ in \TNG and $10^{-1.5}$ in \flamingo. Therefore, our luminosity selection can be safely regarded as probing the Eddington-ratio regime that is considered characteristic of AGNs \citep[see Section 1.6 of][]{Banados2026}.

As alluded to above, we also focus exclusively on the (central) SMBHs of central galaxies, i.e. of galaxies that dominate the potential well of their host haloes. The object-to-object variations that we quantify throughout the paper --  which is to be intended as {\it intrinsic} scatter free from any additional observational uncertainty unless otherwise stated -- is therefore likely a lower limit of the actual intrinsic scatter that would arise if also satellite galaxies, and those more affected by environmental processes, were considered.

\section{AGN Luminosity -- Halo Mass Relation}
\label{sec:results_plane}

Both \TNG and \flamingo predict complex relationships between the bolometric luminosity of AGNs and the halo mass of their hosts before Cosmic Noon, with very large scatter.

\begin{figure*}
\centering
    \includegraphics[width=0.62\textwidth]{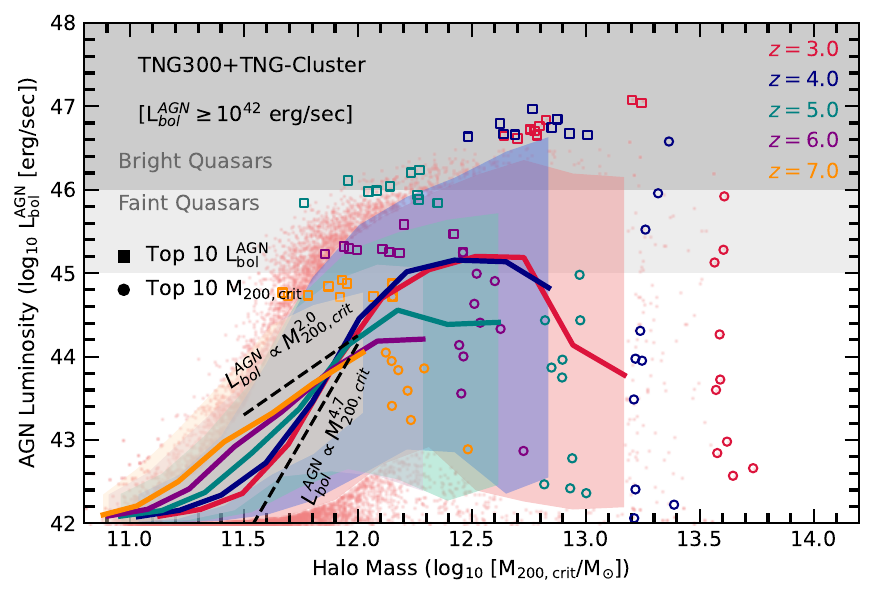}\\
 \includegraphics[width=0.28\linewidth]{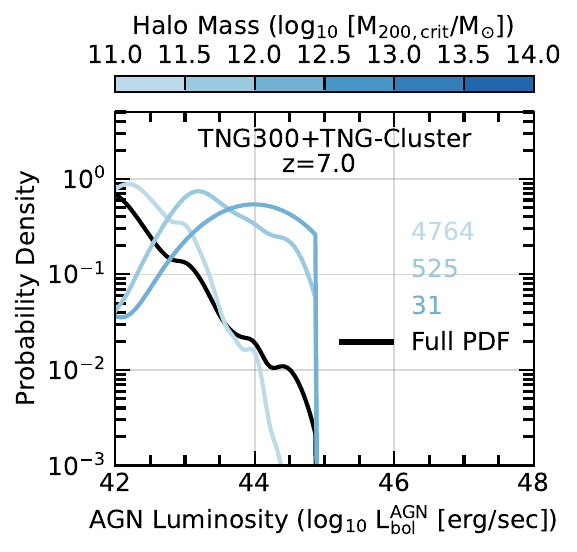}
    \includegraphics[width=0.28\linewidth]{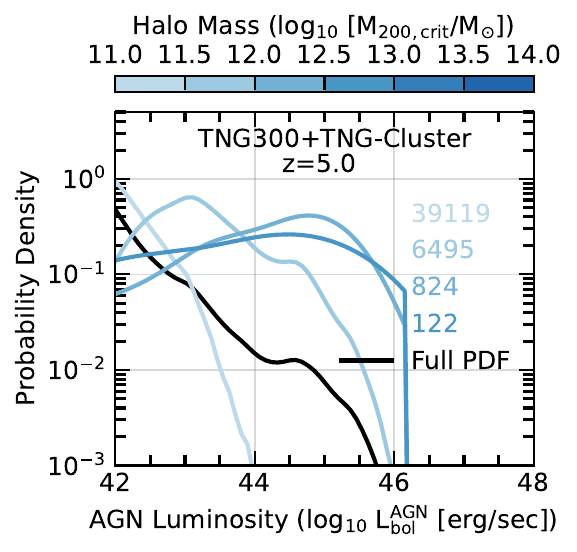}
    \includegraphics[width=0.28\linewidth]{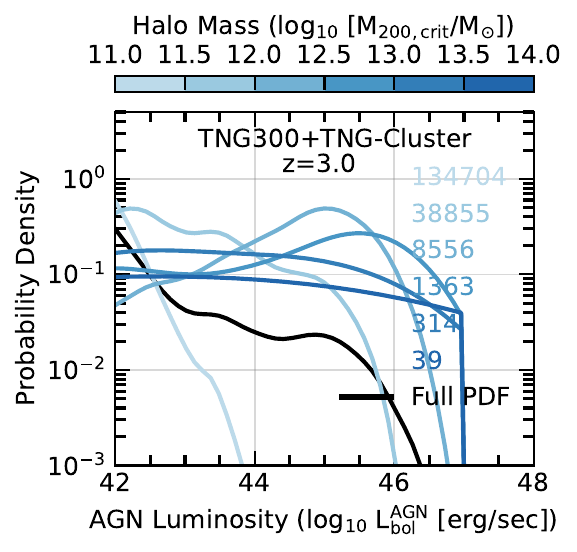}
     \includegraphics[width=0.28\linewidth]{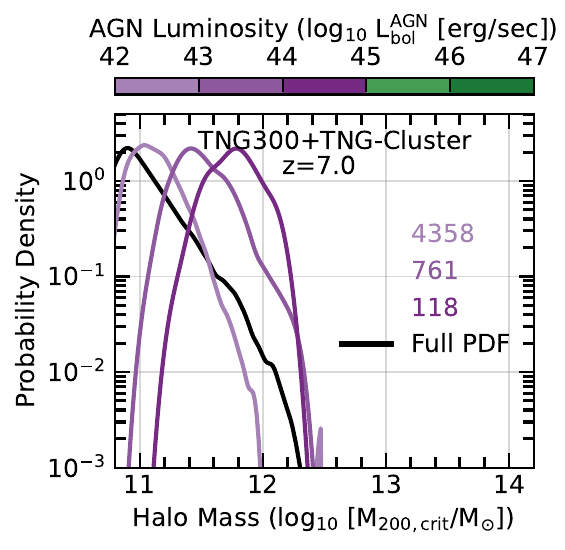}
 \includegraphics[width=0.28\linewidth]{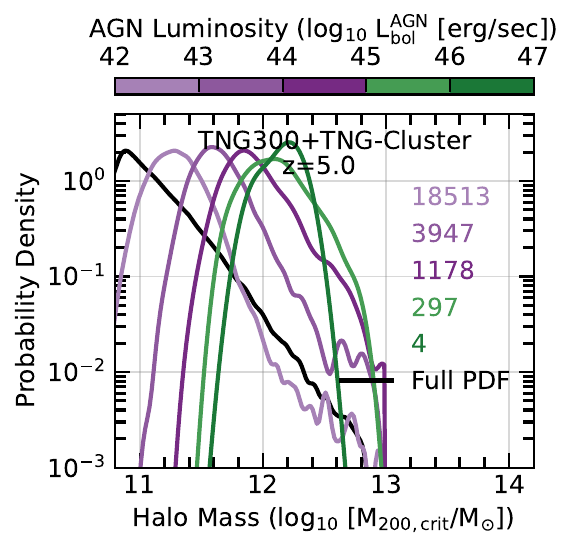}
   \includegraphics[width=0.28\linewidth]{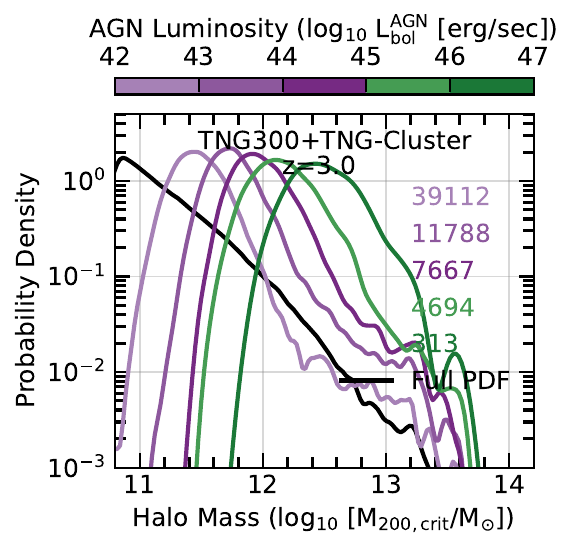}
  \caption{\textbf{Relationship between AGN bolometric luminosity and host halo mass predicted by the \texttt{TNG300} and \texttt{TNG-Cluster} simulations, before Cosmic Noon.} In the top panel, we focus on all simulated objects with AGN bolometric luminosity $\geq 10^{42}$ erg~s$^{-1}$ and plot with solid curves the median AGN bolometric luminosity in bins of halo mass (width 0.2 dex), shown between the 0.1 and 99.9th percentiles of the halo mass distribution and for $z=3-7$ (different colors). Shaded regions represent the 5th to 95th percentile range of AGN bolometric luminosities within each halo mass bin. Individual markers represent individual AGN: we show all AGN at $z=3$ with red small circles and highlight the 10 most massive (circles) and the 10 most luminous (squares) systems at each redshift. In the second row of the figure we quantify the conditional KDE probability of obtaining a certain AGN bolometric luminosity given a certain host halo mass (similar to the conditional luminosity function of quasars in clustering models), and the reversed conditional probability of obtaining a certain host halo mass given a certain AGN bolometric luminosity in the bottom panels. In both cases, we show results from $z=7$ to 3, from left to right. In these panels, no explicit selection is applied to the simulated populations and the black PDFs are the respective probabilities of the full sample, across all possible luminosities or masses. Annotations denote the number of objects in each luminosity or mass bin, at each redshift. For clarity, we put the KDE to zero manually in bins with no simulated objects. Thanks to their volume, \TNG can sample a few bright quasars as luminous as $10^{46}$ erg~s$^{-1}$ as early as $z\sim5$ and hundreds of faint quasars around Cosmic Noon. According to \TNG, the median relations between AGN bolometric luminosity and host halo mass (top panel) evolve only mildly with redshift and show that the luminosity of AGN increases with halo mass before flattening or even dipping down for halo masses $\gtrsim 10^{12}$ (solid curves). This is due to the effects of the AGN feedback in the \illustrisTNG model, which quenches star-formation and also SMBH gas accretion already at these high redshifts. Both the distributions of points in the top panel and the PDFs demonstrate that the object-to-object variations (i.e. the scatter around the median relation) can be very large, from one to a few orders of magnitude: we expand on this in Fig.~\ref{Fig:scatter_tng_flamingo}.  
  All this demonstrates the strong non-linearity and the large scatter of the AGN bolometric luminosity -- host halo mass relation, suggesting that halo mass alone may play only a partial role in setting the {\it instantaneous} luminosity of AGN. However, the shape of the relation at the low-mass end should be interpreted with caution, as its slope and normalization are affected by the imposed threshold on AGN bolometric luminosity.}
\label{fig:lum_vs_halo_tng}
\end{figure*}

\begin{figure*}
\centering
    \includegraphics[width=0.62\textwidth]{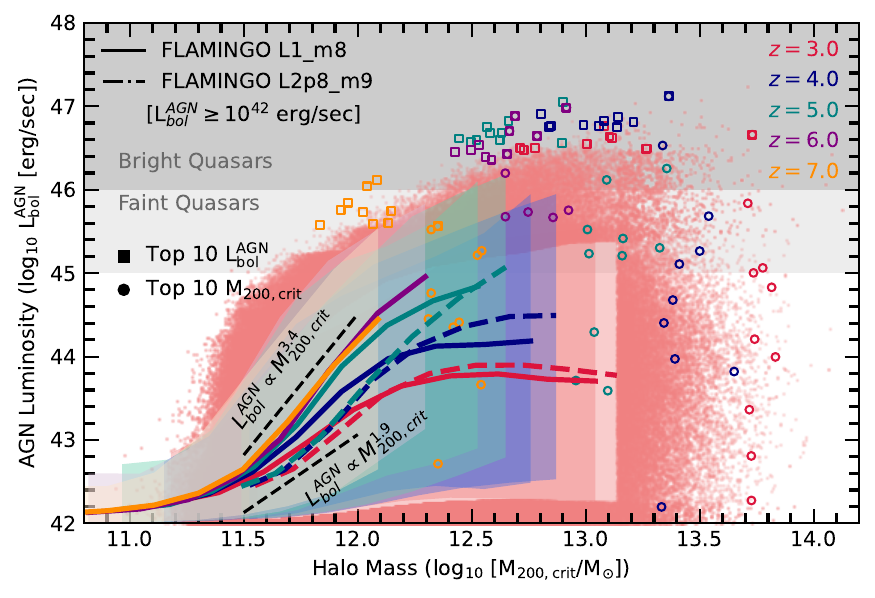}\\
       \includegraphics[width=0.28\linewidth]{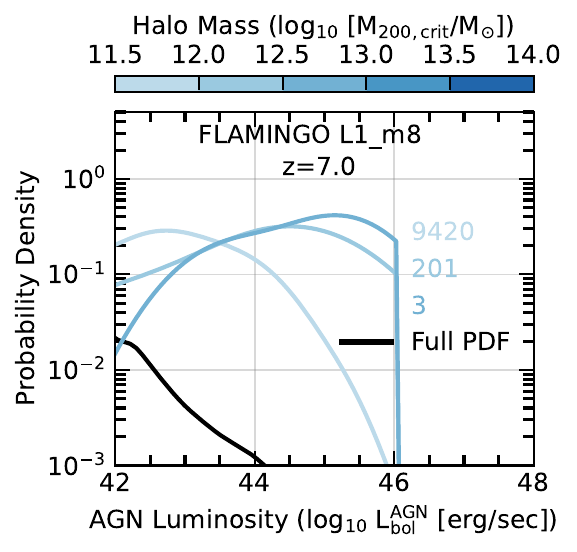}
 \includegraphics[width=0.28\linewidth]{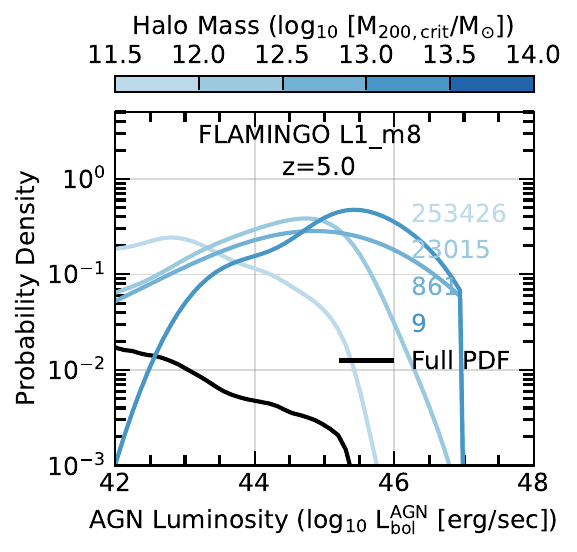}
   \includegraphics[width=0.28\linewidth]{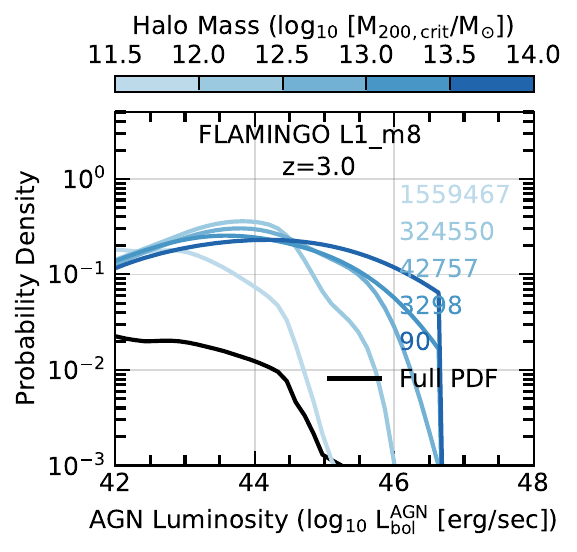}  
        \includegraphics[width=0.28\linewidth]{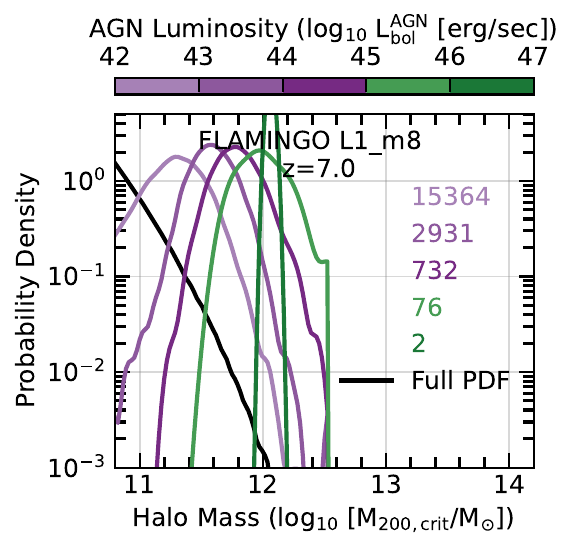}
 \includegraphics[width=0.28\linewidth]{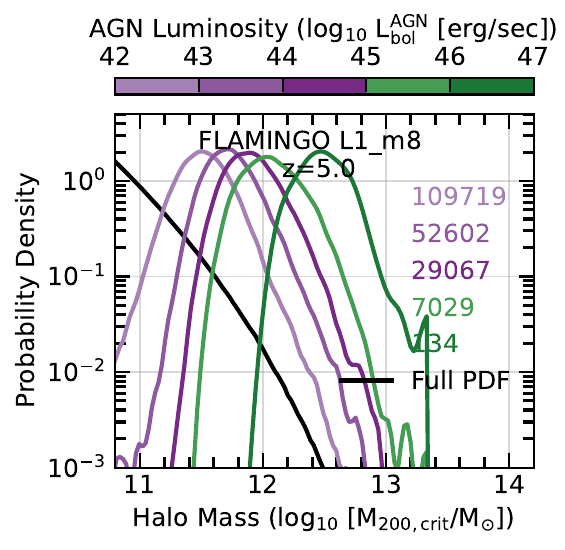}
   \includegraphics[width=0.28\linewidth]{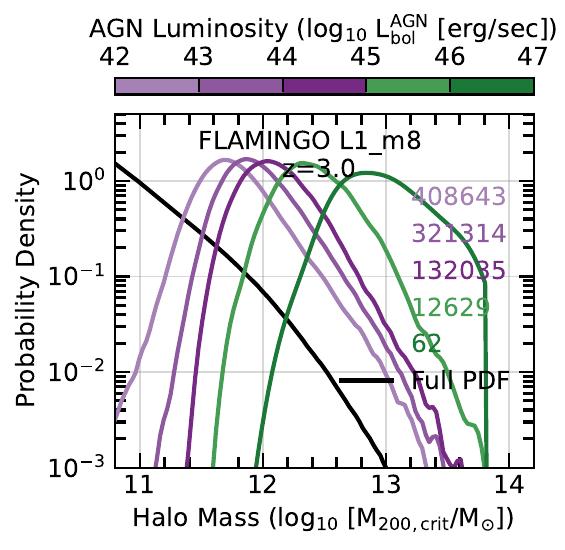} 
 \caption{\textbf{Same as Fig.~\ref{fig:lum_vs_halo_tng}, i.e. relationship between AGN bolometric luminosity and host halo mass before Cosmic Noon, but here predicted by the \texttt{FLAMINGO} simulations.} Annotations and plotting choices are as in Fig.~\ref{fig:lum_vs_halo_tng}, with the difference that the top panel shows median curves for the \flone and \fltwo runs separately (both in the {\it fiducial} \flamingo model): solid and dashed curves, respectively. For \fltwo, only $z=3$, 4, and 5 are available. Markers denote individual objects, here from \flone alone: as in Fig.~\ref{fig:lum_vs_halo_tng} small red circles showcase all AGN but only at $z=3$, for clarity. The middle and bottom panels focus exclusively on the higher-resolution \flone run. The large volumes of the \flamingo simulations provide much-improved statistics, with thousands of quasars sampled at $z\leq5$. This occurs despite the fact that, unlike in \TNG, AGN bolometric luminosities in \flamingo show a somewhat shallower dependence on host halo mass, on average, particularly at $z\lesssim 5$ and around the critical mass scale of $10^{12}$\msun, where AGN are one to two orders of magnitude fainter. As in \TNG, the \flamingo runs predict a flattening of the AGN bolometric luminosity -- host halo mass relation at the high-mass end, but without the drop seen in the highest-mass bins. This is reflected in the large widths of the conditional probability distributions in the lower half of the figure. As in \TNG, the most luminous AGN i.e. quasars rarely reside in the most massive halo at any given studied epoch.}
 \label{fig:lum_vs_halo_fl}
\end{figure*}

This is quantified in Figs.~\ref{fig:lum_vs_halo_tng} and \ref{fig:lum_vs_halo_fl}, where we show how the populations of simulated SMBHs occupy the AGN bolometric luminosity vs. host halo mass plane at $z=3-7$ for \TNG and \flamingo, respectively. In this section, we intentionally explore the relation across a wide range of AGN bolometric luminosities, encompassing observable values and those within the limits of the simulation data ($\Lbol \geq 10^{42}$ and up to about $ 10^{47-47.5}$~erg~s$^{-1}$). In Section~\ref{sec:results_quasars}, we will instead focus on quasars alone ($\geq 10^{45}$~erg~s$^{-1}$) and provide what simulations return in terms of their halo mass redshift evolution. Furthermore, we first focus on the results of \TNG and \flamingo (Sections~\ref{sec:results_plane_relation}, \ref{sec:results_plane_zfits}, \ref{sec:results_plane_trendsofpeaks}, and \ref{sec:results_plane_scatter}), to then contextualize these novel results with what expected from other, less recent cosmological simulations (Section~\ref{sec:results_compare}).

\subsection{A widely and richly populated $\Lbol$ vs. $\Mtwohcrit$ plane}
In particular, Fig.~\ref{fig:lum_vs_halo_tng} presents the AGN bolometric luminosity -- host halo mass relation according to the combined \TNGthreeh and \TNGcluster simulations (details in Section~\ref{sec:methods_sims_big}). The top panel provides the medians of AGN bolometric luminosity in bins of host halo mass (0.2 dex) at redshifts $z=3,4,5,6,$ and 7 (solid curves), with shaded areas encompassing the 5th to 95th percentiles of luminosity in each halo mass bin. Individual markers denote individual simulated objects: red small circles show all AGN at e.g. $z=3$ whereas empty circles and squares are the 10 most massive and most luminous systems, respectively, at each redshift. The bottom two rows show the Kernel Density Estimates (KDE) of the conditional probability density functions (PDFs) of obtaining an AGN bolometric luminosity given a host halo mass (middle) and, conversely, of obtaining a host halo mass given an AGN bolometric luminosity (bottom), at three example redshifts ($z=7, 5, $ and 3, from left to right). It should be noted that, whereas in the top panel the limit of $\Lbol \gtrsim 10^{42}$ erg~s$^{-1}$ is applied, the conditional probabilities are shown without any selection on the simulations data, to convey the width of the whole underlying population. In practice, the middle panels represent the expected conditional luminosity functions of AGN according to \TNG, invoked in many models for the clustering of quasars \citep[e.g.][]{Pizzati2024b}.

Fig.~\ref{fig:lum_vs_halo_fl} is the analog overview for the \flamingo simulations adopted in this work (details in Section~\ref{sec:methods_sims_big}). In the top panel, the \flone and \fltwo results are shown separately (solid and dashed curves, respectively), returning a reasonable consistency at the available overlapping redshifts. The conditional probability plots are shown here only for the higher-resolution run \flone, for brevity. We have checked that results from the \flamingo \fltwo are overall consistent (see Appendix~\ref{app:allsims}): we will comment throughout the text about possible differences.

Remarkably, thanks to their large volumes,\TNG (\flamingo \flone) can sample hundreds (thousands) of quasars ($\Lbol>10^{45}$) as early as at $z\sim5$, with such numbers increasing by a factor of ten (a few) by $z=3$. For the \flamingo \fltwo run, these statistics are even more robust and we count (but do not show) e.g. 2, 6 and 3 quasars as luminous as $\Lbol\geq10^{47}$~erg~s$^{-1}$ at $z=5, 4$, and 3, respectively. All this occurs despite the fact that, as a result of the underlying \illustrisTNG and \flamingo models, the average relationship of AGN bolometric luminosity as a function of host halo mass flattens towards the massive end and amid the fact that the typical \flamingo AGN are less luminous than those from \TNG, in most of the considered redshift and halo mass ranges. 

\subsection{Non-linear, non-monotonic relations and high-mass quenching}
\label{sec:results_plane_relation}

The median relationships between AGN bolometric luminosity and host halo mass of Figs.~\ref{fig:lum_vs_halo_tng} and \ref{fig:lum_vs_halo_fl}, top panels, are clearly non linear and not necessarily strictly monotonic. 

According to both the \TNG and \flamingo suites, more massive haloes host more luminous AGN at their centers: this is the case at all studied redshifts ($z=3-7$) but only up to halo masses of about $10^{12}$\msun. For more massive haloes than that, such an average dependence flattens: this is particularly manifest e.g. for the \flamingo runs at e.g. $z=3$ and 4 (red and blue curves, respectively), whereby the typical AGN luminosity of haloes with $\Mtwohcrit \sim 10^{12.2}$ \msun is the same as those that are almost one order of magnitude more massive. The flattening of the median AGN bolometric luminosity vs. host halo mass relation predicted by the \TNG and \flamingo simulations starts to appear at $z\lesssim 6$ and, in the case of \TNG, such a median relation even turns over, with the most massive haloes of \TNG (e.g. $\Mtwohcrit\gtrsim10^{12.5-13}$\msun) hosting slightly (up to 1 dex) less luminous AGNs on average than those at the transitional mass of about $10^{12}$\msun at $z=4$ (at $z=3$). In fact, as the individual circle and square markers show, the most luminous AGNs i.e. quasars rarely reside in the most massive haloes at any epoch.

Based on previous studies of SMBH feedback in \illustrisTNG and \flamingo galaxies \citep[e.g.][]{Weinberger-feedback, Flamingo-Schaye2023}, our understanding of the \illustrisTNG and \flamingo fiducial models, and our analysis of the SMBH populations presented in Appendix~\ref{app:SMBHpopulations}, we attribute the flattening and turnover of the median AGN bolometric luminosity versus host halo mass at halo masses $\gtrsim10^{12}$~\msun to SMBH feedback. This feedback effectively suppresses gas accretion onto SMBHs, while also reducing star formation in massive galaxies. Energy injection from SMBHs -- kinetic in the case of \illustrisTNG and thermal in the \flamingo fiducial model used here -- drives winds and outflows that expel gas from the central regions of galaxies, thereby limiting further gas accretion onto the SMBHs. We return to this interpretation throughout the analysis and discussion.

In fact, whether the high-mass end of the predicted relationships between AGN bolometric luminosity and host halo mass is conceptually well described by a running median remains a matter for further study and debate. In both the \TNG and \flamingo simulation suites, the bending (and even the downward turnover) of the median relations is driven by a fraction of SMBHs with suppressed gas mass accretion (possibly due to depleted gas supplies) that grows with halo mass and toward lower redshift. However, it is not completely evident from the top panels of Figs.~\ref{fig:lum_vs_halo_tng} and \ref{fig:lum_vs_halo_fl} whether the SMBHs in high-mass haloes form a single population with very diverse AGN bolometric luminosities, or whether a main population with a monotonic trend exists that is skewed by AGN falling off this main sequence. Moreover, the exact extent to which the median relations bend depends on the adopted selection (not shown), redshift, and galaxy-formation model, with debates similar to those concerning the star-forming main sequence at the high-mass end of the galaxy stellar mass function \citep[see e.g.][for a discussion in the context of the \illustrisTNG model]{Donnari2019}. From the shapes of the PDFs in Figs.~\ref{fig:lum_vs_halo_tng} and \ref{fig:lum_vs_halo_fl} and our knowledge of the \TNG and \flamingo galaxy populations, it would appear that a {\it main sequence} of highly-active AGNs is less pronounced (i.e. with less dominant peaks of AGN bolometric luminosity at fixed host halo mass) than for the (specific) star formation rates of galaxies, even if different groupings of highly-accreting and low-accreting AGN clearly populate the high-mass end, at least for \TNG (see Appendix~\ref{app:SMBHpopulations}, Fig.~\ref{fig:app_SMBHs_2}).

It should be noted from the outset that the precise locus (i.e. normalization) of the median AGN bolometric luminosity vs. host halo mass relation naturally depends on the considered sample of SMBHs (and haloes or galaxies). For example, if AGN with bolometric luminosities lower than $\sim 10^{42}$ erg~s$^{-1}$ were included (see Section~\ref{sec:methods_selection}), the {\it median} relations emerging from the \TNG and \flamingo simulations would necessarily shift to lower values, owing to the vast numbers of low-accreting and low-mass SMBHs populating mostly low-mass galaxies and haloes (see Appendix~\ref{app:SMBHpopulations}, Fig.~\ref{fig:app_SMBHs_2} and solid black curves therein). Conversely, higher-luminosity cuts would shift the median relation at the low-mass end to larger values. Similarly, alternative selections applied to the underlying populations of galaxies, SMBHs, or haloes that exclude systems at the low-mass or low-luminosity end would also modify the overall normalization of the median relations. For analogous reasons, the shape of the relation at the low-mass end of Figs.~\ref{fig:lum_vs_halo_tng} and \ref{fig:lum_vs_halo_fl} should be interpreted with caution, as its flattening arises from the imposed AGN bolometric luminosity cut  (with flatter relations with higher thresholds in AGN bolometric luminosity). Toward the high-mass end, by contrast, the presence of the flattening (and of the turnover in \TNG) is less sensitive to the details of the selections, provided sufficiently massive or luminous galaxies, haloes, or AGNs are included. 

\begin{table}
\caption{Fits to the median relation between AGN bolometric luminosity and host halo mass assuming a power-law dependence: $\Lbol = L_0(\Mtwohcrit/10^{12.5}$\msun)$^{\gamma}$. We provide results at different redshifts for the \TNG (upper 5 rows) and the \flamingo \flone runs (bottom 5 rows). The fit is performed exclusively in the halo mass range $\Mtwohcrit = 10^{11.5-12}$, to avoid the regimes affected by the adopted AGN-luminosity selection (towards the low-mass end) and by the high-mass quenching (whereby the shape of the relationship is starkly different). The last two columns provide the spread in AGN bolometric luminosities for this halo mass range, between the 16th-84th and 5th-95th percentiles ranges. According to both \TNG and \flamingo, the relationship between AGN bolometric luminosity and host halo mass is largely non-linear in this halo mass range, but with opposite trends of steepness with decreasing redshift. }
\begin{center}
\begin{tabular}{c c c c c}   
\hline
Redshift &
log$_{\rm 10}$ L$_{0}$& 
$\gamma$
& $\Delta$ log$_{\rm 10} \Lbol$
& $\Delta$ log$_{\rm 10} \Lbol$\\
 &[erg~s$^{-1}$]&&(p84-p16) [dex] &(p95-p5) [dex]\\
\hline
&&&\\
\multicolumn{3}{l}{\TNG} &&\\
&&&\\
3.0 &46.7&4.7&1.8&2.8\\
4.0&45.9&3.6&1.6&2.6\\
5.0&45.7&3.0&1.4&2.4\\
6.0&45.3&2.3&1.3&2.1\\
7.0&45.0&1.0&1.2&1.9\\

&&&\\
\multicolumn{3}{l}{\flamingo \flone} & &\\
&&&\\

3.0 &44.3&1.9&1.4&2.1\\
4.0&44.9&2.4&1.6&2.4\\
5.0&45.7&3.2&1.7&2.6\\
6.0&46.1&3.6&1.7&2.6\\
7.0&45.9&3.4&1.5&2.2\\
\hline

\end{tabular}
\label{Tab:fits}
\end{center}
\end{table}

\subsection{Redshift evolutions and power-law fits}
\label{sec:results_plane_zfits}

So far we deliberately treated the full redshift range together, as the general trends and considerations described above remain broadly consistent across the $\sim2$ billion years of cosmic evolution studied here. Differences between $z=3$ and $z=7$ are overall modest, although the \TNG and \flamingo runs predict somewhat different redshift trends in detail.

Within the adopted selection, \TNG return a steady and monotonic redshift dependence at the low-mass end (host halo masses $\lesssim 10^{11.5-12}$\msun), with lower average AGN bolometric luminosities at fixed halo mass at lower redshifts (different colored curves in the top panel of Fig.~\ref{fig:lum_vs_halo_tng}). In contrast, the \flamingo runs exhibit little-to-no redshift dependence in this regime. At larger host halo masses, the \flamingo runs display stronger redshift dependencies, especially at $z\lesssim 5$, again with lower average AGN bolometric luminosities at lower redshifts — a trend that is inverted in \TNG in the same mass range (colored curves in the top panels of Fig.~\ref{fig:lum_vs_halo_tng} vs. Fig.~\ref{fig:lum_vs_halo_fl}).

As noted above, in both the \TNG and \flamingo runs, the flattening and downward turnover of the AGN bolometric luminosity -- host halo mass relation become progressively more evident toward lower redshift, as larger fractions of AGN and galaxy populations at the high-mass end quench. 

We further summarize the AGN bolometric luminosity -- host halo mass relations and their redshift dependencies in Table~\ref{Tab:fits}: there we provide the best-fit parameters of the power-law description of the medians, i.e. assuming $\Lbol = L_0(\Mtwohcrit/10^{12.5}$\msun)$^{\gamma}$, with $L_0$ and $\gamma$ being the normalization and power-law slope, respectively \citep[see e.g.][]{Pizzati2024b}. In fact, we perform the fit only in the narrow range of host halo masses that is not directly affected by our AGN-luminosity selection and is not shaped by the high-mass quenching, so in the range $\Mtwohcrit=10^{11.5-12}$ \msun. The functional forms reported in Table~\ref{Tab:fits} could be useful to inform future AGN and quasar population models.

The redshift trends of the best-fit values for the normalization $L_0$ confirm what enunciated above for the \flamingo runs: lower AGN bolometric luminosities at lower redshifts for host haloes with $10^{11.5-12}$ \msun. On the other hand, the situation for \TNG is complicated by the inversion of the trend of AGN bolometric luminosities with redshift at fixed halo mass precisely around $\Mtwohcrit\lesssim10^{12}$\msun. By comparing the best-fit values between \TNG and the \flamingo runs, it is also manifest what emerged by visual inspection: at $z=3$ ($z=4$) the typical luminosities of AGNs hosted by haloes of $10^{11.5-12}$ \msun are more than 2 (1) orders of magnitude lower in \flamingo \flone than in \TNG, such difference lessening (and inverting) towards higher redshifts. 

The redshift trends of the best-fit values for the power-law slope $\gamma$ quantify that, at host halo masses immediately before the bending, the increase in AGN bolometric luminosity with halo mass is stronger (i.e. the power-law slopes are larger i.e. the relations steeper) towards lower redshifts for \TNG. However, the opposite is the case for \flamingo \flone. At $z=3$, for example, the median relation between AGN bolometric luminosity and host halo mass is much steeper in \TNG than in \flamingo \flone. The same qualitative results apply also to \flamingo \fltwo.

We emphasize that these power-law fits describe population-level trends and further study would be needed to extract how individual AGNs evolve in the plane of AGN bolometric luminosity vs. host halo mass. We find that, while most AGNs appear to undergo an initial growth phase that follows a power-law relation, they generally deviate from it at the high-mass end, as feedback effects become increasingly important. To illustrate this behaviour, Appendix~\ref{App:tracks} presents the evolutionary histories of the brightest and most massive systems in our $z=3$ sample from \TNG.

\begin{figure*}
\centering
     \includegraphics[width=0.43\linewidth]{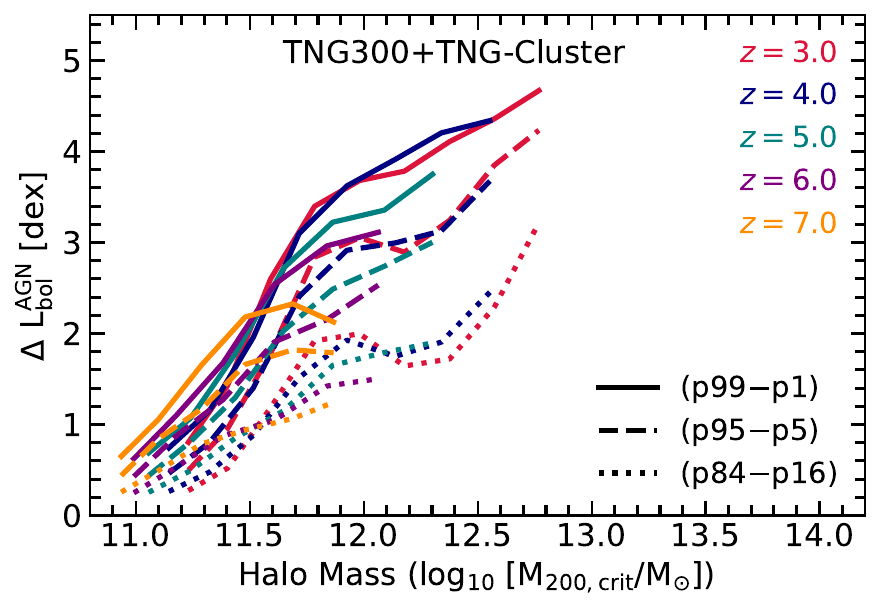}
     \includegraphics[width=0.43\linewidth]{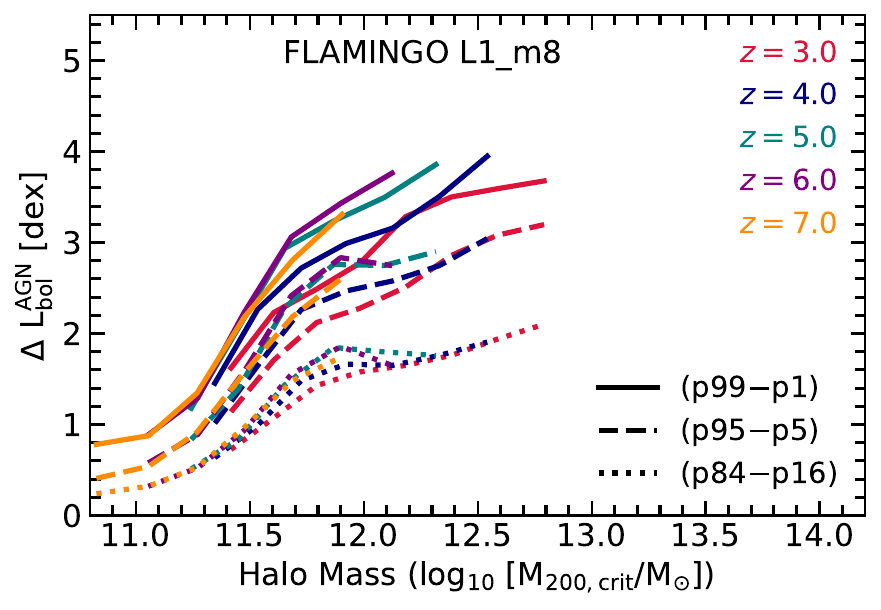}
     \includegraphics[width=0.43\linewidth]{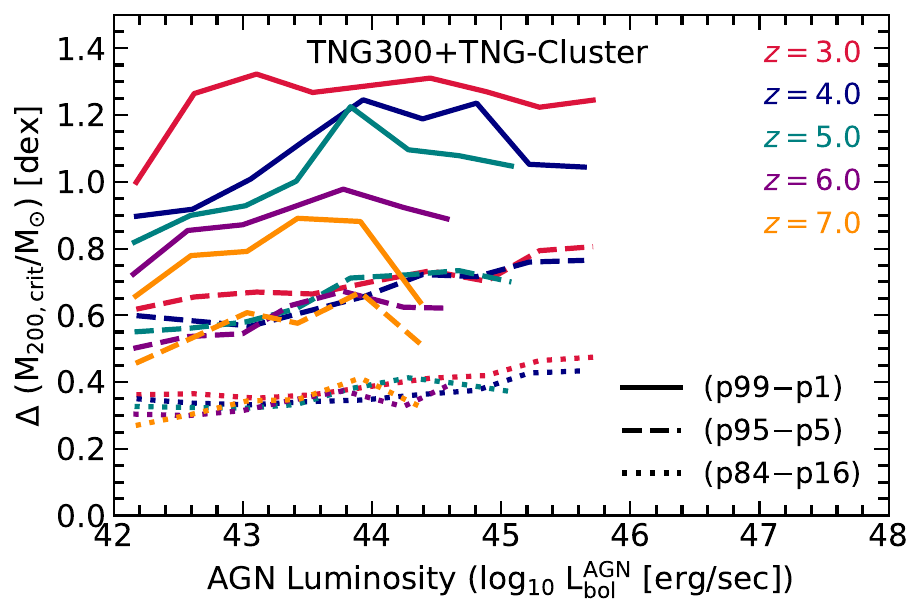}
     \includegraphics[width=0.43\linewidth]{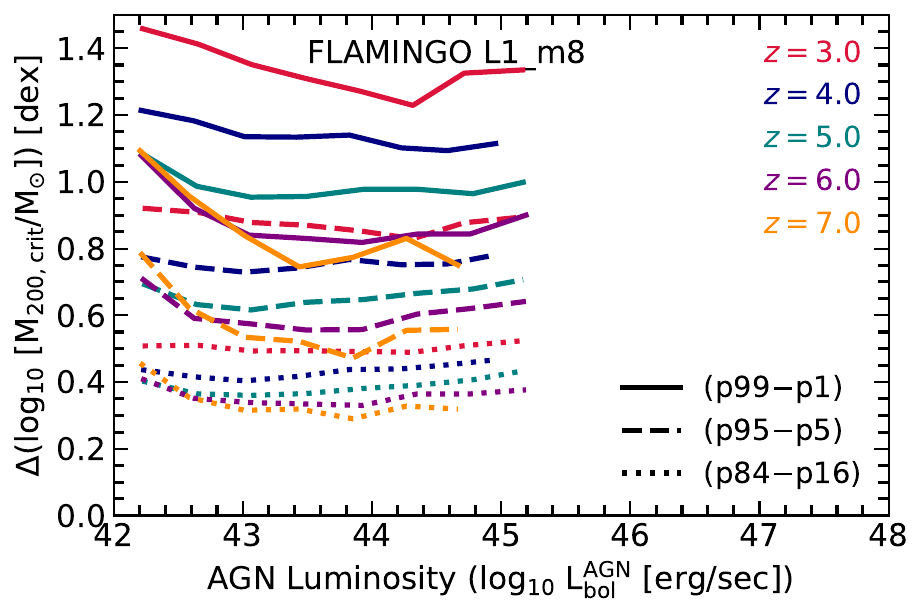}

 \caption{\textbf{Quantification of the scatter(s) between host halo mass and AGN bolometric luminosity according to the \TNG and \flamingo simulations.} With this figure, in practice we provide summary statistics of the relations and PDFs shown in Figs.~\ref{fig:lum_vs_halo_tng} and \ref{fig:lum_vs_halo_fl} for \TNG (left) and the \flamingo \flone run (right), respectively. The top row describes the scatter in AGN luminosity as a function of halo mass while the bottom row shows the scatter in  halo mass as a function of AGN luminosity. The solid, dashed, and dotted curves denote the 99th-1st, 95th-5th, and 84th-16th percentile differences, respectively. According to both simulation suites, the scatter in luminosity, given a host halo mass, is much larger and varies much more strongly with halo mass (top row) than the scatter in halo mass given an AGN luminosity (bottom row). Additionally, there is only a mild evolution with redshift in the \illustrisTNG model but a comparatively stronger redshift evolution in \flamingo. Taken together, these results indicate that AGN luminosity may be highly stochastic and determined by more than just the host halo mass. As a caveat, the apparent low scatter in AGN luminosities at the low halo-mass end arises from the adopted selection, which removes the large population of low-luminosity AGNs in this regime.}
 \label{Fig:scatter_tng_flamingo}
\end{figure*}

\subsection{Trends of AGN luminosities and of halo masses}
\label{sec:results_plane_trendsofpeaks}

What discussed so far, i.e. the median trends of AGN bolometric luminosity as a function of host halo mass at various redshifts, is also visualized and captured in the middle rows of Figs.~\ref{fig:lum_vs_halo_tng} and \ref{fig:lum_vs_halo_fl}, namely in the peaks of the conditional probabilities of obtaining a certain AGN bolometric luminosity given a certain host halo mass. Whereas we quantify the large widths of the distributions in Section~\ref{sec:results_plane_scatter}, here we pause on two considerations. 

First, the PDFs of the full distribution of AGN bolometric luminosities (black curves in the middle rows of Figs.~\ref{fig:lum_vs_halo_tng} and \ref{fig:lum_vs_halo_fl}, with no explicit selection on the simulated and available systems, i.e. essentially the SMBH distribution) show that, for both \TNG and the \flamingo runs, the most frequent AGN bolometric luminosities in the simulations lie far below the adopted observable minimum cut of $10^{42}$ erg~s$^{-1}$ (and below the plotted ranges). 
Second, as anticipated above, there is no dominant peak of AGN bolometric luminosities in the highest host halo mass bins toward lower redshifts (darker blue curves, e.g. in the $z=3$ rightmost middle panels), with flat distributions spanning orders of magnitude in luminosity.

But what about the median trends of host halo mass as a function of AGN bolometric luminosity? Due to the large system-to-system variations across the $\Lbol$–$\Mtwohcrit$ plane, the median relations shown in the top panels of Figs.~\ref{fig:lum_vs_halo_tng} and \ref{fig:lum_vs_halo_fl} do not necessarily coincide with their inverse ones, namely the median host halo mass in bins of AGN bolometric luminosity. We have checked (but do not show, for brevity) that such curves are indeed distinct from the solid curves of Figs.~\ref{fig:lum_vs_halo_tng} and \ref{fig:lum_vs_halo_fl}, and that for both the \TNG and the \flamingo runs they appear steeper (almost vertical) and exhibit stronger redshift dependencies. 
The distributions of host halo masses in the bottom panels of Figs.~\ref{fig:lum_vs_halo_tng} and \ref{fig:lum_vs_halo_fl} indeed suggest a mild increase in the typical (i.e. peak) halo masses hosting AGNs of increasing luminosity, with variations in median halo mass of up to 1.5 dex in both galaxy-formation models for AGNs spanning five orders of magnitude in luminosity. Compared to the PDFs in the middle rows, the conditional probabilities of host halo mass at fixed AGN bolometric luminosity are much narrower; that is, AGNs of given bolometric luminosities reside within specific ranges of the global host halo mass distribution (i.e. the underlying halo mass distribution represented here by the black solid PDFs). In the \flamingo runs, at all redshifts, the luminosity PDFs appear to peak at slightly larger halo masses than in \TNG and, barring the most luminous quasars at $z=3$, these are $\sim10^{11.5-12.5}$\msun. This likely corresponds to the halo mass associated with the \textit{rapid growth phase} of SMBHs in the \flamingo model. Since the transition into this \textit{rapid growth phase} occurs at approximately the same virial temperature, this characteristic halo mass shifts somewhat to higher values with redshift \citep[see Figure~5 of][]{Eagle-rapid-growth2018}. The luminosity dependence of the halo masses of these quasar populations is consistent with the findings of \cite{flamingo_quasar_clustering}.
We further quantify the redshift evolution of the host halo masses of quasars according to \TNG and \flamingo in Section~\ref{sec:results_quasars}, whereas the analysis of the halo, dynamical, and stellar masses of hosts of less luminous high-redshift AGN is postponed to dedicated future work.

Finally, before delving into the extent of the system-to-system variations predicted by the simulations (Section~\ref{sec:results_plane_scatter}), we remind the reader that any adopted selection (here $\Lbol \geq 10^{42}$ erg~s$^{-1}$) will necessarily and differently modulate the distributions and trends presented in Figs.~\ref{fig:lum_vs_halo_tng} and \ref{fig:lum_vs_halo_fl}; this is true both for simulated and observational data. 
Furthermore, the results presented here are inherently constrained by the adopted simulation resolution and the assumptions of the underlying galaxy-formation models. For example, the SMBHs simulated in all runs of Table~\ref{Tab:simulations} cannot capture the expected short-timescale variability in AGN instantaneous accretion rates, which is thought to be driven by small-scale physics that transports gas from the central regions of the galaxy down to the SMBH’s sphere of influence. As shown in Figure~7 of \cite{flamingo_quasar_clustering} and confirmed independently by us (but not shown here for brevity), including an artificial luminosity scatter reduces the luminosity dependence of host halo masses -- more so in the \flamingo runs than in \TNG, and particularly for brighter AGN/quasars. This suggests that the apparent trend may result from an undercounting of bright AGNs in low-mass haloes, also due to the imposed Eddington limit, leading to a preferential selection of sources in more massive haloes. We return to the questions of the Eddington limit and luminosity flickering in Section~\ref{sec:disc}.

\subsection{Broad conditional probabilities and large scatters}
\label{sec:results_plane_scatter}

From Figs.~\ref{fig:lum_vs_halo_tng} and \ref{fig:lum_vs_halo_fl}, particularly the PDFs in the middle and bottom panels, it is evident that, according to both the \TNG and \flamingo simulations, the bolometric luminosity of an AGN hosted by a halo of a given mass -- and, conversely but to a much lesser degree, the halo mass of a given-luminosity AGN -- can vary widely from halo to halo. 

Fig.~\ref{Fig:scatter_tng_flamingo} summarizes the magnitudes and trends in the scatter visualized in Figs.~\ref{fig:lum_vs_halo_tng} and \ref{fig:lum_vs_halo_fl}, for \TNG (left) and \flamingo \flone (right). The top panels show the scatter in AGN bolometric luminosity at fixed host halo mass, while the bottom panels show the scatter in host halo mass at fixed AGN bolometric luminosity, with different colors indicating different redshifts. Scatter is quantified as the difference between the 99th–1st (solid), 95th–5th (dashed), and 84th–16th (dotted) percentiles.

Firstly, the y-axis ranges of the top and bottom panels are largely different: namely, as the comparison of the middle to the bottom PDFs of Figs.~\ref{fig:lum_vs_halo_tng} and \ref{fig:lum_vs_halo_fl} already suggested, the variations in AGN bolometric luminosity given a host halo mass (up to $2-5$ dex across all studied redshifts, regimes and scatter definitions) are much larger than the variations in host halo mass for a given AGN bolometric luminosity ($0.3-1.5$ dex). In other words, host halo mass alone is a poor predictor of AGN bolometric luminosity; on the other hand, AGN bolometric luminosity is a much better predictor, and selector, of host halo mass, with 1-sigma scatters in host halo mass $\lesssim 0.5$ dex across all studied five orders of magnitude in luminosity.

The halo-to-halo variations are predicted to be broadly similar in the \TNG and \flamingo models, with \TNG exhibiting a slightly larger scatter in AGN bolometric luminosity at fixed host halo mass. 

According to both simulations suites, the scatter in AGN bolometric luminosity increases with host halo mass, as a result of gas suppression around SMBHs in high-mass haloes due to AGN feedback, as discussed earlier. In fact, we have checked that the steep increase at the low-mass end is mostly driven by the adopted observational cut of $\Lbol \geq 10^{42}$ erg~s$^{-1}$, i.e. the low-mass variations are \textit{artificially} underestimated due to the exclusion of lower-luminosity SMBHs. The dependence on redshift is mild, and maximal for \TNG (but not \flamingo) towards Cosmic Noon. 
On the other hand, the scatter in host halo mass varies only mildly (and negligibly) with AGN bolometric luminosity, for both models. In fact, we know that, at least for \TNG, such a scatter increases to a few dex for AGNs with luminosities lower than the adopted cut, because of the effects of SMBH quenching (see also Appendix~\ref{app:SMBHpopulations}). Moreover, the scatter in host halo mass increases towards lower redshifts, consistently according to \flamingo and especially towards the tails of the mass distributions for \TNG.

To quote some reference values, at the transitional mass scale of $\sim10^{12}$\msun, both \TNG and \flamingo \flone predict a 2-sigma scatter in AGN bolometric luminosity of about 2.5 dex (see also values reported in Table~\ref{Tab:fits}), which increases at the 1st-99th percentile level to $3-4$ dex and $2.8-3.5$ dex (across redshifts) for \TNG and \flamingo \flone, respectively. On the other hand, the 2-sigma scatter variations of the host halo masses of quasars is about 0.7 dex for \TNG and $0.5-0.9$ dex (across redshifts) for \flamingo \flone.

We remind the reader that the intra-population variations predicted by \TNG and \flamingo, and quantified throughout this paper, are {\it intrinsic}, i.e., free from e.g. observational uncertainty. Since we focus on SMBHs in central galaxies -- rather than including satellites, see Section~\ref{sec:methods_selection} -- these estimates likely represent a lower bound on the true intrinsic scatter of the full galaxy population. In Section~\ref{sec:disc}, we further discuss the possibility that the intrinsic halo-to-halo variations in the Universe may be even larger.

\begin{figure*}
\centering
 \includegraphics[width=0.95\columnwidth]{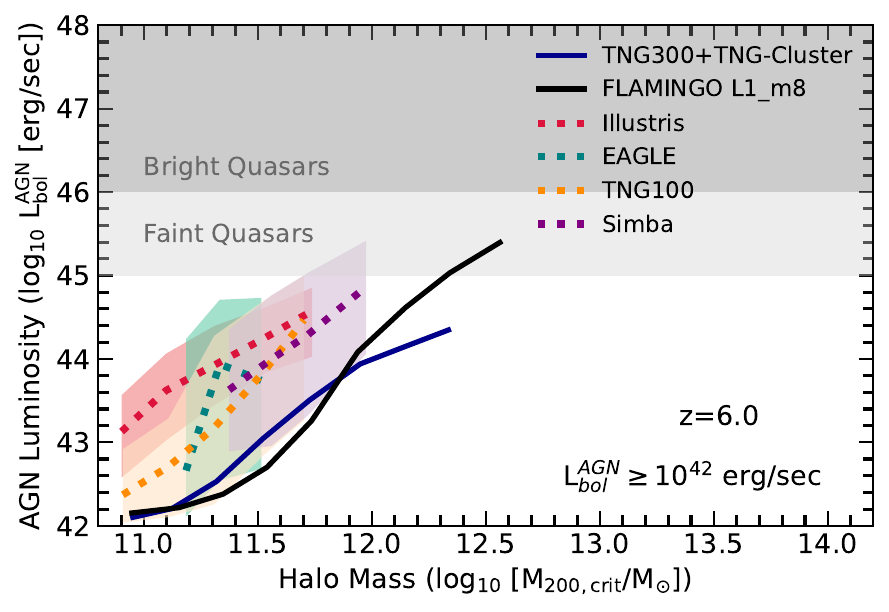}
    \includegraphics[width=0.95\columnwidth]{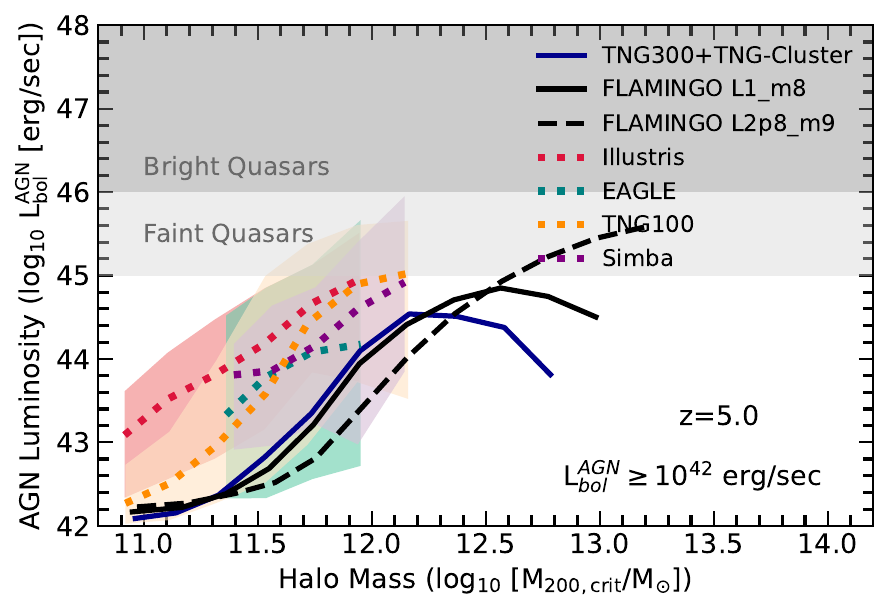}
    \includegraphics[width=0.95\columnwidth]{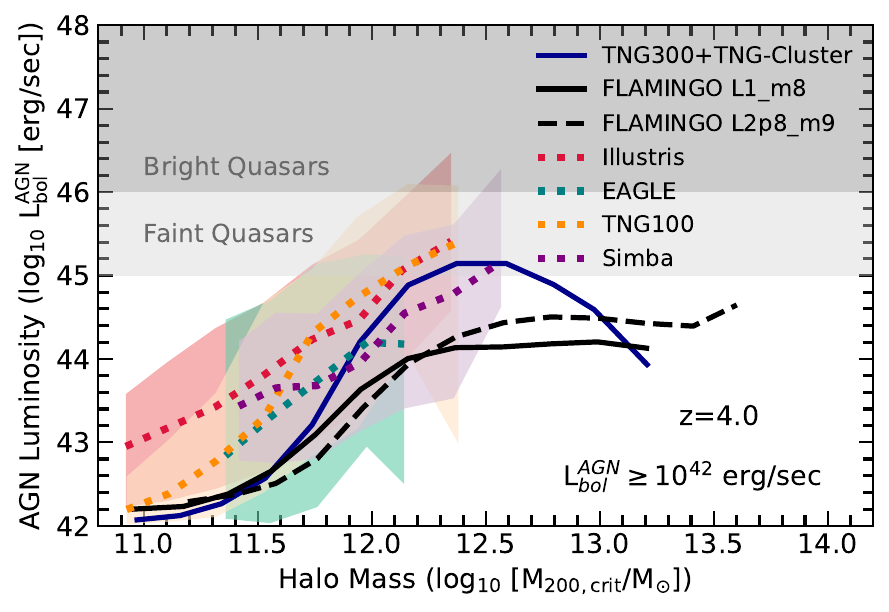}
    \includegraphics[width=0.95\columnwidth]
    {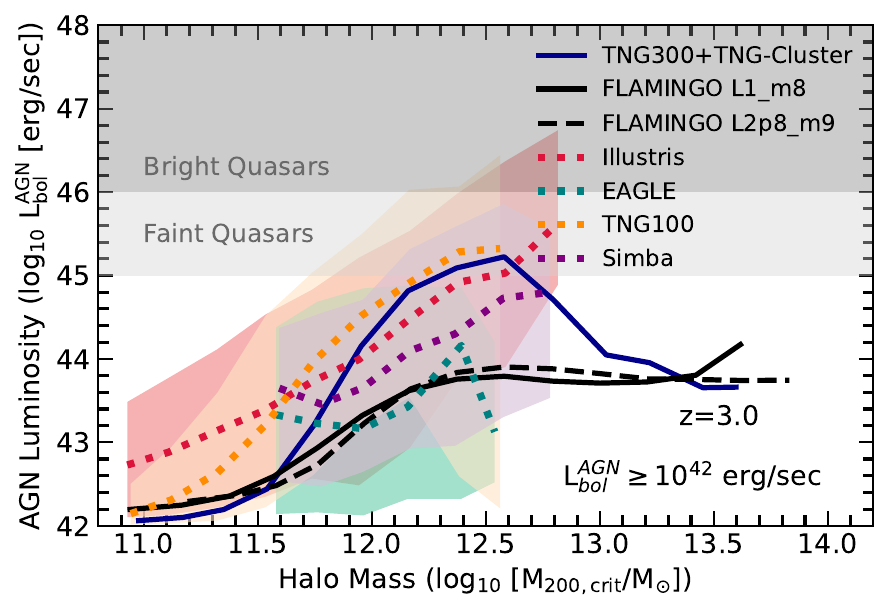}
\caption{\textbf{Relation between bolometric AGN luminosity and host halo mass predicted by the \illustris, \texttt{TNG100}, \eagle and \simba simulations, at various epochs before Cosmic Noon.} The AGN definition and selection ($\Lbol \geq 10^{42}$ erg~s$^{-1}$) and annotations are the same as for Fig.~\ref{fig:lum_vs_halo_tng}. 
The panels show the median bolometric luminosity trends at $z=6, 5, 4$, and $3$ (from top left to bottom right). For all runs consistently, the solid curves show the median AGN luminosity vs. the median host halo mass in bins of the latter of width 0.2 dex between $10^{10.8-14.2}$~\msun with at least 10 haloes per bin in the bin. The shaded regions represent the 5th to 95th percentile of $\Lbol$ range within each halo mass bin. The blue and black curves for the \TNG and the \flamingo boxes are similar to those of Figs.~\ref{fig:lum_vs_halo_tng} and \ref{fig:lum_vs_halo_fl}, top panels. Despite differences in seeding, accretion, and feedback models (and numerical resolution), all simulations show similar trends with varying normalizations. The halo-to-halo variations are large in all models, and particularly so in \eagle. Only the \TNG and \flamingo simulations, chiefly thanks to their larger volumes, reveal the high mass-end flattening.}
  \label{fig:lum_vs_halo_all_median}
\end{figure*}

\subsection{Comparing \TNG and \flamingo to additional cosmological simulations}
\label{sec:results_compare}

A comparison across all simulations of Table~\ref{Tab:simulations} immediately highlights that robustly probing the high-mass end of the AGN bolometric luminosity -- host halo mass relation requires sufficiently large cosmological volumes.

Fig.~\ref{fig:lum_vs_halo_all_median} shows the median AGN bolometric luminosities in 0.2~dex bins of host halo mass for the main simulation suites discussed so far (\TNG and \flamingo \flone and \fltwo in blue and black, respectively) together with analog results from \illustris, \eagle, \TNGoneh, and \simba, encompassing $(\sim100\ \mathrm{Mpc})^3$ comoving volumes (dotted curves). The median curves are shown for the same percentile ranges for all runs (1st to 99.5th percentiles). Shaded regions indicate the scatter in AGN bolometric luminosity (5th to 95th percentiles).

As manifest from Fig.~\ref{fig:lum_vs_halo_all_median}, only simulations such as \TNG and \flamingo sample halo masses above $10^{12-12.5}$~\msun in a statistically-meaningful way, whereas smaller-volume simulations do not adequately populate this regime at early times in cosmic evolution. In fact, we count 517, 149, 617, and 51 AGNs ($\Lbol \geq 10^{42}$ erg~s$^{-1}$) at $z=7$ in \illustris, \eagle, \TNGoneh, and \simba, respectively, in comparison to 5237 and 19105 in \TNG and \flamingo \flone, respectively. However, the majority of the AGNs of the smaller-volume simulations (but for \illustris) are at the low-luminosity end. Here we hence restrict our highest redshift to $z = 6$ unlike the previous analyses in the previous Sections. 

The behavior at the high-mass end and the manifestation of AGN quenching visible in \TNG and \flamingo (Section~\ref{sec:results_plane_relation}) cannot be assessed in these smaller simulations: namely, the flattening or turnover suggested by the large-volume runs is not clearly captured here. For the \illustrisTNG model, this is certainly due to the more limited volume of e.g. \TNGoneh compared to \TNG, as they share the same underlying modeling choices. Similarly is the case for \eagle compared to \flamingo, given the similarity of their underlying galaxy-formation model, and indeed some high-mass end flattening seems to be present, amid low number statistics, in \eagle at $z\lesssim4$.

Nevertheless, in the host halo mass range of $10^{11-12}$~\msun, across all runs, the median AGN bolometric luminosity -- host halo mass relation remains non linear, with similar power-law slopes across all runs especially at $z\gtrsim5$. According to all simulations, AGN bolometric luminosity increases by $2-3$~dex on average within 1~dex of host halo mass.

The redshift evolution between $z=6$ and $z=3$ remains overall mild, again in line with the trends discussed earlier, although differences in the detailed evolution persist across models. The differences in normalization are not negligible (up to $1-1.5$~dex) and yet within the large halo-to-halo variations -- the shaded regions (especially for \eagle and \simba and towards $z\lesssim4$) are very broad. 

The differences in scatter and median relation of Fig.~\ref{fig:lum_vs_halo_all_median}, and across redshift, naturally ensue from the different seeding prescriptions, growth and feedback of SMBHs, and overall star-formation and feedback models adopted in these simulations, which in turn lead to different AGN populations \citep[see e.g. also Fig.~2 of][]{Habouzit2022a}. Despite these variations, and variations across three orders of magnitudes in mass resolution and a factor of 30’000 in volume, the behavior of the AGN bolometric luminosity -- host halo mass relation is nevertheless broadly consistent across current state-of-the-art cosmological galaxy formation simulations.

More generally, as emphasized in the previous Sections, the median relations capture only a limited aspect of the underlying distributions. The substantial intrinsic halo-to-halo variations, only partially reflected in the median trends, are quantified and compared in detail in Section~\ref{sec:results_plane_scatter} and Appendix~\ref{app:allsims}.

\section{The halo masses of quasars according to \TNG and \flamingo}
\label{sec:results_quasars} %

\begin{figure*}
\centering
   \includegraphics[width=0.84\textwidth,  height=0.48\textheight,trim=0 0 0 0.5cm]{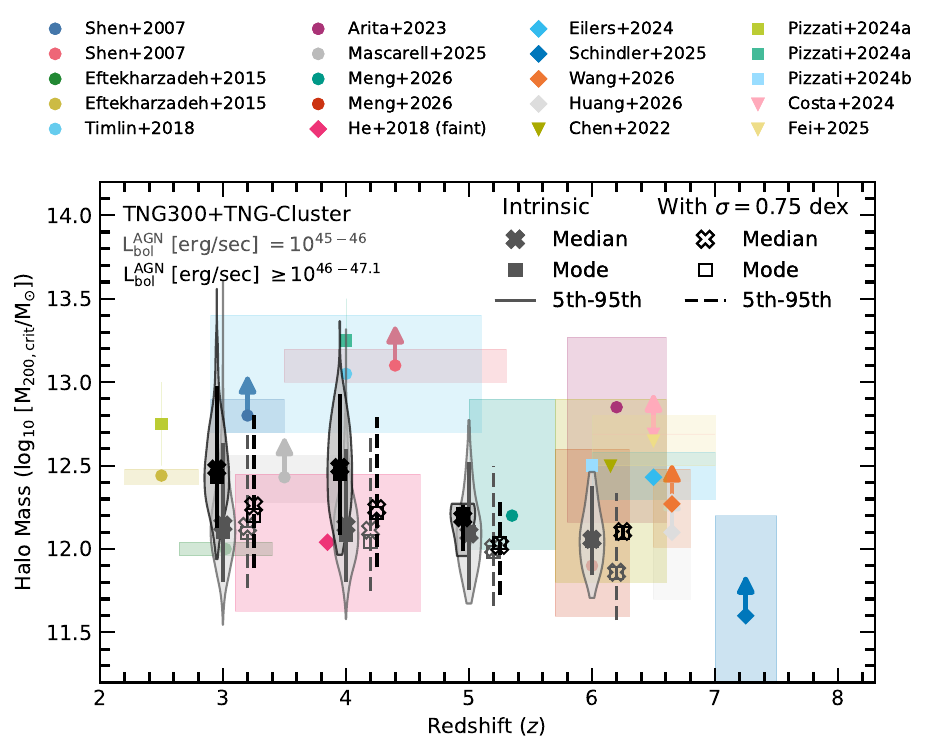}
  
 \includegraphics[width=0.84\textwidth, height=0.38\textheight,trim=0 0.2cm 0 3cm,
  clip]{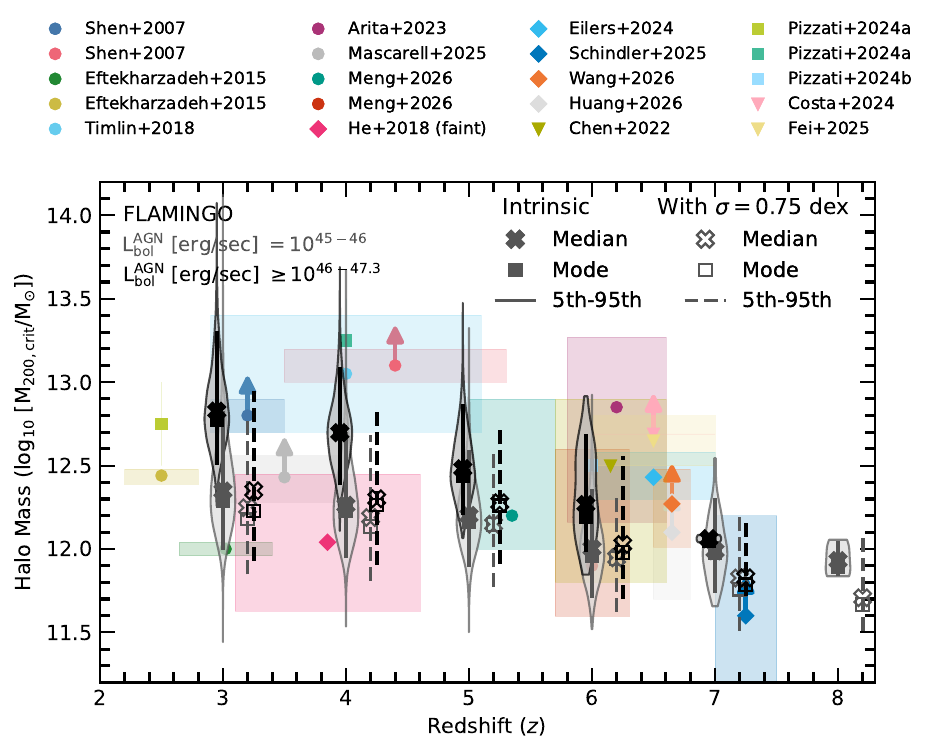}

\caption{\textbf{Evolution of the host halo masses of quasars according to the \TNG and \flamingo simulations before Cosmic Noon}. We focus on quasars in two bolometric luminosity bins, $\Lbol = 10^{45-46}$ erg~s$^{-1}$ (teal) and $\Lbol \geq 10^{46}$ erg~s$^{-1}$ (red), simulated by the \TNG (top) and \flamingo simulations (bottom panel, \flone and \fltwo combined). The distributions of the host halo masses of such simulated quasars are represented by violins at a few sampled redshifts, with filled square and the cross markers denoting the mode and the median values, respectively; the vertical lines in the violins extend from the 5th to 95th percentiles of the distribution. Analog empty markers and dashed vertical lines denote results from the simulations obtained by including an additional log-normal scatter of $\sigma = 0.75$ to the quasars bolometric luminosities. Colored markers surrounded by shaded regions represent the observationally-inferred host halo masses listed in Table~\ref{Tab:observations}, with upward arrows indicating {\it minimum} values; the heights (widths) of the shaded rectangles represent uncertainties in host halo mass (the observed redshift range) of the given sample. 
Most (see text) observational inferences as well as the \TNG and \flamingo cosmological galaxy simulations agree in that there is a only a mild redshift evolution of the host halo masses of quasars, which typically reside in haloes of $10^{12-13}$~\msun at at $z=3-7$, but with large quasar-to-quasar variations.}
   \label{fig:quasarmasses}
\end{figure*}

Given the complexity of the AGN bolometric luminosity -- host halo mass plane discussed so far, what do the \TNG and \flamingo simulations predict for the evolution of the host halo masses of the most luminous AGNs, i.e. quasars, across time?

As anticipated in Section~\ref{sec:intro} most observational measurements at $z=0-6$ are broadly consistent with luminous AGNs and quasars residing in hosts with total masses of $10^{12-13}$~\msun. We here focus on the redshift range $z=3-7$ and on simulated objects with $\Lbol \geq 10^{45}$~ erg~s$^{-1}$ (our definition of quasars; Section~\ref{sec:methods_defs}).We summarize the empirical landscape of quasar host halo masses in this redshift range in Appendix~\ref{App:obs}. Fig.~\ref{fig:quasarmasses} show that state-of-the-art cosmological galaxy simulations such as \TNG (top) and \flamingo (\flone and \fltwo combined; bottom) also find quasars to reside in haloes of typical mass $\sim10^{12-13}$~\msun, in agreement with observations (Table~\ref{Tab:observations}). The characteristic halo mass evolves only weakly across the explored redshift range (in \flamingo), although the underlying distributions are broad and extend to both lower and higher halo masses.

In particular, in Fig.~\ref{fig:quasarmasses}, violin markers show the simulation results, split in two bins of AGN bolometric luminosity, as per the gray bands in the top panels of Figs.~\ref{fig:lum_vs_halo_tng} and \ref{fig:lum_vs_halo_fl} and Fig.~\ref{fig:lum_vs_halo_all_median}: faint and bright quasars shown in gray and black, respectively (see also the green curves of the bottom panels of Figs.~\ref{fig:lum_vs_halo_tng} and \ref{fig:lum_vs_halo_fl}). 
For each violin at different redshift we show the median (filled cross markers), the mode (filled square markers), and the 5th-95th percentile range (solid vertical bars), to highlight both the most frequent and typical host halo masses but also the wide quasar-to-quasar scatter. 
The shaded rectangles and different colored markers show the empirically-inferred host halo masses of quasars, i.e. based on observational measurements -- those reported as a minimum mass are indicated by upward arrows. The height of the rectangles indicates the uncertainty in the observationally-inferred halo masses, the length indicates the redshift range over which the observed quasar sample is distributed. All mass values have been converted to the same cosmology as used in the \illustrisTNG (which is similar to that used in \flamingo) : see note of Table~\ref{Tab:observations}. 
 
As manifest in Fig.~\ref{fig:quasarmasses}, top, the \TNG simulations predict that the bulk of quasars (square markers) reside in haloes with $\Mtwohcrit = 10^{12-12.5}$~\msun across all studied epochs, $z=3-6$, in both bins of quasar bolometric luminosity. The more luminous quasars (black) typically exhibits a modestly higher median host halo mass than the lower-luminosity population (gray), with a difference of $0.2-0.4$ dex across this redshift range; however, the overlap between the full distributions remains large.

In the bottom panel of Fig.~\ref{fig:quasarmasses}, we show results from the combination of \flamingo \flone and \fltwo, since differences are negligible unless otherwise noted. Compared to \illustrisTNG, the quasars in the \flamingo model occupy a somewhat wider and larger range of median halo masses ($\Mtwohcrit = 10^{11.8-12.8}$\msun across the $z=3-8$ range). Moreover, the typical halo masses show a stronger redshift evolution: the bulk of bright quasars shift from $10^{12}$\msun halo masses at $z=7$ to $10^{12.8}$\msun at $z=3$. The luminosity dependence of the typical host halo masses is also somewhat stronger in \flamingo than in \TNG (and is stronger in \flamingo \flone alone, not shown, compared to \fltwo, which dominates the low-redshift results). 

As mentioned in Section~\ref{sec:results_plane_trendsofpeaks}, the luminosity dependence of the typical host halo masses of simulated quasars visible in Fig.~\ref{fig:quasarmasses} may be plausibly exacerbated by an underestimation of the variability in quasars' instantaneous bolometric luminosity (see Section~\ref{sec:disc}). This is shown by comparing the filled markers and violins discussed so far, which represent the intrinsic outcome of the simulations, with empty cross and square markers (and associated 5th-95th percentile ranges indicated as dashed vertical bars). The latter are obtained by assigning to each simulated system a bolometric luminosity randomly drawn from a log-normal distribution centered at the simulated value (logarithmic) and with standard deviation (i.e. log-normal scatter) of $\sigma = 0.75$ dex. This value is inspired by \cite{flamingo_quasar_clustering} and is meant to merely give an idea of what the impact of enhanced quasar variability could be. With the additional scatter in quasar bolometric luminosities, both simulation suites return somewhat smaller typical host halo masses with reduced redshift evolution, i.e. $\Mtwohcrit\sim10^{12-12.5}$~\msun across the $z=3-7$ redshift range for $~10^{45-47}$ erg~s$^{-1}$ quasars. This illustrative exercise is meant to remind that the true width and shape of the bolometric luminosity distribution of AGNs and quasars necessarily impact the median values (or summary statistics) inferred for the underlying host halo masses (see also Section~\ref{sec:intro}).

We pause here to forewarn that the picture summarized in Fig.~\ref{fig:quasarmasses} and predicted by \TNG and \flamingo only applies to quasars: the situation is rather different for lower-luminosity AGNs, as evincible from the previous subsections. 

Although an apples-to-apples comparison with observations is beyond the scope of this paper, Fig.~\ref{fig:quasarmasses} shows that the \TNG and \flamingo simulations are generally consistent with observational inferences (violins and markers vs. shaded rectangles): namely, both simulations and most observations support a picture whereby luminous quasars are predominantly hosted by haloes of similar mass across $z=3-7$. In fact, as Table~\ref{Tab:observations} summarizes, almost all observational results at $z=2-4$ are inferred from clustering and cross-correlation measurements for quasars much brighter than $\Lbol \sim 10^{46}$ erg~s$^{-1}$, except for the photometric sample of \cite{He2018}, which may include a substantial number of faint quasars. At higher redshifts, most observational constraints actually probe quasars as luminous as $10^{47-48}$ erg~s$^{-1}$, or more. On the other hand, the most luminous quasar in \TNG (\flamingo \fltwo) reaches $\Lbol = 10^{47.1}$ erg~s$^{-1}$ at $z=3$ ($\Lbol = 10^{47.3}$~erg~s$^{-1}$ at $z=5$). Given the weak (or even negligible) dependence of host halo mass on quasars' bolometric luminosity suggested by the simulations and discussed above, we consider this comparison to remain qualitatively, if not quantitatively, sound.

Looking at the observational findings of Table~\ref{Tab:observations} in more detail, in Fig.~\ref{fig:quasarmasses} we distinguish among inferences based on quasar-quasar clustering (shown as colored circles), quasar-galaxy clustering (colored diamonds) as well as other diverse methods (triangles and squares). For example, among the latter, the host halo mass values from the halo occupation model of \cite{Pizzati2024} are based on the clustering measurements of \cite{Shen2007} at $z=4$ (cyan square) and on the sample of \cite{Sarah2015} at $z\sim2.5$ (green square), whereas, at $z=6$, the \cite{Pizzati2024b} values (pink square) use the clustering measurements of \cite{Eilers} -- in all cases combined with quasar luminosity function measurements at the respective redshifts.  
As anticipated in Section~\ref{sec:intro} and now clearly visualized in Fig.~\ref{fig:quasarmasses}, across $z = 2.5-7.25$, the majority of the observationally-inferred host halo masses of quasars lie in the range $\sim10^{12}-10^{12.5}$\msun. There are exceptions to this trend at $z \sim 3-4$ where measurements of \cite{Shen2007} and \cite{Timlin2018} report a higher characteristic halo mass ($\gtrsim 10^{13}$\msun) as well as at $z \sim 6$ according to \cite{Arita2023}\footnote{As discussed in \cite{Sarah2015}, the higher host halo masses inferred by \cite{Shen2007} are likely driven by their use of clustering fits on larger scales, where the measurements may be affected by observational systematics. At higher redshift, as discussed in Appendix C of \citet{Pizzati2024b}, re-interpreting the clustering measurements of \citet{Arita2023} with a different, more physically-motivated choice for the correlation function shape and normalization favors a somewhat lower (but poorly constrained) minimum host halo mass value (of $\lesssim 10^{12.5}$\msun).}. Additionally, the cross-correlation measurement at $z \sim 7.25$ by \cite{Schindler2023} point to a lower minimum value of host halo mass ($\sim 10^{11.5}$\msun) for their two quasars than most estimates at lower redshifts.

From lower to higher redshift, aside from the measurements of \citet{Shen2007} and \citet{Timlin2018}, all  observational estimates of quasars' host halo masses at $z \sim 2.5-4$ are broadly consistent with those predicted by both the \TNG and \flamingo simulations within their 5th-95th percentile scatters, i.e. also thanks to the large scatter in host halo mass at fixed simulated quasar bolometric luminosity. At $z \gtrsim 6$ the observational constraints are sparser and the quasars samples are very small (of the order of a few to a few tens) in both simulations and observations \citep[except for the sample of][]{Meng2026}. While there are no bright quasars in \illustrisTNG at this redshift, host halo masses inferred from the cross-correlation measurements of \cite{Eilers}, \cite{Wang2026} and \cite{Huang2026}, the measurement of density fields of \cite{Chen2022}, gas kinematic measurements of \cite{Qinyue2025} as well as the model of \cite{Pizzati2024b} based on \cite{Eilers} data with luminosity function measurements are all consistent with those from the bright quasar sample of \flamingo. The $z=5-6.3$ clustering measurements of \cite{Meng2026} return characteristic host halo masses almost identical to those from \TNG and \flamingo faint quasars. Even the results based on the faint sample of \citet{Arita2023}, albeit higher in the bulk, do overlap with the high-mass end of the faint quasar distributions in both \TNG and \flamingo. 

To summarize, the \TNG and \flamingo \flone and \fltwo simulations predict host halo masses of $z=3-7$ quasars that are consistent with most observational studies, barring a couple of observational studies at $z\sim4$: namely, quasars typically reside in haloes with $\Mtwohcrit \sim 10^{12-12.8}$~\msun also before Cosmic Noon. However, importantly, the quasar-to-quasar variation is large, with 5th–95th percentiles spanning about 1 dex in host halo mass (possibly underestimated and yet much smaller than the corresponding scatter in AGN bolometric luminosity at fixed host halo mass; see Table~\ref{Tab:fits} and previous and Discussion Sections). The little-to-no redshift dependence of the typical host halo mass, seen in both simulations and most observations, supports the idea that the typical quasars reside in progressively rarer haloes at higher redshifts. In fact, according to the simulations studied here, the most luminous AGNs generally do not inhabit the most massive haloes at any given epoch all the way up to $z\lesssim7$ (see also individual markers in Figs.~\ref{fig:lum_vs_halo_tng} and \ref{fig:lum_vs_halo_fl}). On the other hand, the large diversity predicted by the simulations cautions that this is not always the case: namely, individual quasars may reside in haloes that are significantly more or less massive than the typical value, without altering the overall picture. 

The presence of quasars predominantly in similar-mass haloes, as inferred from observations and predicted by simulations at any given redshift, suggests that although quasar activity may be stochastic (as indicated by the large scatter in luminosity at fixed host halo mass), it nevertheless shows a preference for a characteristic halo-mass range, perhaps a {\it sweet spot} between low-mass haloes, where stellar feedback suppresses SMBH growth \citep[e.g.][]{Habouzit2019, Truong2021}, and high-mass haloes, where AGN feedback inhibits further SMBH fueling \citep[e.g.][]{Fanidakis2013,Bower2017}. In the \illustrisTNG model, this characteristic, redshift-independent halo mass reflects the efficiency of kinetic AGN feedback \citep{Weinberger-feedback}. In \flamingo, the characteristic halo masses are likely a consequence of the rapid SMBH accretion phase \citep[e.g.][]{Bower2017,McAlpine2017}, which naturally arises in these simulations.

\begin{figure*}
\centering
  \includegraphics[width=0.9\columnwidth]{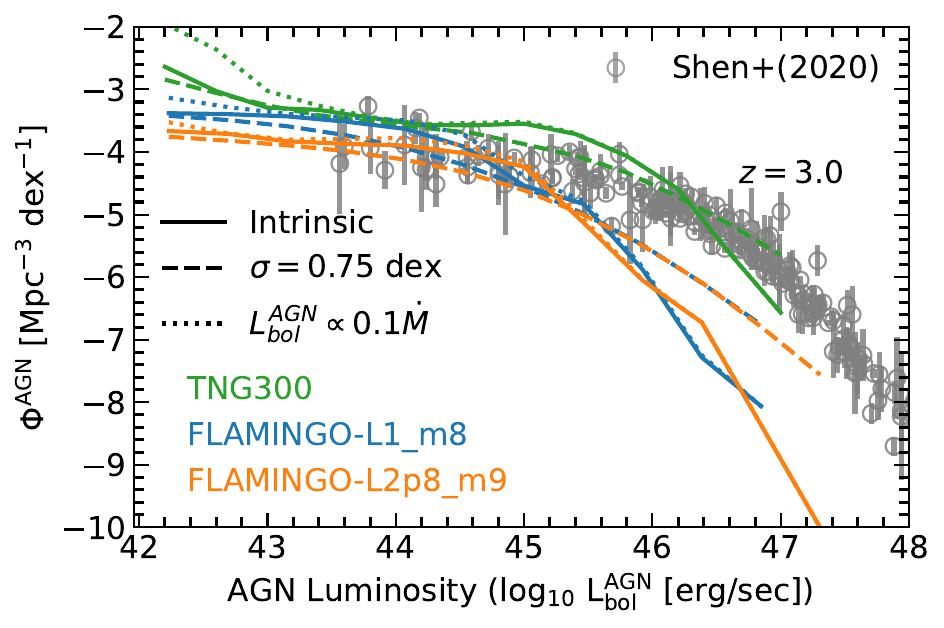}
  \includegraphics[width=0.9\columnwidth]{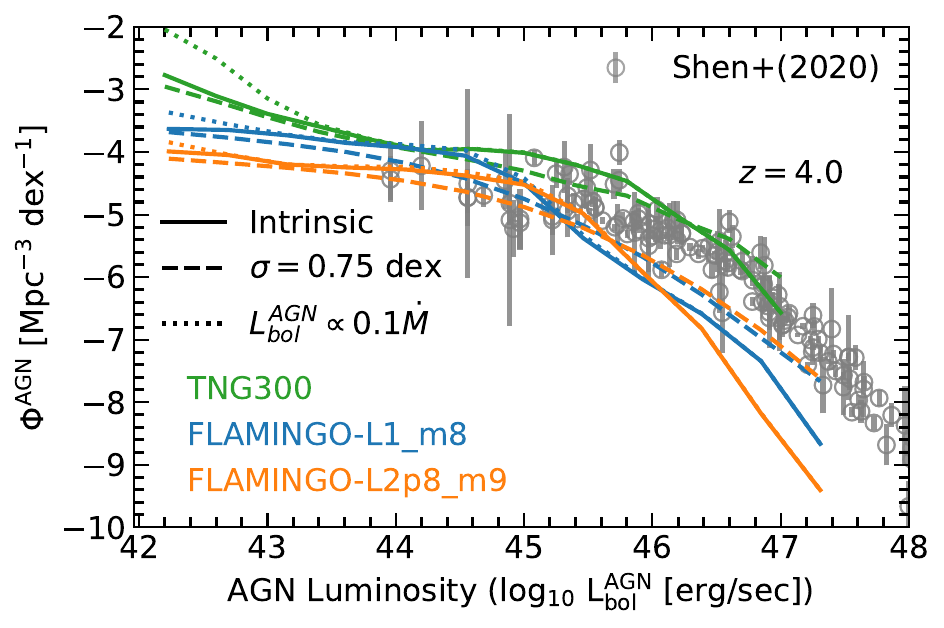}
  \includegraphics[width=0.9\columnwidth]{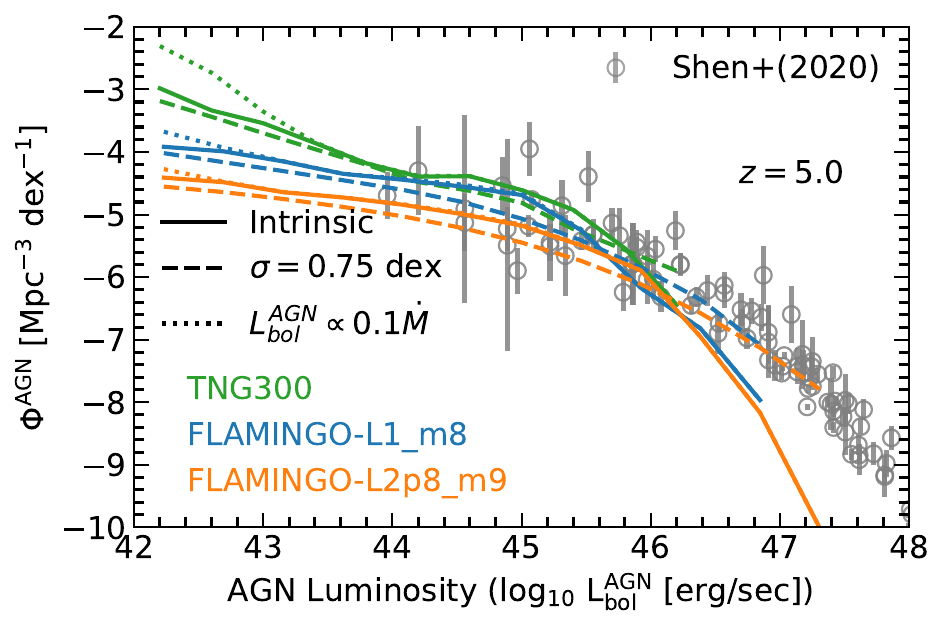}
  \includegraphics[width=0.9\columnwidth]{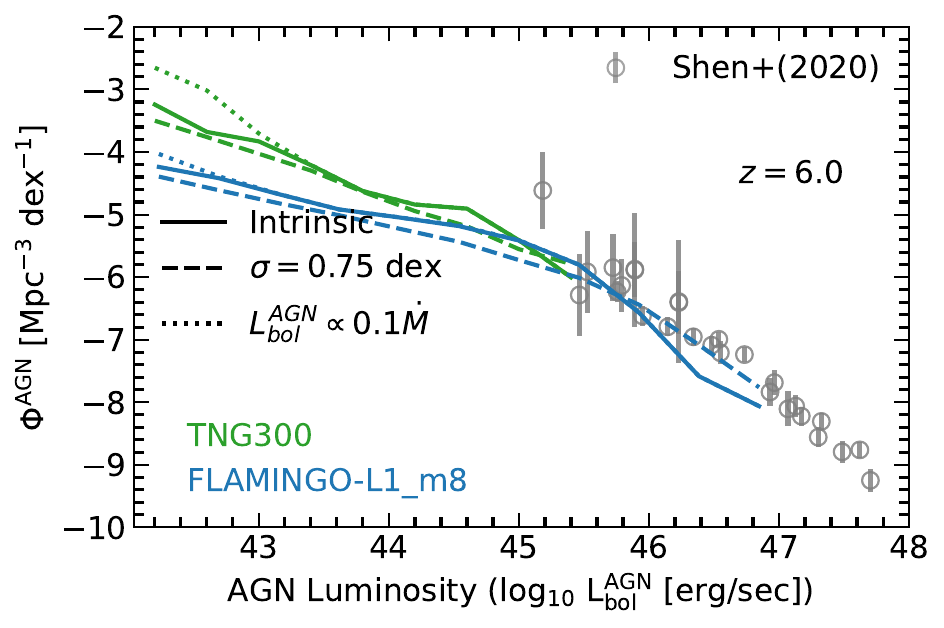}

  \caption{\textbf{Bolometric luminosity function of AGNs and quasars from the \flamingo and the \TNGthreeh simulations, juxtaposed to observational measurements, at $z=3-6$.} Different curves represent different ways to assign bolometric luminosities to the simulated systems (see text for details). Observational results compiled by \citealt{Shen2020} are given for mere reference, as the comparison is by no mean like-for-like and we do not attempt, for example, to account for possible obscuration effects. With these caveats in mind, the figure shows that the AGN populations simulated with the \illustrisTNG and \flamingo galaxy formation models are in the ball park of observed number densities of AGNs and quasars at Cosmic Noon and earlier.}
  \label{fig:luminosityfunctions}
\end{figure*}

\section{Discussion} 
\label{sec:disc}

\subsection{On the reliability of the simulations results}
\label{sec:disc_reliability}

According to all cosmological hydrodynamical simulations of galaxies examined in this paper, chiefly the \TNG suite and the \flamingo \flone and \fltwo simulations, more luminous AGNs reside on average in more massive haloes across the $z=3-7$ range, but only up to host halo masses of about $10^{12}$~\msun, after which the relation flattens. The relationship between AGN bolometric luminosity and host halo mass exhibits broadly similar median trends across a number of other simulations, from \illustris to \simba, but with intrinsic scatter that substantially exceeds both any variation of the median with host halo mass within a single simulation and any differences in the median normalizations and trends among models.

Despite differences in detail, such an agreement across models is significant given the diversity of simulation setups  -- including modeling choices, resolution, and volume -- and lends support to the robustness of the results presented throughout.

Focusing on luminous AGNs (i.e. quasars), we also showed that simulations and observations are in agreement, at least at face value, in the sense that quasars typically reside in haloes of approximately constant mass across cosmic epochs, of about $10^{12-12.5}$~\msun (Fig.~\ref{fig:quasarmasses}). This agreement is non-trivial, especially because these simulations were primarily designed to reproduce galaxy properties at low redshift, and high-redshift quasar observables were certainly not among the criteria used to define any of the underlying fiducial galaxy-formation model.

The qualitative agreement with observational findings is even more interesting given the fundamental constraints of the simulations -- whereby, first and foremost, AGN luminosity is not a direct output of the models -- as well as relevant modeling choices and unavoidable limitations. For example, in all simulations considered here, SMBHs are allowed to grow within an Eddington limit. Moreover, most large-volume simulations are limited by (spatial and mass) resolution and cannot capture short-timescale bursts of AGN accretion and, hence, luminosity, i.e. shorter than the time steps of the simulations \citep[at best tens of thousands of years in the best-resolution runs of the \illustrisTNG project][]{Pillepich2021} . Such variability a priori contributes to the intrinsic scatter in AGN bolometric luminosity e.g. at fixed host halo mass.

In relation to the underestimated short-timescale variability of AGN bolometric luminosities in the cosmological galaxy simulations above, an introduction of an artificial scatter in the quasar luminosities of \flamingo \fltwo enabled \cite{flamingo_quasar_clustering} to match both the observed quasar luminosity function up to $z \sim 5$ as well as the observed clustering of bright quasars ($\gtrsim 10^{45}$ erg~s$^{-1}$) at $z \lesssim 3$. In previous Sections, we have embraced the argument of \cite{flamingo_quasar_clustering}  and shown, for example, that introducing an artificial scatter of 0.75 dex in bolometric luminosities reduces the dependence of host halo mass on quasar luminosity, more so in the \flamingo rather than in the \TNG runs (Fig.~\ref{fig:quasarmasses}). 

In Fig.~\ref{fig:luminosityfunctions}, and amid the caveats thoroughly discussed by \cite{flamingo_quasar_clustering}\footnote{As noted in \cite{flamingo_quasar_clustering}, while introducing an artificial scatter can improve agreement with integrated quasar observables such as the luminosity function at the high-luminosity end and clustering statistics, it can also assign unrealistically high luminosities to relatively low-mass SMBHs in simulations, a behaviour that is not typically seen in observations.}, we confirm that such an artificial scatter can improve the agreement of the \flamingo output with demographics observables, particularly the quasar luminosity function. There we present the bolometric luminosity functions of AGNs and quasars from the simulations at $z=3, 4, 5,$ and 6, one redshift per panel. For the \illustrisTNG simulations, only \TNGthreeh is shown, since defining an effective volume for the \TNGcluster zoom-in sample is not at all straightforward. Observational results combined by \cite{Shen2020} are reported, for reference, as gray data points with errorbars, whereas for the simulations we show results adopting three different sets of assumptions. The intrinsic outcome of the simulations based on the fiducial conversion of SMBH mass accretion rates into AGN bolometric luminosities of Eq.~\ref{eq:luminosities} and adopted throughout is shown by solid curves; simulations results including an additional log-normal scatter of $\sigma = 0.75$ dex to the AGN bolometric luminosities are shown via dashed curves, whereas, for comparison, we also include the case where SMBH mass accretion rates are converted into AGN bolometric luminosities as in \cite{flamingo_quasar_clustering}, i.e. without the distinction between high- and low-accreting SMBHs. This set of varying curves is meant to bracket at least some of the known systematic uncertainties when extracting AGNs and quasars luminosity functions from cosmological galaxy simulations and comparing them to observations. In fact, here we neglect obscuration effects as well as any observational selection-function effects. 

Although Fig.~\ref{fig:luminosityfunctions} shall be intended merely qualitatively, as it does not at all provide a like-for-like comparison, it shows that the AGN populations simulated with the \illustrisTNG and \flamingo galaxy formation models are in the ball park of the observed number densities of AGNs and quasars at Cosmic Noon and earlier. Particularly, the luminosity functions of AGNs and quasars of \TNGthreeh are, at face value, consistent with those from observations between $z=3$ and $z\lesssim6$, whereas those from \flamingo are somewhat lower, consistently with the lower overall normalizations of the AGN bolometric luminosity -- host halo mass relations seen in the previous Sections.

\subsection{On the large scatter of the AGN bolometric luminosity --  host halo mass relation, and its implications}
\label{sec:disc_scatter}

A key finding of this paper is that, in spite of different modeling choices and simulation setups, all the modern cosmological simulations of galaxies studied here predict very large system-to-system variations around the median relations between AGN bolometric luminosity and host halo mass at $z=3-7$, i.e. of {\it orders of magnitude}. More specifically, both the \TNG and \flamingo simulations return a much larger scatter in AGN bolometric luminosity for a given host halo mass (up to $\sim3$ dex for the 5th-95th percentile ranges, across models, redshifts, and halo masses) than in host halo mass at given bolometric luminosity (up to about $1$ dex). 

Overall, our findings are qualitatively consistent with previous results reporting large scatter in AGN luminosity or SMBH mass accretion rate at fixed star formation rate of the host galaxy, e.g. in \eagle \citep{McAlpine2017}, as well as at fixed host galaxy stellar mass or SMBH mass across models \citep{Habouzit2022a}. 

Noticeably, the magnitude of the scatter quantified in this paper may even be a lower estimate, for three reasons. Firstly, we have a priori adopted a threshold of $\Lbol \geq 10^{42}$ erg~s$^{-1}$, which, while below observational limits at these redshifts, still neglects the vastly more numerous objects in the simulations (and in the Universe) with smaller halo, galaxy, SMBH masses. Secondly, throughout our analysis we have exclusively focused on central galaxies, whereas halo-scale environmental effects as those suffered by satellite galaxies may enhance diversity, possibly also at the considered redshifts \citep[see e.g. ][for the case of low-redshift with \illustrisTNG]{Kurinchi-Vendhan2025}. Finally, as discussed in the previous Section, the simulations adopted here plausibly fail to capture short-timescale AGN luminosity variability.

From a conceptual viewpoint, the existence of average trends characterized by large scatter in AGN bolometric luminosity at fixed host halo mass suggests that, as expected, SMBH mass accretion -- and hence AGN luminosity --  is only partially determined by the mass of the host halo. We speculate that additional drivers of AGN luminosity, i.e. host properties, could further constrain (i.e. reduce the spread of) the distributions. These may be related to the availability of gas in the inner regions of galaxies, as proxied by, e.g., star formation rate or galaxy-wide gas fractions, as well as to the position of the host halo within the large-scale structure. We also expect that a significant fraction of the scatter may remain ``unexplained'', pointing to an intrinsic stochasticity in SMBH activity. 

However, we can already distinguish different regimes for the relevance of such stochasticity. For example, at the high halo-mass end, we know from the behavior of the \TNG and \flamingo simulations (see also Appendix~\ref{app:SMBHpopulations}) that the increasing halo-to-halo variation with host halo mass is driven by star formation and SMBH accretion quenching. In this regime, additional properties, such as SMBH mass or galaxy star-formation state, may help further constrain the AGN luminosity.

We note that the above discussion applies to {\it instantaneous} quantities only. In contrast, SMBH mass accretion rates integrated over their lifetime -- which, modulo SMBH-SMBH mergers, correspond to the SMBH masses -- are expected and known to be significantly less variable, i.e. with tighter correlations. 

From a practical point of view, the large scatter in the simulated AGN/quasar luminosity -- host halo mass relation suggests that characteristic halo masses alone may be insufficient for comparison with clustering-based halo mass estimates. In fact, the wide scatter in AGN luminosity has important implications for halo occupation models, which often assume or infer a relatively small scatter ($\lesssim 1$ dex) in quasar luminosity at fixed halo mass \citep{White2008,Wyithe2009,Shankar2010,Pizzati2024}. While some of these findings were motivated by the strong quasar clustering measured by \cite{Shen2007} at $z \sim 4$ -- which is largely inconsistent with most physical models -- subsequent work by \cite{Pizzati2024,Pizzati2024b} has shown that the same framework yields a larger scatter ($0.5$–$0.6$ dex) in luminosity at fixed halo mass when applied to clustering measurements at $z \sim 2.5$ rather than 4. Beyond details, the halo-to-halo variations predicted by the \TNG and \flamingo simulations, but also by smaller-volume simulations (see Appendix~\ref{app:allsims}), significantly exceeds the scatter assumed or inferred in halo occupation models, even those calibrated to quasar clustering and luminosity function measurements. Thus, while the host halo masses inferred from clustering are broadly consistent with those returned by simulations, the scatter in luminosity at fixed halo mass is much larger than that typically applicable to HOD frameworks based on similar observational constraints. The origin of this apparent tension remains unclear. As outlined above, a coherent reconciliation between simulations and observations may require either probing a broader range of AGN luminosities observationally, or more carefully accounting for limitations arising from finite volume, resolution, and SMBH accretion modelling in current galaxy-formation simulations.

\section{Summary and conclusions} \label{sec:conclusions}

In this paper we have extracted and analyzed the relation between AGN bolometric luminosity and host halo mass predicted by large volume cosmological hydrodynamical simulations of galaxies at high redshift, i.e. prior to Cosmic Noon ($\gtrsim 3$). We have focused mainly on two suites of simulations: the \TNGthreeh run of the \illustrisTNG project combined with the more recent \TNGcluster set of zoom-in simulations of massive galaxy clusters, and the \flone and \fltwo \flamingo simulations, in their fiducial galaxy-formation model implementation (Section~\ref{sec:methods_sims_big}). Both suites self-consistently evolve cold DM, gas, stars, and SMBHs under gravity, (magneto-)hydrodynamics, and a number of differently-implemented astrophysics i.e. galaxy-physics processes, enabling predictions of halo, galaxy, and SMBH populations across cosmic time and across large cosmological volumes spanning comoving gigaparsecs on a side. 

We have adopted standard post-processing conversions to translate the instantaneous mass accretion rates of SMBHs (which is a direct outcome of the simulations) into bolometric luminosities (Eq.~\ref{eq:luminosities}) and consider only simulated AGNs with bolometric luminosity $\Lbol \gtrsim 10^{42}$ erg~s$^{-1}$ (Section~\ref{sec:methods_selection}). This allows us to focus on regimes that are less directly affected by modeling choices such as SMBH seeding (Section~\ref{sec:methods_selection} and Appendix~\ref{app:SMBHpopulations}) while simultaneously probing a wide range of SMBH activity levels. 
The adopted luminosity limit also includes sources with luminosities below those currently accessible observationally at the considered redshifts.

Our selected AGN sample spans, a priori, wide ranges of underlying SMBH masses and host galaxy stellar masses (Appendix~\ref{app:SMBHpopulations}). In particular, thanks to the large volumes of the \TNG and \flamingo simulations -- compared to previous-generation projects such as \illustris or \eagle -- we can access haloes with total masses of $\Mtwohcrit = 10^{11-14}$~\msun\ even at these redshifts. Importantly, \TNG (\flamingo \flone) return hundreds (thousands) of luminous AGNs, i.e. quasars ($\Lbol \geq 10^{45}$ erg~s$^{-1}$), as early as $z\sim5$, enabling robust statistical studies of quasar populations in cosmological simulations.\\

Our three main quantitative findings are summarized below:
\begin{itemize}
    \item According to all cosmological hydrodynamical simulations of galaxies examined in this paper (Table~\ref{Tab:simulations}), more luminous AGNs reside, on average, in more massive haloes across the $z = 3-7$ range (Figs.~\ref{fig:lum_vs_halo_tng}, \ref{fig:lum_vs_halo_fl}, and \ref{fig:lum_vs_halo_all_median}). However, as clearly revealed by \TNG and \flamingo, this trend holds only up to host halo masses of $\sim 10^{12}$\msun, beyond which the median relation flattens (according to \flamingo, Fig.~\ref{fig:lum_vs_halo_fl}) or even turns over (\TNG, Fig.~\ref{fig:lum_vs_halo_tng}), at least at $z \lesssim 5-6$. Before the flattening, the median relations between AGN bolometric luminosity and host halo mass are well described by a power-law, with slopes ranging from 1 to 5 depending on redshift and simulation (Table~\ref{Tab:fits}).\\
    
    \item Despite different modeling choices and simulation setups, all the modern cosmological simulations of galaxies considered here return very large system-to-system variations around the median relations between AGN bolometric luminosity and host halo mass at $z = 3-7$, i.e. spanning orders of magnitude (Figs.~\ref{fig:lum_vs_halo_tng} and \ref{fig:lum_vs_halo_fl}, middle and bottom). More specifically, both the \TNG and \flamingo simulation suites return a much larger scatter in AGN bolometric luminosity at fixed host halo mass than in host halo mass at fixed AGN bolometric luminosity (Fig.~\ref{Fig:scatter_tng_flamingo} top vs. bottom: up to about 3 dex compared to up to about 1 dex, for the 5th–95th percentile ranges, across models, redshifts, and halo masses), with the former increasing with increasing host halo mass and decreasing redshift. \\ 
      
    \item Focusing on the most luminous AGNs, \TNG predict that quasars (here $\Lbol \sim 10^{45-47}$ erg~s$^{-1}$) typically reside in haloes of median mass $\Mtwohcrit = 10^{12-12.5}$\msun across the $z=3-6$ range. Similarly, the typical quasars in the \flamingo simulations also reside in haloes of $\Mtwohcrit = 10^{11.8-12.8}$\msun at $z=3-8$, with somewhat larger but still negligible redshift and luminosity dependencies than in \TNG (Fig.~\ref{fig:quasarmasses}). These trends and values are consistent with the largest majority of observationally-inferred results -- except for the findings of \cite{Shen2007, Timlin2018} at $z\sim4$; see Table~\ref{Tab:observations}. This consistency, however, is facilitated by the large quasar-to-quasar variations predicted by the simulations ($0.5-1$~dex scatter in host halo mass, 5th-95th percentiles).
    
\end{itemize}
These quantitative results have broader scientific implications.
The existence of average trends characterized by large scatter in AGN bolometric luminosity at fixed host halo mass suggests that, as expected, SMBH mass accretion -- and hence AGN luminosity -- is only partially determined by the mass of the host halo (Sections~\ref{sec:results_plane_scatter} and \ref{sec:disc_scatter}). Some level of intrinsic stochasticity is expected in the instantaneous properties of SMBHs, but further investigation is needed to identify additional physical drivers of AGN luminosity, whether related to their halo or galaxy hosts, or to the large-scale environment. We emphasize that all our findings apply to {\it instantaneous} SMBH properties, whereas integrated accretion rates (i.e. SMBH masses) are known to be less variable.

We interpret the flattening and turnover at the high-mass end of the median AGN bolometric luminosity -- host halo mass relation as a manifestation of SMBH feedback, which suppresses gas accretion onto SMBHs effectively quenching their AGN luminosity (as well as star formation in their massive host galaxies). We speculate that the same feedback, at least in the \illustrisTNG model, is responsible for the emergence of a characteristic host halo-mass range for quasars (Section~\ref{sec:results_quasars}). Both simulations and most observations consistently support a picture of little-to-no redshift dependence of the typical host halo mass of quasars, and this implies that quasars generally reside in increasingly rarer (but indeed not necessarily more massive) haloes at earlier cosmic epochs, all the way up to $z\lesssim7$. Another implication of this high-mass AGN flattening is that, in both the \TNG and \flamingo simulations, the most luminous quasars rarely reside in the most massive haloes at any of the epochs considered (see e.g. individual markers in the top panels of Figs.~\ref{fig:lum_vs_halo_tng} and \ref{fig:lum_vs_halo_fl}). However, the large diversity predicted by the simulations cautions that individual quasars may reside in haloes that are significantly more or less massive than the typical {\it characteristic} value, without altering the overall picture and without diminishing the interest in recent findings of two bright quasars at $z\sim 7$ possibly residing in haloes with a minimum mass below $10^{12}$~\msun\ \citep[][although with wide uncertainties]{Schindler2026}. A fundamental question remains: at which high(er) redshift does the characteristic host halo mass of quasars begin to deviate from the $10^{12-13}$ \msun transitional scale, if at all?

Our findings based on cosmological galaxy simulations may be of inspiration and utility for the next generations of AGN halo occupation models and for future interpretations of clustering measurements, particularly the fact that the median relation between AGN bolometric luminosity and host halo mass can be highly non-linear, non-strictly monotonic, and characterized by large levels of scatter. Even though we believe that the scatter in AGN luminosity at fixed host halo mass may plausibly be underestimated by simulations (Section~\ref{sec:disc_scatter}), it is already significantly larger than the constraints from halo occupation distribution models of quasars, which are derived from quasar clustering and luminosity function measurements across redshifts. This discrepancy is possibly driven by the limited sensitivity of observations to only the very luminous quasars and their association to correspondingly-high values of a minimum halo mass, leading to an implicit halo-mass threshold in correlation analyses. However, such a tension will require further investigation.

Finally, the robustness of the results presented here is supported by the fact that, aside from model-dependent differences in absolute normalisations and some (weak) redshift trends (e.g. \flamingo AGNs being typically less luminous than those from \TNG at fixed host halo mass in some mass and redshift regimes), the overall qualitative picture is consistent across a range of galaxy-formation models and numerical resolutions. The AGN and quasar luminosity functions from the \TNGthreeh and \flamingo simulations are also broadly consistent with observational constraints at $z=3$–6 (Fig.~\ref{fig:luminosityfunctions}), albeit with important caveats in this comparison that warrant further investigations and discussions \citep[see in the meantime][for \flamingo]{flamingo_quasar_clustering}. 

Modulo the issue of scatter discussed above, the agreement between simulations and observations in terms of typical host halo masses of high-redshift quasars is also encouraging. Mild discrepancies remain between observations and simulations, pointing to two main avenues forward. On the observational side, larger, uniformly-selected samples with consistent luminosity thresholds and improved clustering measurements at e.g. $z \gtrsim 4$ would reduce uncertainties and cosmic variance. On the modeling side, our findings support and motivate more detailed and accurate data comparisons, including forward modeling of mock surveys -- by applying consistent selection functions, luminosity conversions, obscuration, and duty-cycle modeling as in the observations --  to simultaneously test both observational inference methods and the modeling of SMBH and galaxy physics implemented in large-volume hydrodynamical simulations.

\section*{Acknowledgements}

We thank the other builders of the \flamingo simulations, John Helly, Matthieu Schaller, and Roi Kugel, for granting us preliminary access to the simulations output prior to its public release. This project has received funding from the European Research Council (ERC) under the European Union’s Horizon 2020 research and innovation programme. In particular, AK, AP, and JB acknowledge funding from the European Union (ERC, COSMIC-KEY, 101087822, PI: Pillepich). DN acknowledges funding from the Deutsche Forschungsgemeinschaft (DFG) through an Emmy Noether Research Group (grant number NE 2441/1-1). 

The \texttt{TNG-Cluster} simulation suite has been executed on several machines: with compute time awarded under the TNG-Cluster project on the HoreKa supercomputer, funded by the Ministry of Science, Research and the Arts Baden-Württemberg and by the Federal Ministry of Education and Research; the bwForCluster Helix supercomputer, supported by the state of Baden-Württemberg through bwHPC and the German Research Foundation (DFG) through grant INST 35/1597-1 FUGG; the Vera cluster of the Max Planck Institute for Astronomy (MPIA), as well as the Cobra and Raven clusters, all three operated by the Max Planck Computational Data Facility (MPCDF); and the BinAC cluster, supported by the High Performance and Cloud Computing Group at the Zentrum für Datenverarbeitung of the University of Tübingen, the state of Baden-Württemberg through bwHPC and the German Research Foundation (DFG) through grant no INST 37/935-1 FUGG. 

We acknowledge the Virgo Consortium for granting access to their data repositories. The FLAMINGO simulations were performed using the Durham Memory Intensive system managed by the Institute for Computational Cosmology on behalf of the STFC DiRAC facility (www.dirac.ac.uk). The equipment was funded by BEIS capital funding via STFC capital grants ST/K00042X/1, ST/P002293/1,
ST/R002371/1 and ST/S002502/1, Durham University and STFC
operations grant ST/R000832/1. DiRAC is part of the National e-
Infrastructure. 

All the analysis of the simulations and computations associated to this paper have been realized on the Vera cluster of the MPCDF.
We acknowledge the use of ChatGPT for assistance with proofreading and improving the language and style of selected sections of the manuscript. All scientific content, analysis, and interpretation were developed independently by the authors, and no large language model was used in the production of the scientific results. This work made use of the \texttt{Matplotlib} \citep{Matplotlib} and \texttt{Tol-colors} \footnote{https://tol-colors.readthedocs.io/en/latest/} packages for plotting.

\section*{Data Availability}
The data underlying the plots within this article are available on reasonable request to the corresponding author. 
All the output of the TNG-Cluster simulation suite is now publicly available, as are the \illustrisTNG simulations -- see \url{www.tng-project.org/data} and \citet{Nelson-TNG-data} -- and the Illustris simulation -- see \url{www.illustris-project.org/data} and \citet{Nelson-illustris-data}. The FLAMINGO simulation data was also recently made public \citep{Helly2026}. We acknowledge the Virgo Consortium for making their simulation data available -- see \url{https://dataweb.cosma.dur.ac.uk:8443/flamingo/introduction.html}.

\bibliographystyle{mnras}
\bibliography{AGN_halo}

@ARTICLE{Moreno2025,
       author = {{Forouhar Moreno}, Victor J. and {Helly}, John and {McGibbon}, Robert and {Schaye}, Joop and {Schaller}, Matthieu and {Han}, Jiaxiní and {Kugel}, Roi and {Bah{\'e}}, Yannick M.},
        title = "{Assessing subhalo finders in cosmological hydrodynamical simulations}",
      journal = {\mnras},
         year = 2025,
        month = oct,
       volume = {543},
       number = {2},
        pages = {1339-1372},
          doi = {10.1093/mnras/staf1478},
archivePrefix = {arXiv},
       eprint = {2502.06932},
 primaryClass = {astro-ph.CO},
       adsurl = {https://ui.adsabs.harvard.edu/abs/2025MNRAS.543.1339F}
}

@ARTICLE{HBT,
       author = {{Han}, Jiaxin and {Cole}, Shaun and {Frenk}, Carlos S. and {Benitez-Llambay}, Alejandro and {Helly}, John},
        title = "{HBT+: an improved code for finding subhaloes and building merger trees in cosmological simulations}",
      journal = {\mnras},
         year = 2018,
        month = feb,
       volume = {474},
       number = {1},
        pages = {604-617},
          doi = {10.1093/mnras/stx2792},
archivePrefix = {arXiv},
       eprint = {1708.03646},
 primaryClass = {astro-ph.CO},
       adsurl = {https://ui.adsabs.harvard.edu/abs/2018MNRAS.474..604H}
}

@ARTICLE{Vogelsberger_review,
       author = {{Vogelsberger}, Mark and {Marinacci}, Federico and {Torrey}, Paul and {Puchwein}, Ewald},
        title = "{Cosmological simulations of galaxy formation}",
      journal = {Nature Reviews Physics},
         year = 2020,
        month = jan,
       volume = {2},
       number = {1},
        pages = {42-66},
          doi = {10.1038/s42254-019-0127-2},
archivePrefix = {arXiv},
       eprint = {1909.07976},
 primaryClass = {astro-ph.GA},
       adsurl = {https://ui.adsabs.harvard.edu/abs/2020NatRP...2...42V}
}

@ARTICLE{Vogelsberger-illustris-model,
       author = {{Vogelsberger}, Mark and {Genel}, Shy and {Sijacki}, Debora and {Torrey}, Paul and {Springel}, Volker and {Hernquist}, Lars},
        title = "{A model for cosmological simulations of galaxy formation physics}",
      journal = {\mnras},
         year = 2013,
        month = dec,
       volume = {436},
       number = {4},
        pages = {3031-3067},
          doi = {10.1093/mnras/stt1789},
archivePrefix = {arXiv},
       eprint = {1305.2913},
 primaryClass = {astro-ph.CO},
       adsurl = {https://ui.adsabs.harvard.edu/abs/2013MNRAS.436.3031V}
}

@ARTICLE{Vogelsberger2014-illustris,
       author = {{Vogelsberger}, M. and {Genel}, S. and {Springel}, V. and {Torrey}, P. and {Sijacki}, D. and {Xu}, D. and {Snyder}, G. and {Bird}, S. and {Nelson}, D. and {Hernquist}, L.},
        title = "{Properties of galaxies reproduced by a hydrodynamic simulation}",
      journal = {\nat},
         year = 2014,
        month = may,
       volume = {509},
       number = {7499},
        pages = {177-182},
          doi = {10.1038/nature13316},
archivePrefix = {arXiv},
       eprint = {1405.1418},
 primaryClass = {astro-ph.CO},
       adsurl = {https://ui.adsabs.harvard.edu/abs/2014Natur.509..177V}
}

@ARTICLE{Genel-Illustris,
       author = {{Genel}, Shy and {Vogelsberger}, Mark and {Springel}, Volker and {Sijacki}, Debora and {Nelson}, Dylan and {Snyder}, Greg and {Rodriguez-Gomez}, Vicente and {Torrey}, Paul and {Hernquist}, Lars},
        title = "{Introducing the Illustris project: the evolution of galaxy populations across cosmic time}",
      journal = {\mnras},
         year = 2014,
        month = nov,
       volume = {445},
       number = {1},
        pages = {175-200},
          doi = {10.1093/mnras/stu1654},
archivePrefix = {arXiv},
       eprint = {1405.3749},
 primaryClass = {astro-ph.CO},
       adsurl = {https://ui.adsabs.harvard.edu/abs/2014MNRAS.445..175G}
}

@ARTICLE{Nelson-illustris-data,
      author = {{Nelson}, D. and {Pillepich}, A. and {Genel}, S. and {Vogelsberger}, M. and {Springel}, V. and {Torrey}, P. and {Rodriguez-Gomez}, V. and {Sijacki}, D. and {Snyder}, G.~F. and {Griffen}, B. and {Marinacci}, F. and {Blecha}, L. and {Sales}, L. and {Xu}, D. and {Hernquist}, L.},
        title = "{The illustris simulation: Public data release}",
      journal = {Astronomy and Computing},
         year = 2015,
        month = nov,
       volume = {13},
        pages = {12-37},
          doi = {10.1016/j.ascom.2015.09.003},
archivePrefix = {arXiv},
       eprint = {1504.00362},
 primaryClass = {astro-ph.CO},
       adsurl = {https://ui.adsabs.harvard.edu/abs/2015A&C....13...12N}
}

@ARTICLE{Pillepich-TNG50,
       author = {{Pillepich}, Annalisa and {Nelson}, Dylan and {Springel}, Volker and {Pakmor}, R{\"u}diger and {Torrey}, Paul and {Weinberger}, Rainer and {Vogelsberger}, Mark and {Marinacci}, Federico and {Genel}, Shy and {van der Wel}, Arjen and {Hernquist}, Lars},
        title = "{First results from the TNG50 simulation: the evolution of stellar and gaseous discs across cosmic time}",
      journal = {\mnras},
         year = 2019,
        month = dec,
       volume = {490},
       number = {3},
        pages = {3196-3233},
          doi = {10.1093/mnras/stz2338},
archivePrefix = {arXiv},
       eprint = {1902.05553},
 primaryClass = {astro-ph.GA},
       adsurl = {https://ui.adsabs.harvard.edu/abs/2019MNRAS.490.3196P}
}

@ARTICLE{Nelson-TNG50,
       author = {{Nelson}, Dylan and {Pillepich}, Annalisa and {Springel}, Volker and {Pakmor}, R{\"u}diger and {Weinberger}, Rainer and {Genel}, Shy and {Torrey}, Paul and {Vogelsberger}, Mark and {Marinacci}, Federico and {Hernquist}, Lars},
        title = "{First results from the TNG50 simulation: galactic outflows driven by supernovae and black hole feedback}",
      journal = {\mnras},
         year = 2019,
        month = dec,
       volume = {490},
       number = {3},
        pages = {3234-3261},
          doi = {10.1093/mnras/stz2306},
archivePrefix = {arXiv},
       eprint = {1902.05554},
 primaryClass = {astro-ph.GA},
       adsurl = {https://ui.adsabs.harvard.edu/abs/2019MNRAS.490.3234N}
}

@ARTICLE{Pillepich-TNG,
       author = {{Pillepich}, Annalisa and {Springel}, Volker and {Nelson}, Dylan and {Genel}, Shy and {Naiman}, Jill and {Pakmor}, R{\"u}diger and {Hernquist}, Lars and {Torrey}, Paul and {Vogelsberger}, Mark and {Weinberger}, Rainer and {Marinacci}, Federico},
        title = "{Simulating galaxy formation with the IllustrisTNG model}",
      journal = {\mnras},
         year = 2018,
        month = jan,
       volume = {473},
       number = {3},
        pages = {4077-4106},
          doi = {10.1093/mnras/stx2656},
archivePrefix = {arXiv},
       eprint = {1703.02970},
 primaryClass = {astro-ph.GA},
       adsurl = {https://ui.adsabs.harvard.edu/abs/2018MNRAS.473.4077P}
}

@ARTICLE{Pillepich2021,
       author = {{Pillepich}, Annalisa and {Nelson}, Dylan and {Truong}, Nhut and {Weinberger}, Rainer and {Martin-Navarro}, Ignacio and {Springel}, Volker and {Faber}, Sandy M. and {Hernquist}, Lars},
        title = "{X-ray bubbles in the circumgalactic medium of TNG50 Milky Way- and M31-like galaxies: signposts of supermassive black hole activity}",
      journal = {\mnras},
         year = 2021,
        month = dec,
       volume = {508},
       number = {4},
        pages = {4667-4695},
          doi = {10.1093/mnras/stab2779},
archivePrefix = {arXiv},
       eprint = {2105.08062},
 primaryClass = {astro-ph.GA},
       adsurl = {https://ui.adsabs.harvard.edu/abs/2021MNRAS.508.4667P}
}

@ARTICLE{Weinberger-TNG,
       author = {{Weinberger}, Rainer and {Springel}, Volker and {Hernquist}, Lars and {Pillepich}, Annalisa and {Marinacci}, Federico and {Pakmor}, R{\"u}diger and {Nelson}, Dylan and {Genel}, Shy and {Vogelsberger}, Mark and {Naiman}, Jill and {Torrey}, Paul},
        title = "{Simulating galaxy formation with black hole driven thermal and kinetic feedback}",
      journal = {\mnras},
         year = 2017,
        month = mar,
       volume = {465},
       number = {3},
        pages = {3291-3308},
          doi = {10.1093/mnras/stw2944},
archivePrefix = {arXiv},
       eprint = {1607.03486},
 primaryClass = {astro-ph.GA},
       adsurl = {https://ui.adsabs.harvard.edu/abs/2017MNRAS.465.3291W}
}

@ARTICLE{Weinberger-feedback,
       author = {{Weinberger}, Rainer and {Springel}, Volker and {Pakmor}, R{\"u}diger and {Nelson}, Dylan and {Genel}, Shy and {Pillepich}, Annalisa and {Vogelsberger}, Mark and {Marinacci}, Federico and {Naiman}, Jill and {Torrey}, Paul and {Hernquist}, Lars},
        title = "{Supermassive black holes and their feedback effects in the IllustrisTNG simulation}",
      journal = {\mnras},
         year = 2018,
        month = sep,
       volume = {479},
       number = {3},
        pages = {4056-4072},
          doi = {10.1093/mnras/sty1733},
archivePrefix = {arXiv},
       eprint = {1710.04659},
 primaryClass = {astro-ph.GA},
       adsurl = {https://ui.adsabs.harvard.edu/abs/2018MNRAS.479.4056W}
}

@ARTICLE{Nelson-TNG-Cluster,
       author = {{Nelson}, Dylan and {Pillepich}, Annalisa and {Ayromlou}, Mohammadreza and {Lee}, Wonki and {Lehle}, Katrin and {Rohr}, Eric and {Truong}, Nhut},
        title = "{Introducing the TNG-Cluster simulation: Overview and the physical properties of the gaseous intracluster medium}",
      journal = {\aap},
         year = 2024,
        month = jun,
       volume = {686},
          eid = {A157},
        pages = {A157},
          doi = {10.1051/0004-6361/202348608},
archivePrefix = {arXiv},
       eprint = {2311.06338},
 primaryClass = {astro-ph.GA},
       adsurl = {https://ui.adsabs.harvard.edu/abs/2024A&A...686A.157N}
}

@ARTICLE{Nelson-TNG-data,
       author = {{Nelson}, Dylan and {Springel}, Volker and {Pillepich}, Annalisa and {Rodriguez-Gomez}, Vicente and {Torrey}, Paul and {Genel}, Shy and {Vogelsberger}, Mark and {Pakmor}, Ruediger and {Marinacci}, Federico and {Weinberger}, Rainer and {Kelley}, Luke and {Lovell}, Mark and {Diemer}, Benedikt and {Hernquist}, Lars},
        title = "{The IllustrisTNG simulations: public data release}",
      journal = {Computational Astrophysics and Cosmology},
         year = 2019,
        month = may,
       volume = {6},
       number = {1},
          eid = {2},
        pages = {2},
          doi = {10.1186/s40668-019-0028-x},
archivePrefix = {arXiv},
       eprint = {1812.05609},
 primaryClass = {astro-ph.GA},
       adsurl = {https://ui.adsabs.harvard.edu/abs/2019ComAC...6....2N}
}

@ARTICLE{Pillepich2018,
       author = {{Pillepich}, Annalisa and {Nelson}, Dylan and {Hernquist}, Lars and {Springel}, Volker and {Pakmor}, R{\"u}diger and {Torrey}, Paul and {Weinberger}, Rainer and {Genel}, Shy and {Naiman}, Jill P. and {Marinacci}, Federico and {Vogelsberger}, Mark},
        title = "{First results from the IllustrisTNG simulations: the stellar mass content of groups and clusters of galaxies}",
      journal = {\mnras},
         year = 2018,
        month = mar,
       volume = {475},
       number = {1},
        pages = {648-675},
          doi = {10.1093/mnras/stx3112},
archivePrefix = {arXiv},
       eprint = {1707.03406},
 primaryClass = {astro-ph.GA},
       adsurl = {https://ui.adsabs.harvard.edu/abs/2018MNRAS.475..648P}
}

@ARTICLE{Springel-TNG-clustering,
       author = {{Springel}, Volker and {Pakmor}, R{\"u}diger and {Pillepich}, Annalisa and {Weinberger}, Rainer and {Nelson}, Dylan and {Hernquist}, Lars and {Vogelsberger}, Mark and {Genel}, Shy and {Torrey}, Paul and {Marinacci}, Federico and {Naiman}, Jill},
        title = "{First results from the IllustrisTNG simulations: matter and galaxy clustering}",
      journal = {\mnras},
         year = 2018,
        month = mar,
       volume = {475},
       number = {1},
        pages = {676-698},
          doi = {10.1093/mnras/stx3304},
archivePrefix = {arXiv},
       eprint = {1707.03397},
 primaryClass = {astro-ph.GA},
       adsurl = {https://ui.adsabs.harvard.edu/abs/2018MNRAS.475..676S}
}

@ARTICLE{Nelson-TNG-color,
       author = {{Nelson}, Dylan and {Pillepich}, Annalisa and {Springel}, Volker and {Weinberger}, Rainer and {Hernquist}, Lars and {Pakmor}, R{\"u}diger and {Genel}, Shy and {Torrey}, Paul and {Vogelsberger}, Mark and {Kauffmann}, Guinevere and {Marinacci}, Federico and {Naiman}, Jill},
        title = "{First results from the IllustrisTNG simulations: the galaxy colour bimodality}",
      journal = {\mnras},
         year = 2018,
        month = mar,
       volume = {475},
       number = {1},
        pages = {624-647},
          doi = {10.1093/mnras/stx3040},
archivePrefix = {arXiv},
       eprint = {1707.03395},
 primaryClass = {astro-ph.GA},
       adsurl = {https://ui.adsabs.harvard.edu/abs/2018MNRAS.475..624N}
}

@ARTICLE{Naiman-TNG,
       author = {{Naiman}, Jill P. and {Pillepich}, Annalisa and {Springel}, Volker and {Ramirez-Ruiz}, Enrico and {Torrey}, Paul and {Vogelsberger}, Mark and {Pakmor}, R{\"u}diger and {Nelson}, Dylan and {Marinacci}, Federico and {Hernquist}, Lars and {Weinberger}, Rainer and {Genel}, Shy},
        title = "{First results from the IllustrisTNG simulations: a tale of two elements - chemical evolution of magnesium and europium}",
      journal = {\mnras},
         year = 2018,
        month = jun,
       volume = {477},
       number = {1},
        pages = {1206-1224},
          doi = {10.1093/mnras/sty618},
archivePrefix = {arXiv},
       eprint = {1707.03401},
 primaryClass = {astro-ph.GA},
       adsurl = {https://ui.adsabs.harvard.edu/abs/2018MNRAS.477.1206N}
}

@ARTICLE{Marinacci-TNG,
       author = {{Marinacci}, Federico and {Vogelsberger}, Mark and {Pakmor}, R{\"u}diger and {Torrey}, Paul and {Springel}, Volker and {Hernquist}, Lars and {Nelson}, Dylan and {Weinberger}, Rainer and {Pillepich}, Annalisa and {Naiman}, Jill and {Genel}, Shy},
        title = "{First results from the IllustrisTNG simulations: radio haloes and magnetic fields}",
      journal = {\mnras},
         year = 2018,
        month = nov,
       volume = {480},
       number = {4},
        pages = {5113-5139},
          doi = {10.1093/mnras/sty2206},
archivePrefix = {arXiv},
       eprint = {1707.03396},
 primaryClass = {astro-ph.CO},
       adsurl = {https://ui.adsabs.harvard.edu/abs/2018MNRAS.480.5113M}
}

@ARTICLE{Truong2021,
       author = {{Truong}, Nhut and {Pillepich}, Annalisa and {Werner}, Norbert},
        title = "{Correlations between supermassive black holes and hot gas atmospheres in IllustrisTNG and X-ray observations}",
      journal = {\mnras},
         year = 2021,
        month = feb,
       volume = {501},
       number = {2},
        pages = {2210-2230},
          doi = {10.1093/mnras/staa3880},
archivePrefix = {arXiv},
       eprint = {2009.06634},
 primaryClass = {astro-ph.CO},
       adsurl = {https://ui.adsabs.harvard.edu/abs/2021MNRAS.501.2210T}
}

@ARTICLE{Davies2020,
       author = {{Davies}, Jonathan J. and {Crain}, Robert A. and {Oppenheimer}, Benjamin D. and {Schaye}, Joop},
        title = "{The quenching and morphological evolution of central galaxies is facilitated by the feedback-driven expulsion of circumgalactic gas}",
      journal = {\mnras},
         year = 2020,
        month = jan,
       volume = {491},
       number = {3},
        pages = {4462-4480},
          doi = {10.1093/mnras/stz3201},
archivePrefix = {arXiv},
       eprint = {1908.11380},
 primaryClass = {astro-ph.GA},
       adsurl = {https://ui.adsabs.harvard.edu/abs/2020MNRAS.491.4462D}
}

@ARTICLE{Zinger2020,
       author = {{Zinger}, Elad and {Pillepich}, Annalisa and {Nelson}, Dylan and {Weinberger}, Rainer and {Pakmor}, R{\"u}diger and {Springel}, Volker and {Hernquist}, Lars and {Marinacci}, Federico and {Vogelsberger}, Mark},
        title = "{Ejective and preventative: the IllustrisTNG black hole feedback and its effects on the thermodynamics of the gas within and around galaxies}",
      journal = {\mnras},
         year = 2020,
        month = nov,
       volume = {499},
       number = {1},
        pages = {768-792},
          doi = {10.1093/mnras/staa2607},
archivePrefix = {arXiv},
       eprint = {2004.06132},
 primaryClass = {astro-ph.GA},
       adsurl = {https://ui.adsabs.harvard.edu/abs/2020MNRAS.499..768Z}
}

@ARTICLE{Kurinchi-Vendhan2024,
       author = {{Kurinchi-Vendhan}, Shalini and {Farcy}, Marion and {Hirschmann}, Michaela and {Valentino}, Francesco},
        title = "{On the origin of star formation quenching in massive galaxies at z {\ensuremath{\gtrsim}} 3 in the cosmological simulations IllustrisTNG}",
      journal = {\mnras},
         year = 2024,
        month = nov,
       volume = {534},
       number = {4},
        pages = {3974-3988},
          doi = {10.1093/mnras/stae2297},
archivePrefix = {arXiv},
       eprint = {2310.03083},
 primaryClass = {astro-ph.GA},
       adsurl = {https://ui.adsabs.harvard.edu/abs/2024MNRAS.534.3974K}
}

@ARTICLE{Kurinchi-Vendhan2025,
       author = {{Kurinchi-Vendhan}, Shalini and {Rohr}, Eric and {Pillepich}, Annalisa and {Zinger}, Elad and {Ayromlou}, Mohammadreza and {Joshi}, Gandhali D.},
        title = "{Jellyfish galaxies with the IllustrisTNG simulations {\textendash} Supermassive black hole activity in dense environments with ram-pressure stripped satellites}",
      journal = {\mnras},
         year = 2025,
        month = sep,
       volume = {542},
       number = {3},
        pages = {1901-1922},
          doi = {10.1093/mnras/staf1280},
archivePrefix = {arXiv},
       eprint = {2506.05474},
 primaryClass = {astro-ph.GA},
       adsurl = {https://ui.adsabs.harvard.edu/abs/2025MNRAS.542.1901K}
}

@ARTICLE{Eagle-Crain,
       author = {{Crain}, Robert A. and {Schaye}, Joop and {Bower}, Richard G. and {Furlong}, Michelle and {Schaller}, Matthieu and {Theuns}, Tom and {Dalla Vecchia}, Claudio and {Frenk}, Carlos S. and {McCarthy}, Ian G. and {Helly}, John C. and {Jenkins}, Adrian and {Rosas-Guevara}, Yetli M. and {White}, Simon D.~M. and {Trayford}, James W.},
        title = "{The EAGLE simulations of galaxy formation: calibration of subgrid physics and model variations}",
      journal = {\mnras},
         year = 2015,
        month = jun,
       volume = {450},
       number = {2},
        pages = {1937-1961},
          doi = {10.1093/mnras/stv725},
archivePrefix = {arXiv},
       eprint = {1501.01311},
 primaryClass = {astro-ph.GA},
       adsurl = {https://ui.adsabs.harvard.edu/abs/2015MNRAS.450.1937C}
}

@ARTICLE{Eagle-Schaye,
       author = {{Schaye}, Joop and {Crain}, Robert A. and {Bower}, Richard G. and {Furlong}, Michelle and {Schaller}, Matthieu and {Theuns}, Tom and {Dalla Vecchia}, Claudio and {Frenk}, Carlos S. and {McCarthy}, I.~G. and {Helly}, John C. and {Jenkins}, Adrian and {Rosas-Guevara}, Y.~M. and {White}, Simon D.~M. and {Baes}, Maarten and {Booth}, C.~M. and {Camps}, Peter and {Navarro}, Julio F. and {Qu}, Yan and {Rahmati}, Alireza and {Sawala}, Till and {Thomas}, Peter A. and {Trayford}, James},
        title = "{The EAGLE project: simulating the evolution and assembly of galaxies and their environments}",
      journal = {\mnras},
         year = 2015,
        month = jan,
       volume = {446},
       number = {1},
        pages = {521-554},
          doi = {10.1093/mnras/stu2058},
archivePrefix = {arXiv},
       eprint = {1407.7040},
 primaryClass = {astro-ph.GA},
       adsurl = {https://ui.adsabs.harvard.edu/abs/2015MNRAS.446..521S}
}

@ARTICLE{Eagle-data-release,
       author = {{McAlpine}, S. and {Helly}, J.~C. and {Schaller}, M. and {Trayford}, J.~W. and {Qu}, Y. and {Furlong}, M. and {Bower}, R.~G. and {Crain}, R.~A. and {Schaye}, J. and {Theuns}, T. and {Dalla Vecchia}, C. and {Frenk}, C.~S. and {McCarthy}, I.~G. and {Jenkins}, A. and {Rosas-Guevara}, Y. and {White}, S.~D.~M. and {Baes}, M. and {Camps}, P. and {Lemson}, G.},
        title = "{The EAGLE simulations of galaxy formation: Public release of halo and galaxy catalogues}",
      journal = {Astronomy and Computing},
         year = 2016,
        month = apr,
       volume = {15},
        pages = {72-89},
          doi = {10.1016/j.ascom.2016.02.004},
archivePrefix = {arXiv},
       eprint = {1510.01320},
 primaryClass = {astro-ph.GA},
       adsurl = {https://ui.adsabs.harvard.edu/abs/2016A&C....15...72M}
}

@ARTICLE{Eagle-rapid-growth2018,
       author = {{McAlpine}, Stuart and {Bower}, Richard G. and {Rosario}, David J. and {Crain}, Robert A. and {Schaye}, Joop and {Theuns}, Tom},
        title = "{The rapid growth phase of supermassive black holes}",
      journal = {\mnras},
         year = 2018,
        month = dec,
       volume = {481},
       number = {3},
        pages = {3118-3128},
          doi = {10.1093/mnras/sty2489},
archivePrefix = {arXiv},
       eprint = {1805.08293},
 primaryClass = {astro-ph.GA},
       adsurl = {https://ui.adsabs.harvard.edu/abs/2018MNRAS.481.3118M}
}

@ARTICLE{McAlpine2017,
       author = {{McAlpine}, Stuart and {Bower}, Richard G. and {Harrison}, Chris M. and {Crain}, Robert A. and {Schaller}, Matthieu and {Schaye}, Joop and {Theuns}, Tom},
        title = "{The link between galaxy and black hole growth in the eagle simulation}",
      journal = {\mnras},
         year = 2017,
        month = jul,
       volume = {468},
       number = {3},
        pages = {3395-3407},
          doi = {10.1093/mnras/stx658},
archivePrefix = {arXiv},
       eprint = {1701.01122},
 primaryClass = {astro-ph.GA},
       adsurl = {https://ui.adsabs.harvard.edu/abs/2017MNRAS.468.3395M}
}

@ARTICLE{Simba-Dave,
       author = {{Dav{\'e}}, Romeel and {Angl{\'e}s-Alc{\'a}zar}, Daniel and {Narayanan}, Desika and {Li}, Qi and {Rafieferantsoa}, Mika H. and {Appleby}, Sarah},
        title = "{SIMBA: Cosmological simulations with black hole growth and feedback}",
      journal = {\mnras},
         year = 2019,
        month = jun,
       volume = {486},
       number = {2},
        pages = {2827-2849},
          doi = {10.1093/mnras/stz937},
archivePrefix = {arXiv},
       eprint = {1901.10203},
 primaryClass = {astro-ph.GA},
       adsurl = {https://ui.adsabs.harvard.edu/abs/2019MNRAS.486.2827D}
}

@ARTICLE{Simba-Wu,
       author = {{Wu}, Xiaohan and {Dav{\'e}}, Romeel and {Tacchella}, Sandro and {Lotz}, Jennifer},
        title = "{Photometric properties of reionization-epoch galaxies in the SIMBA simulations}",
      journal = {\mnras},
         year = 2020,
        month = jun,
       volume = {494},
       number = {4},
        pages = {5636-5651},
          doi = {10.1093/mnras/staa1044},
archivePrefix = {arXiv},
       eprint = {1911.06330},
 primaryClass = {astro-ph.GA},
       adsurl = {https://ui.adsabs.harvard.edu/abs/2020MNRAS.494.5636W}
}

@ARTICLE{Davis_fof,
       author = {{Davis}, M. and {Efstathiou}, G. and {Frenk}, C.~S. and {White}, S.~D.~M.},
        title = "{The evolution of large-scale structure in a universe dominated by cold dark matter}",
      journal = {\apj},
         year = 1985,
        month = may,
       volume = {292},
        pages = {371-394},
          doi = {10.1086/163168},
       adsurl = {https://ui.adsabs.harvard.edu/abs/1985ApJ...292..371D}
}

@ARTICLE{Springel2001,
       author = {{Springel}, Volker and {Yoshida}, Naoki and {White}, Simon D.~M.},
        title = "{GADGET: a code for collisionless and gasdynamical cosmological simulations}",
      journal = {\na},
         year = 2001,
        month = apr,
       volume = {6},
       number = {2},
        pages = {79-117},
          doi = {10.1016/S1384-1076(01)00042-2},
archivePrefix = {arXiv},
       eprint = {astro-ph/0003162},
 primaryClass = {astro-ph},
       adsurl = {https://ui.adsabs.harvard.edu/abs/2001NewA....6...79S}
}

@ARTICLE{Eagle-feedback2009,
       author = {{Booth}, C.~M. and {Schaye}, Joop},
        title = "{Cosmological simulations of the growth of supermassive black holes and feedback from active galactic nuclei: method and tests}",
      journal = {\mnras},
         year = 2009,
        month = sep,
       volume = {398},
       number = {1},
        pages = {53-74},
          doi = {10.1111/j.1365-2966.2009.15043.x},
archivePrefix = {arXiv},
       eprint = {0904.2572},
 primaryClass = {astro-ph.CO},
       adsurl = {https://ui.adsabs.harvard.edu/abs/2009MNRAS.398...53B}
}

@ARTICLE{Flamingo-Schaye2023,
       author = {{Schaye}, Joop and {Kugel}, Roi and {Schaller}, Matthieu and {Helly}, John C. and {Braspenning}, Joey and {Elbers}, Willem and {McCarthy}, Ian G. and {van Daalen}, Marcel P. and {Vandenbroucke}, Bert and {Frenk}, Carlos S. and {Kwan}, Juliana and {Salcido}, Jaime and {Bah{\'e}}, Yannick M. and {Borrow}, Josh and {Chaikin}, Evgenii and {Hahn}, Oliver and {Hu{\v{s}}ko}, Filip and {Jenkins}, Adrian and {Lacey}, Cedric G. and {Nobels}, Folkert S.~J.},
        title = "{The FLAMINGO project: cosmological hydrodynamical simulations for large-scale structure and galaxy cluster surveys}",
      journal = {\mnras},
         year = 2023,
        month = dec,
       volume = {526},
       number = {4},
        pages = {4978-5020},
          doi = {10.1093/mnras/stad2419},
archivePrefix = {arXiv},
       eprint = {2306.04024},
 primaryClass = {astro-ph.CO},
       adsurl = {https://ui.adsabs.harvard.edu/abs/2023MNRAS.526.4978S}
}

@ARTICLE{Flamingo-Roi2023,
       author = {{Kugel}, Roi and {Schaye}, Joop and {Schaller}, Matthieu and {Helly}, John C. and {Braspenning}, Joey and {Elbers}, Willem and {Frenk}, Carlos S. and {McCarthy}, Ian G. and {Kwan}, Juliana and {Salcido}, Jaime and {van Daalen}, Marcel P. and {Vandenbroucke}, Bert and {Bah{\'e}}, Yannick M. and {Borrow}, Josh and {Chaikin}, Evgenii and {Hu{\v{s}}ko}, Filip and {Jenkins}, Adrian and {Lacey}, Cedric G. and {Nobels}, Folkert S.~J. and {Vernon}, Ian},
        title = "{FLAMINGO: calibrating large cosmological hydrodynamical simulations with machine learning}",
      journal = {\mnras},
         year = 2023,
        month = dec,
       volume = {526},
       number = {4},
        pages = {6103-6127},
          doi = {10.1093/mnras/stad2540},
archivePrefix = {arXiv},
       eprint = {2306.05492},
 primaryClass = {astro-ph.CO},
       adsurl = {https://ui.adsabs.harvard.edu/abs/2023MNRAS.526.6103K}
}

@ARTICLE{Helly2026,
       author = {{Helly}, John C. and {McGibbon}, Robert J. and {Schaye}, Joop and {Schaller}, Matthieu and {McDonald}, William and {Braspenning}, Joey and {Broxterman}, Jeger C. and {Costello}, Emily E. and {Elbers}, Willem and {Forouhar Moreno}, Victor J. and {Frenk}, Carlos S. and {Jenkins}, Adrian and {Kugel}, Roi and {McCarthy}, Ian G. and {Salcido}, Jaime and {van Daalen}, Marcel P. and {Vandenbroucke}, Bert and {Yang}, Tianyi},
        title = "{The FLAMINGO simulations data release}",
      journal = {arXiv e-prints},
         year = 2026,
        month = apr,
          eid = {arXiv:2604.24324},
        pages = {arXiv:2604.24324},
          doi = {10.48550/arXiv.2604.24324},
archivePrefix = {arXiv},
       eprint = {2604.24324},
 primaryClass = {astro-ph.CO},
       adsurl = {https://ui.adsabs.harvard.edu/abs/2026arXiv260424324H}
}

@ARTICLE{Bower2017,
       author = {{Bower}, Richard G. and {Schaye}, Joop and {Frenk}, Carlos S. and {Theuns}, Tom and {Schaller}, Matthieu and {Crain}, Robert A. and {McAlpine}, Stuart},
        title = "{The dark nemesis of galaxy formation: why hot haloes trigger black hole growth and bring star formation to an end}",
      journal = {\mnras},
         year = 2017,
        month = feb,
       volume = {465},
       number = {1},
        pages = {32-44},
          doi = {10.1093/mnras/stw2735},
archivePrefix = {arXiv},
       eprint = {1607.07445},
 primaryClass = {astro-ph.GA},
       adsurl = {https://ui.adsabs.harvard.edu/abs/2017MNRAS.465...32B}
}

@ARTICLE{Dubois2015,
       author = {{Dubois}, Yohan and {Volonteri}, Marta and {Silk}, Joseph and {Devriendt}, Julien and {Slyz}, Adrianne and {Teyssier}, Romain},
        title = "{Black hole evolution - I. Supernova-regulated black hole growth}",
      journal = {\mnras},
         year = 2015,
        month = sep,
       volume = {452},
       number = {2},
        pages = {1502-1518},
          doi = {10.1093/mnras/stv1416},
archivePrefix = {arXiv},
       eprint = {1504.00018},
 primaryClass = {astro-ph.GA},
       adsurl = {https://ui.adsabs.harvard.edu/abs/2015MNRAS.452.1502D}
}

@BOOK{Peebles,
       author = {{Peebles}, P.~J.~E.},
        title = "{The large-scale structure of the universe}",
     publisher = "Princeton University Press",
         year = 1980,
       adsurl = {https://ui.adsabs.harvard.edu/abs/1980lssu.book.....P}
}

@ARTICLE{Eddington,
       author = {{Eddington}, A.~S.},
        title = "{Das Strahlungsgleichgewicht der Sterne}",
      journal = {Zeitschrift fur Physik},
         year = 1921,
        month = dec,
       volume = {7},
       number = {1},
        pages = {351-397},
          doi = {10.1007/BF01332806},
       adsurl = {https://ui.adsabs.harvard.edu/abs/1921ZPhy....7..351E}
}

@ARTICLE{LyndenBell69,
       author = {{Lynden-Bell}, D.},
        title = "{Galactic Nuclei as Collapsed Old Quasars}",
      journal = {\nat},
         year = 1969,
        month = aug,
       volume = {223},
       number = {5207},
        pages = {690-694},
          doi = {10.1038/223690a0},
       adsurl = {https://ui.adsabs.harvard.edu/abs/1969Natur.223..690L}
}

@ARTICLE{Schindler2023,
       author = {{Schindler}, Jan-Torge and {Ba{\~n}ados}, Eduardo and {Connor}, Thomas and {Decarli}, Roberto and {Fan}, Xiaohui and {Farina}, Emanuele Paolo and {Mazzucchelli}, Chiara and {Nanni}, Riccardo and {Rix}, Hans-Walter and {Stern}, Daniel and {Venemans}, Bram P. and {Walter}, Fabian},
        title = "{The Pan-STARRS1 z > 5.6 Quasar Survey. III. The z {\ensuremath{\approx}} 6 Quasar Luminosity Function}",
      journal = {\apj},
         year = 2023,
        month = jan,
       volume = {943},
       number = {1},
          eid = {67},
        pages = {67},
          doi = {10.3847/1538-4357/aca7ca},
archivePrefix = {arXiv},
       eprint = {2212.04179},
 primaryClass = {astro-ph.GA},
       adsurl = {https://ui.adsabs.harvard.edu/abs/2023ApJ...943...67S}
}

@ARTICLE{Press_and_davis_1982,
       author = {{Press}, W.~H. and {Davis}, M.},
        title = "{How to identify and weigh virialized clusters of galaxies in a complete redshift catalog}",
      journal = {\apj},
         year = 1982,
        month = aug,
       volume = {259},
        pages = {449-473},
          doi = {10.1086/160183},
       adsurl = {https://ui.adsabs.harvard.edu/abs/1982ApJ...259..449P}
}

@ARTICLE{Mascarell2025,
       author = {{Giner Mascarell}, Mariona and {Eilers}, Anna-Christina and {Storey-Fisher}, Kate},
        title = "{Quasar clustering and duty cycle measurements at $0\leq z\leq 4$ with the Gaia-unWISE Catalog}",
      journal = {arXiv e-prints},
         year = 2025,
        month = nov,
          eid = {arXiv:2511.17413},
        pages = {arXiv:2511.17413},
          doi = {10.48550/arXiv.2511.17413},
archivePrefix = {arXiv},
       eprint = {2511.17413},
 primaryClass = {astro-ph.GA},
       adsurl = {https://ui.adsabs.harvard.edu/abs/2025arXiv251117413G}
}

@ARTICLE{sdss2000,
       author = {{York}, Donald G. and {Adelman}, J. and {Anderson}, Jr., John E. and {Anderson}, Scott F. and {Annis}, James and {Bahcall}, Neta A. and {Bakken}, J.~A. and {Barkhouser}, Robert and {Bastian}, Steven and {Berman}, Eileen and {Boroski}, William N. and {Bracker}, Steve and {Briegel}, Charlie and {Briggs}, John W. and {Brinkmann}, J. and {Brunner}, Robert and {Burles}, Scott and {Carey}, Larry and {Carr}, Michael A. and {Castander}, Francisco J. and {Chen}, Bing and {Colestock}, Patrick L. and {Connolly}, A.~J. and {Crocker}, J.~H. and {Csabai}, Istv{\'a}n and {Czarapata}, Paul C. and {Davis}, John Eric and {Doi}, Mamoru and {Dombeck}, Tom and {Eisenstein}, Daniel and {Ellman}, Nancy and {Elms}, Brian R. and {Evans}, Michael L. and {Fan}, Xiaohui and {Federwitz}, Glenn R. and {Fiscelli}, Larry and {Friedman}, Scott and {Frieman}, Joshua A. and {Fukugita}, Masataka and {Gillespie}, Bruce and {Gunn}, James E. and {Gurbani}, Vijay K. and {de Haas}, Ernst and {Haldeman}, Merle and {Harris}, Frederick H. and {Hayes}, J. and {Heckman}, Timothy M. and {Hennessy}, G.~S. and {Hindsley}, Robert B. and {Holm}, Scott and {Holmgren}, Donald J. and {Huang}, Chi-hao and {Hull}, Charles and {Husby}, Don and {Ichikawa}, Shin-Ichi and {Ichikawa}, Takashi and {Ivezi{\'c}}, {\v{Z}}eljko and {Kent}, Stephen and {Kim}, Rita S.~J. and {Kinney}, E. and {Klaene}, Mark and {Kleinman}, A.~N. and {Kleinman}, S. and {Knapp}, G.~R. and {Korienek}, John and {Kron}, Richard G. and {Kunszt}, Peter Z. and {Lamb}, D.~Q. and {Lee}, B. and {Leger}, R. French and {Limmongkol}, Siriluk and {Lindenmeyer}, Carl and {Long}, Daniel C. and {Loomis}, Craig and {Loveday}, Jon and {Lucinio}, Rich and {Lupton}, Robert H. and {MacKinnon}, Bryan and {Mannery}, Edward J. and {Mantsch}, P.~M. and {Margon}, Bruce and {McGehee}, Peregrine and {McKay}, Timothy A. and {Meiksin}, Avery and {Merelli}, Aronne and {Monet}, David G. and {Munn}, Jeffrey A. and {Narayanan}, Vijay K. and {Nash}, Thomas and {Neilsen}, Eric and {Neswold}, Rich and {Newberg}, Heidi Jo and {Nichol}, R.~C. and {Nicinski}, Tom and {Nonino}, Mario and {Okada}, Norio and {Okamura}, Sadanori and {Ostriker}, Jeremiah P. and {Owen}, Russell and {Pauls}, A. George and {Peoples}, John and {Peterson}, R.~L. and {Petravick}, Donald and {Pier}, Jeffrey R. and {Pope}, Adrian and {Pordes}, Ruth and {Prosapio}, Angela and {Rechenmacher}, Ron and {Quinn}, Thomas R. and {Richards}, Gordon T. and {Richmond}, Michael W. and {Rivetta}, Claudio H. and {Rockosi}, Constance M. and {Ruthmansdorfer}, Kurt and {Sandford}, Dale and {Schlegel}, David J. and {Schneider}, Donald P. and {Sekiguchi}, Maki and {Sergey}, Gary and {Shimasaku}, Kazuhiro and {Siegmund}, Walter A. and {Smee}, Stephen and {Smith}, J. Allyn and {Snedden}, S. and {Stone}, R. and {Stoughton}, Chris and {Strauss}, Michael A. and {Stubbs}, Christopher and {SubbaRao}, Mark and {Szalay}, Alexander S. and {Szapudi}, Istvan and {Szokoly}, Gyula P. and {Thakar}, Anirudda R. and {Tremonti}, Christy and {Tucker}, Douglas L. and {Uomoto}, Alan and {Vanden Berk}, Dan and {Vogeley}, Michael S. and {Waddell}, Patrick and {Wang}, Shu-i. and {Watanabe}, Masaru and {Weinberg}, David H. and {Yanny}, Brian and {Yasuda}, Naoki and {SDSS Collaboration}},
        title = "{The Sloan Digital Sky Survey: Technical Summary}",
      journal = {\aj},
         year = 2000,
        month = sep,
       volume = {120},
       number = {3},
        pages = {1579-1587},
          doi = {10.1086/301513},
archivePrefix = {arXiv},
       eprint = {astro-ph/0006396},
 primaryClass = {astro-ph},
       adsurl = {https://ui.adsabs.harvard.edu/abs/2000AJ....120.1579Y}
}

@ARTICLE{2df2004,
       author = {{Croom}, S.~M. and {Smith}, R.~J. and {Boyle}, B.~J. and {Shanks}, T. and {Miller}, L. and {Outram}, P.~J. and {Loaring}, N.~S.},
        title = "{The 2dF QSO Redshift Survey - XII. The spectroscopic catalogue and luminosity function}",
      journal = {\mnras},
         year = 2004,
        month = apr,
       volume = {349},
       number = {4},
        pages = {1397-1418},
          doi = {10.1111/j.1365-2966.2004.07619.x},
archivePrefix = {arXiv},
       eprint = {astro-ph/0403040},
 primaryClass = {astro-ph},
       adsurl = {https://ui.adsabs.harvard.edu/abs/2004MNRAS.349.1397C}
}

@ARTICLE{Croom2005,
       author = {{Croom}, Scott M. and {Boyle}, B.~J. and {Shanks}, T. and {Smith}, R.~J. and {Miller}, L. and {Outram}, P.~J. and {Loaring}, N.~S. and {Hoyle}, F. and {da {\^A}ngela}, J.},
        title = "{The 2dF QSO Redshift Survey - XIV. Structure and evolution from the two-point correlation function}",
      journal = {\mnras},
         year = 2005,
        month = jan,
       volume = {356},
       number = {2},
        pages = {415-438},
          doi = {10.1111/j.1365-2966.2004.08379.x},
archivePrefix = {arXiv},
       eprint = {astro-ph/0409314},
 primaryClass = {astro-ph},
       adsurl = {https://ui.adsabs.harvard.edu/abs/2005MNRAS.356..415C}
}

@ARTICLE{Ross2009,
       author = {{Ross}, Nicholas P. and {Shen}, Yue and {Strauss}, Michael A. and {Vanden Berk}, Daniel E. and {Connolly}, Andrew J. and {Richards}, Gordon T. and {Schneider}, Donald P. and {Weinberg}, David H. and {Hall}, Patrick B. and {Bahcall}, Neta A. and {Brunner}, Robert J.},
        title = "{Clustering of Low-redshift (z <= 2.2) Quasars from the Sloan Digital Sky Survey}",
      journal = {\apj},
         year = 2009,
        month = jun,
       volume = {697},
       number = {2},
        pages = {1634-1655},
          doi = {10.1088/0004-637X/697/2/1634},
archivePrefix = {arXiv},
       eprint = {0903.3230},
 primaryClass = {astro-ph.CO},
       adsurl = {https://ui.adsabs.harvard.edu/abs/2009ApJ...697.1634R}
}

@ARTICLE{White2012,
       author = {{White}, Martin and {Myers}, Adam D. and {Ross}, Nicholas P. and {Schlegel}, David J. and {Hennawi}, Joseph F. and {Shen}, Yue and {McGreer}, Ian and {Strauss}, Michael A. and {Bolton}, Adam S. and {Bovy}, Jo and {Fan}, X. and {Miralda-Escude}, Jordi and {Palanque-Delabrouille}, N. and {Paris}, I. and {Petitjean}, P. and {Schneider}, D.~P. and {Viel}, M. and {Weinberg}, David H. and {Yeche}, Ch. and {Zehavi}, I. and {Pan}, K. and {Snedden}, S. and {Bizyaev}, D. and {Brewington}, H. and {Brinkmann}, J. and {Malanushenko}, V. and {Malanushenko}, E. and {Oravetz}, D. and {Simmons}, A. and {Sheldon}, A. and {Weaver}, Benjamin A.},
        title = "{The clustering of intermediate-redshift quasars as measured by the Baryon Oscillation Spectroscopic Survey}",
      journal = {\mnras},
         year = 2012,
        month = aug,
       volume = {424},
       number = {2},
        pages = {933-950},
          doi = {10.1111/j.1365-2966.2012.21251.x},
archivePrefix = {arXiv},
       eprint = {1203.5306},
 primaryClass = {astro-ph.CO},
       adsurl = {https://ui.adsabs.harvard.edu/abs/2012MNRAS.424..933W}
}

@ARTICLE{Sarah2015,
       author = {{Eftekharzadeh}, Sarah and {Myers}, Adam D. and {White}, Martin and {Weinberg}, David H. and {Schneider}, Donald P. and {Shen}, Yue and {Font-Ribera}, Andreu and {Ross}, Nicholas P. and {Paris}, Isabelle and {Streblyanska}, Alina},
        title = "{Clustering of intermediate redshift quasars using the final SDSS III-BOSS sample}",
      journal = {\mnras},
         year = 2015,
        month = nov,
       volume = {453},
       number = {3},
        pages = {2779-2798},
          doi = {10.1093/mnras/stv1763},
archivePrefix = {arXiv},
       eprint = {1507.08380},
 primaryClass = {astro-ph.CO},
       adsurl = {https://ui.adsabs.harvard.edu/abs/2015MNRAS.453.2779E}
}

@ARTICLE{Shen2007,
       author = {{Shen}, Yue and {Strauss}, Michael A. and {Oguri}, Masamune and {Hennawi}, Joseph F. and {Fan}, Xiaohui and {Richards}, Gordon T. and {Hall}, Patrick B. and {Gunn}, James E. and {Schneider}, Donald P. and {Szalay}, Alexander S. and {Thakar}, Anirudda R. and {Vanden Berk}, Daniel E. and {Anderson}, Scott F. and {Bahcall}, Neta A. and {Connolly}, Andrew J. and {Knapp}, Gillian R.},
        title = "{Clustering of High-Redshift (z >= 2.9) Quasars from the Sloan Digital Sky Survey}",
      journal = {\aj},
         year = 2007,
        month = may,
       volume = {133},
       number = {5},
        pages = {2222-2241},
          doi = {10.1086/513517},
archivePrefix = {arXiv},
       eprint = {astro-ph/0702214},
 primaryClass = {astro-ph},
       adsurl = {https://ui.adsabs.harvard.edu/abs/2007AJ....133.2222S}
}

@ARTICLE{Pizzati2024b,
       author = {{Pizzati}, Elia and {Hennawi}, Joseph F. and {Schaye}, Joop and {Schaller}, Matthieu and {Eilers}, Anna-Christina and {Wang}, Feige and {Frenk}, Carlos S. and {Elbers}, Willem and {Helly}, John C. and {Mackenzie}, Ruari and {Matthee}, Jorryt and {Bordoloi}, Rongmon and {Kashino}, Daichi and {Naidu}, Rohan P. and {Yue}, Minghao},
        title = "{A unified model for the clustering of quasars and galaxies at z {\ensuremath{\approx}} 6}",
      journal = {\mnras},
         year = 2024,
        month = nov,
       volume = {534},
       number = {4},
        pages = {3155-3175},
          doi = {10.1093/mnras/stae2307},
archivePrefix = {arXiv},
       eprint = {2403.12140},
 primaryClass = {astro-ph.GA},
       adsurl = {https://ui.adsabs.harvard.edu/abs/2024MNRAS.534.3155P}
}

@ARTICLE{Eilers,
       author = {{Eilers}, Anna-Christina and {Mackenzie}, Ruari and {Pizzati}, Elia and {Matthee}, Jorryt and {Hennawi}, Joseph F. and {Zhang}, Haowen and {Bordoloi}, Rongmon and {Kashino}, Daichi and {Lilly}, Simon J. and {Naidu}, Rohan P. and {Simcoe}, Robert A. and {Yue}, Minghao and {Frenk}, Carlos S. and {Helly}, John C. and {Schaller}, Matthieu and {Schaye}, Joop},
        title = "{EIGER. VI. The Correlation Function, Host Halo Mass, and Duty Cycle of Luminous Quasars at z {\ensuremath{\gtrsim}} 6}",
      journal = {\apj},
         year = 2024,
        month = oct,
       volume = {974},
       number = {2},
          eid = {275},
        pages = {275},
          doi = {10.3847/1538-4357/ad778b},
archivePrefix = {arXiv},
       eprint = {2403.07986},
 primaryClass = {astro-ph.GA},
       adsurl = {https://ui.adsabs.harvard.edu/abs/2024ApJ...974..275E}
}

@ARTICLE{Arita2023,
       author = {{Arita}, Junya and {Kashikawa}, Nobunari and {Matsuoka}, Yoshiki and {He}, Wanqiu and {Ito}, Kei and {Liang}, Yongming and {Ishimoto}, Rikako and {Yoshioka}, Takehiro and {Takeda}, Yoshihiro and {Iwasawa}, Kazushi and {Onoue}, Masafusa and {Toba}, Yoshiki and {Imanishi}, Masatoshi},
        title = "{Subaru High-z Exploration of Low-luminosity Quasars (SHELLQs). XVIII. The Dark Matter Halo Mass of Quasars at z   6}",
      journal = {\apj},
         year = 2023,
        month = sep,
       volume = {954},
       number = {2},
          eid = {210},
        pages = {210},
          doi = {10.3847/1538-4357/ace43a},
archivePrefix = {arXiv},
       eprint = {2307.02531},
 primaryClass = {astro-ph.GA},
       adsurl = {https://ui.adsabs.harvard.edu/abs/2023ApJ...954..210A}
}

@ARTICLE{Timlin2018,
       author = {{Timlin}, John D. and {Ross}, Nicholas P. and {Richards}, Gordon T. and {Myers}, Adam D. and {Pellegrino}, Andrew and {Bauer}, Franz E. and {Lacy}, Mark and {Schneider}, Donald P. and {Wollack}, Edward J. and {Zakamska}, Nadia L.},
        title = "{The Clustering of High-redshift (2.9 {\ensuremath{\leq}} z {\ensuremath{\leq}} 5.1) Quasars in SDSS Stripe 82}",
      journal = {\apj},
         year = 2018,
        month = may,
       volume = {859},
       number = {1},
          eid = {20},
        pages = {20},
          doi = {10.3847/1538-4357/aab9ac},
archivePrefix = {arXiv},
       eprint = {1712.03128},
 primaryClass = {astro-ph.GA},
       adsurl = {https://ui.adsabs.harvard.edu/abs/2018ApJ...859...20T}
}

@ARTICLE{He2018,
       author = {{He}, Wanqiu and {Akiyama}, Masayuki and {Bosch}, James and {Enoki}, Motohiro and {Harikane}, Yuichi and {Ikeda}, Hiroyuki and {Kashikawa}, Nobunari and {Kawaguchi}, Toshihiro and {Komiyama}, Yutaka and {Lee}, Chien-Hsiu and {Matsuoka}, Yoshiki and {Miyazaki}, Satoshi and {Nagao}, Tohru and {Nagashima}, Masahiro and {Niida}, Mana and {Nishizawa}, Atsushi J. and {Oguri}, Masamune and {Onoue}, Masafusa and {Oogi}, Taira and {Ouchi}, Masami and {Schulze}, Andreas and {Shirasaki}, Yuji and {Silverman}, John D. and {Tanaka}, Manobu M. and {Tanaka}, Masayuki and {Toba}, Yoshiki and {Uchiyama}, Hisakazu and {Yamashita}, Takuji},
        title = "{Clustering of quasars in a wide luminosity range at redshift 4 with Subaru Hyper Suprime-Cam Wide-field imaging}",
      journal = {\pasj},
         year = 2018,
        month = jan,
       volume = {70},
          eid = {S33},
        pages = {S33},
          doi = {10.1093/pasj/psx129},
archivePrefix = {arXiv},
       eprint = {1704.08461},
 primaryClass = {astro-ph.GA},
       adsurl = {https://ui.adsabs.harvard.edu/abs/2018PASJ...70S..33H}
}

@ARTICLE{Chen2022,
       author = {{Chen}, Huanqing and {Eilers}, Anna-Christina and {Bosman}, Sarah E.~I. and {Gnedin}, Nickolay Y. and {Fan}, Xiaohui and {Wang}, Feige and {Yang}, Jinyi and {D'Odorico}, Valentina and {Becker}, George D. and {Bischetti}, Manuela and {Mazzucchelli}, Chiara and {Mesinger}, Andrei and {Pallottini}, Andrea},
        title = "{Measuring the Density Fields around Bright Quasars at z   6 with XQR-30 Spectra}",
      journal = {\apj},
         year = 2022,
        month = may,
       volume = {931},
       number = {1},
          eid = {29},
        pages = {29},
          doi = {10.3847/1538-4357/ac658d},
archivePrefix = {arXiv},
       eprint = {2110.13917},
 primaryClass = {astro-ph.GA},
       adsurl = {https://ui.adsabs.harvard.edu/abs/2022ApJ...931...29C}
}

@ARTICLE{Qinyue2025,
       author = {{Fei}, Qinyue and {Silverman}, John D. and {Fujimoto}, Seiji and {Wang}, Ran and {Ho}, Luis C. and {Bischetti}, Manuela and {Carniani}, Stefano and {Ginolfi}, Michele and {Jones}, Gareth and {Maiolino}, Roberto and {Rujopakarn}, Wiphu and {F{\"o}rster Schreiber}, N.~M. and {Espejo Salcedo}, Juan M. and {Lee}, L.~L.},
        title = "{Assessing the Dark Matter Content of Two Quasar Host Galaxies at z {\ensuremath{\sim}} 6 through Gas Kinematics}",
      journal = {\apj},
         year = 2025,
        month = feb,
       volume = {980},
       number = {1},
          eid = {84},
        pages = {84},
          doi = {10.3847/1538-4357/ada145},
archivePrefix = {arXiv},
       eprint = {2501.09077},
 primaryClass = {astro-ph.GA},
       adsurl = {https://ui.adsabs.harvard.edu/abs/2025ApJ...980...84F}
}

@ARTICLE{Leauthaud2015,
       author = {{Leauthaud}, Alexie and {J. Benson}, Andrew and {Civano}, Francesca and {L. Coil}, Alison and {Bundy}, Kevin and {Massey}, Richard and {Schramm}, Malte and {Schulze}, Andreas and {Capak}, Peter and {Elvis}, Martin and {Kulier}, Andrea and {Rhodes}, Jason},
        title = "{The dark matter haloes of moderate luminosity X-ray AGN as determined from weak gravitational lensing and host stellar masses}",
      journal = {\mnras},
         year = 2015,
        month = jan,
       volume = {446},
       number = {2},
        pages = {1874-1888},
          doi = {10.1093/mnras/stu2210},
archivePrefix = {arXiv},
       eprint = {1410.5817},
 primaryClass = {astro-ph.GA},
       adsurl = {https://ui.adsabs.harvard.edu/abs/2015MNRAS.446.1874L}
}

@ARTICLE{Viitanen2019,
       author = {{Viitanen}, A. and {Allevato}, V. and {Finoguenov}, A. and {Bongiorno}, A. and {Cappelluti}, N. and {Gilli}, R. and {Miyaji}, T. and {Salvato}, M.},
        title = "{The XMM-Newton wide field survey in the COSMOS field: Clustering dependence of X-ray selected AGN on host galaxy properties}",
      journal = {\aap},
         year = 2019,
        month = sep,
       volume = {629},
          eid = {A14},
        pages = {A14},
          doi = {10.1051/0004-6361/201935186},
archivePrefix = {arXiv},
       eprint = {1906.07911},
 primaryClass = {astro-ph.GA},
       adsurl = {https://ui.adsabs.harvard.edu/abs/2019A&A...629A..14V}
}

@ARTICLE{Comparat2023,
       author = {{Comparat}, Johan and {Luo}, Wentao and {Merloni}, Andrea and {More}, Surhud and {Salvato}, Mara and {Krumpe}, Mirko and {Miyaji}, Takamitsu and {Brandt}, William and {Georgakakis}, Antonis and {Akiyama}, Masayuki and {Buchner}, Johannes and {Dwelly}, Tom and {Kawaguchi}, Toshihiro and {Liu}, Teng and {Nagao}, Tohru and {Nandra}, Kirpal and {Silverman}, John and {Toba}, Yoshiki and {Anderson}, Scott F. and {Kollmeier}, Juna},
        title = "{The cosmic web of X-ray active galactic nuclei seen through the eROSITA Final Equatorial Depth Survey (eFEDS)}",
      journal = {\aap},
         year = 2023,
        month = may,
       volume = {673},
          eid = {A122},
        pages = {A122},
          doi = {10.1051/0004-6361/202245726},
archivePrefix = {arXiv},
       eprint = {2301.01388},
 primaryClass = {astro-ph.GA},
       adsurl = {https://ui.adsabs.harvard.edu/abs/2023A&A...673A.122C}
}

@ARTICLE{Mountrichas2026,
       author = {{Mountrichas}, G. and {Carrera}, F.~J. and {Shankar}, F. and {Georgakakis}, A.},
        title = "{Large-scale environments of star-forming active galactic nuclei: How black hole mass, accretion rate, and luminosity connect to dark matter halos}",
      journal = {arXiv e-prints},
         year = 2026,
        month = mar,
          eid = {arXiv:2603.11169},
        pages = {arXiv:2603.11169},
          doi = {10.48550/arXiv.2603.11169},
archivePrefix = {arXiv},
       eprint = {2603.11169},
 primaryClass = {astro-ph.GA},
       adsurl = {https://ui.adsabs.harvard.edu/abs/2026arXiv260311169M}
}

@ARTICLE{Schindler2026,
       author = {{Schindler}, Jan-Torge and {Hennawi}, Joseph F. and {Davies}, Frederick B. and {Bosman}, Sarah E.~I. and {Wang}, Feige and {Yang}, Jinyi and {Eilers}, Anna-Christina and {Fan}, Xiaohui and {Kakiichi}, Koki and {Pizzati}, Elia and {Nanni}, Riccardo},
        title = "{A first look at quasar-galaxy clustering at z ≃ 7.3}",
      journal = {\aap},
         year = 2026,
        month = apr,
       volume = {708},
          eid = {A160},
        pages = {A160},
          doi = {10.1051/0004-6361/202557623},
archivePrefix = {arXiv},
       eprint = {2510.08455},
 primaryClass = {astro-ph.GA},
       adsurl = {https://ui.adsabs.harvard.edu/abs/2026A&A...708A.160S}
}

@ARTICLE{Meng2026,
       author = {{Meng}, Hao and {Zhang}, Huanian and {Ye}, Guangping},
        title = "{Probing The Dark Matter Halo of High-redshift Quasar from Wide-Field Clustering Analysis}",
      journal = {arXiv e-prints},
         year = 2026,
        month = feb,
          eid = {arXiv:2602.02778},
        pages = {arXiv:2602.02778},
          doi = {10.48550/arXiv.2602.02778},
archivePrefix = {arXiv},
       eprint = {2602.02778},
 primaryClass = {astro-ph.GA},
       adsurl = {https://ui.adsabs.harvard.edu/abs/2026arXiv260202778M}
}

@ARTICLE{Magorrian98,
       author = {{Magorrian}, John and {Tremaine}, Scott and {Richstone}, Douglas and {Bender}, Ralf and {Bower}, Gary and {Dressler}, Alan and {Faber}, S.~M. and {Gebhardt}, Karl and {Green}, Richard and {Grillmair}, Carl and {Kormendy}, John and {Lauer}, Tod},
        title = "{The Demography of Massive Dark Objects in Galaxy Centers}",
      journal = {\aj},
         year = 1998,
        month = jun,
       volume = {115},
       number = {6},
        pages = {2285-2305},
          doi = {10.1086/300353},
archivePrefix = {arXiv},
       eprint = {astro-ph/9708072},
 primaryClass = {astro-ph},
       adsurl = {https://ui.adsabs.harvard.edu/abs/1998AJ....115.2285M}
}

@ARTICLE{KormendyRichstone95,
       author = {{Kormendy}, John and {Richstone}, Douglas},
        title = "{Inward Bound---The Search For Supermassive Black Holes In Galactic Nuclei}",
      journal = {\araa},
         year = 1995,
        month = jan,
       volume = {33},
        pages = {581},
          doi = {10.1146/annurev.aa.33.090195.003053},
       adsurl = {https://ui.adsabs.harvard.edu/abs/1995ARA&A..33..581K}
}

@ARTICLE{Ferrarese2000,
       author = {{Ferrarese}, Laura and {Merritt}, David},
        title = "{A Fundamental Relation between Supermassive Black Holes and Their Host Galaxies}",
      journal = {\apjl},
         year = 2000,
        month = aug,
       volume = {539},
       number = {1},
        pages = {L9-L12},
          doi = {10.1086/312838},
archivePrefix = {arXiv},
       eprint = {astro-ph/0006053},
 primaryClass = {astro-ph},
       adsurl = {https://ui.adsabs.harvard.edu/abs/2000ApJ...539L...9F}
}

@ARTICLE{Gebhardt2000,
       author = {{Gebhardt}, Karl and {Bender}, Ralf and {Bower}, Gary and {Dressler}, Alan and {Faber}, S.~M. and {Filippenko}, Alexei V. and {Green}, Richard and {Grillmair}, Carl and {Ho}, Luis C. and {Kormendy}, John and {Lauer}, Tod R. and {Magorrian}, John and {Pinkney}, Jason and {Richstone}, Douglas and {Tremaine}, Scott},
        title = "{A Relationship between Nuclear Black Hole Mass and Galaxy Velocity Dispersion}",
      journal = {\apjl},
         year = 2000,
        month = aug,
       volume = {539},
       number = {1},
        pages = {L13-L16},
          doi = {10.1086/312840},
archivePrefix = {arXiv},
       eprint = {astro-ph/0006289},
 primaryClass = {astro-ph},
       adsurl = {https://ui.adsabs.harvard.edu/abs/2000ApJ...539L..13G}
}

@ARTICLE{KormendyHo2013,
       author = {{Kormendy}, John and {Ho}, Luis C.},
        title = "{Coevolution (Or Not) of Supermassive Black Holes and Host Galaxies}",
      journal = {\araa},
         year = 2013,
        month = aug,
       volume = {51},
       number = {1},
        pages = {511-653},
          doi = {10.1146/annurev-astro-082708-101811},
archivePrefix = {arXiv},
       eprint = {1304.7762},
 primaryClass = {astro-ph.CO},
       adsurl = {https://ui.adsabs.harvard.edu/abs/2013ARA&A..51..511K}
}

@ARTICLE{Rees1984
,
       author = {{Rees}, Martin J.},
        title = "{Black Hole Models for Active Galactic Nuclei}",
      journal = {\araa},
         year = 1984,
        month = jan,
       volume = {22},
        pages = {471-506},
          doi = {10.1146/annurev.aa.22.090184.002351},
       adsurl = {https://ui.adsabs.harvard.edu/abs/1984ARA&A..22..471R}
}

@ARTICLE{Heckman2014,
       author = {{Heckman}, Timothy M. and {Best}, Philip N.},
        title = "{The Coevolution of Galaxies and Supermassive Black Holes: Insights from Surveys of the Contemporary Universe}",
      journal = {\araa},
         year = 2014,
        month = aug,
       volume = {52},
        pages = {589-660},
          doi = {10.1146/annurev-astro-081913-035722},
archivePrefix = {arXiv},
       eprint = {1403.4620},
 primaryClass = {astro-ph.GA},
       adsurl = {https://ui.adsabs.harvard.edu/abs/2014ARA&A..52..589H}
}

@ARTICLE{Hopkins2006,
       author = {{Hopkins}, Philip F. and {Hernquist}, Lars and {Cox}, Thomas J. and {Di Matteo}, Tiziana and {Robertson}, Brant and {Springel}, Volker},
        title = "{A Unified, Merger-driven Model of the Origin of Starbursts, Quasars, the Cosmic X-Ray Background, Supermassive Black Holes, and Galaxy Spheroids}",
      journal = {\apjs},
         year = 2006,
        month = mar,
       volume = {163},
       number = {1},
        pages = {1-49},
          doi = {10.1086/499298},
archivePrefix = {arXiv},
       eprint = {astro-ph/0506398},
 primaryClass = {astro-ph},
       adsurl = {https://ui.adsabs.harvard.edu/abs/2006ApJS..163....1H}
}

@ARTICLE{Conroy2013,
       author = {{Conroy}, Charlie and {White}, Martin},
        title = "{A Simple Model for Quasar Demographics}",
      journal = {\apj},
         year = 2013,
        month = jan,
       volume = {762},
       number = {2},
          eid = {70},
        pages = {70},
          doi = {10.1088/0004-637X/762/2/70},
archivePrefix = {arXiv},
       eprint = {1208.3198},
 primaryClass = {astro-ph.CO},
       adsurl = {https://ui.adsabs.harvard.edu/abs/2013ApJ...762...70C}
}

@ARTICLE{Soltan1982,
       author = {{Soltan}, A.},
        title = "{Masses of quasars.}",
      journal = {\mnras},
         year = 1982,
        month = jul,
       volume = {200},
        pages = {115-122},
          doi = {10.1093/mnras/200.1.115},
       adsurl = {https://ui.adsabs.harvard.edu/abs/1982MNRAS.200..115S}
}

@ARTICLE{Lbol-Churazov,
       author = {{Churazov}, E. and {Sazonov}, S. and {Sunyaev}, R. and {Forman}, W. and {Jones}, C. and {B{\"o}hringer}, H.},
        title = "{Supermassive black holes in elliptical galaxies: switching from very bright to very dim}",
      journal = {\mnras},
         year = 2005,
        month = oct,
       volume = {363},
       number = {1},
        pages = {L91-L95},
          doi = {10.1111/j.1745-3933.2005.00093.x},
archivePrefix = {arXiv},
       eprint = {astro-ph/0507073},
 primaryClass = {astro-ph},
       adsurl = {https://ui.adsabs.harvard.edu/abs/2005MNRAS.363L..91C}
}

@ARTICLE{Weinberger25,
       author = {{Weinberger}, Rainer and {Bhowmick}, Aklant and {Blecha}, Laura and {Bryan}, Greg and {Buchner}, Johannes and {Hernquist}, Lars and {Hlavacek-Larrondo}, Julie and {Springel}, Volker},
        title = "{Accretion onto supermassive and intermediate mass black holes in cosmological simulations}",
      journal = {arXiv e-prints},
         year = 2025,
        month = feb,
          eid = {arXiv:2502.13241},
        pages = {arXiv:2502.13241},
          doi = {10.48550/arXiv.2502.13241},
archivePrefix = {arXiv},
       eprint = {2502.13241},
 primaryClass = {astro-ph.GA},
       adsurl = {https://ui.adsabs.harvard.edu/abs/2025arXiv250213241W}
}

@ARTICLE{Hirschmann2014,
       author = {{Hirschmann}, Michaela and {Dolag}, Klaus and {Saro}, Alexandro and {Bachmann}, Lisa and {Borgani}, Stefano and {Burkert}, Andreas},
        title = "{Cosmological simulations of black hole growth: AGN luminosities and downsizing}",
      journal = {\mnras},
         year = 2014,
        month = aug,
       volume = {442},
       number = {3},
        pages = {2304-2324},
          doi = {10.1093/mnras/stu1023},
archivePrefix = {arXiv},
       eprint = {1308.0333},
 primaryClass = {astro-ph.CO},
       adsurl = {https://ui.adsabs.harvard.edu/abs/2014MNRAS.442.2304H}
}

@ARTICLE{Steinborn2015,
       author = {{Steinborn}, Lisa K. and {Dolag}, Klaus and {Hirschmann}, Michaela and {Prieto}, M. Almudena and {Remus}, Rhea-Silvia},
        title = "{A refined sub-grid model for black hole accretion and AGN feedback in large cosmological simulations}",
      journal = {\mnras},
         year = 2015,
        month = apr,
       volume = {448},
       number = {2},
        pages = {1504-1525},
          doi = {10.1093/mnras/stv072},
archivePrefix = {arXiv},
       eprint = {1409.3221},
 primaryClass = {astro-ph.GA},
       adsurl = {https://ui.adsabs.harvard.edu/abs/2015MNRAS.448.1504S}
}

@ARTICLE{Schulze2015,
       author = {{Schulze}, A. and {Bongiorno}, A. and {Gavignaud}, I. and {Schramm}, M. and {Silverman}, J. and {Merloni}, A. and {Zamorani}, G. and {Hirschmann}, M. and {Mainieri}, V. and {Wisotzki}, L. and {Shankar}, F. and {Fiore}, F. and {Koekemoer}, A.~M. and {Temporin}, G.},
        title = "{The cosmic growth of the active black hole population at 1 <z <2 in zCOSMOS, VVDS and SDSS}",
      journal = {\mnras},
         year = 2015,
        month = mar,
       volume = {447},
       number = {3},
        pages = {2085-2111},
          doi = {10.1093/mnras/stu2549},
archivePrefix = {arXiv},
       eprint = {1412.0754},
 primaryClass = {astro-ph.GA},
       adsurl = {https://ui.adsabs.harvard.edu/abs/2015MNRAS.447.2085S}
}

@ARTICLE{Astrid,
       author = {{Ni}, Yueying and {Di Matteo}, Tiziana and {Bird}, Simeon and {Croft}, Rupert and {Feng}, Yu and {Chen}, Nianyi and {Tremmel}, Michael and {DeGraf}, Colin and {Li}, Yin},
        title = "{The ASTRID simulation: the evolution of supermassive black holes}",
      journal = {\mnras},
         year = 2022,
        month = jun,
       volume = {513},
       number = {1},
        pages = {670-692},
          doi = {10.1093/mnras/stac351},
archivePrefix = {arXiv},
       eprint = {2110.14154},
 primaryClass = {astro-ph.GA},
       adsurl = {https://ui.adsabs.harvard.edu/abs/2022MNRAS.513..670N}
}

@ARTICLE{Habouzit2021,
       author = {{Habouzit}, M{\'e}lanie and {Li}, Yuan and {Somerville}, Rachel S. and {Genel}, Shy and {Pillepich}, Annalisa and {Volonteri}, Marta and {Dav{\'e}}, Romeel and {Rosas-Guevara}, Yetli and {McAlpine}, Stuart and {Peirani}, S{\'e}bastien and {Hernquist}, Lars and {Angl{\'e}s-Alc{\'a}zar}, Daniel and {Reines}, Amy and {Bower}, Richard and {Dubois}, Yohan and {Nelson}, Dylan and {Pichon}, Christophe and {Vogelsberger}, Mark},
        title = "{Supermassive black holes in cosmological simulations I: M$_{BH}$ - M$_{{\ensuremath{\star}}}$ relation and black hole mass function}",
      journal = {\mnras},
         year = 2021,
        month = may,
       volume = {503},
       number = {2},
        pages = {1940-1975},
          doi = {10.1093/mnras/stab496},
archivePrefix = {arXiv},
       eprint = {2006.10094},
 primaryClass = {astro-ph.GA},
       adsurl = {https://ui.adsabs.harvard.edu/abs/2021MNRAS.503.1940H}
}

@ARTICLE{Habouzit2022a,
       author = {{Habouzit}, M{\'e}lanie and {Somerville}, Rachel S. and {Li}, Yuan and {Genel}, Shy and {Aird}, James and {Angl{\'e}s-Alc{\'a}zar}, Daniel and {Dav{\'e}}, Romeel and {Georgiev}, Iskren Y. and {McAlpine}, Stuart and {Rosas-Guevara}, Yetli and {Dubois}, Yohan and {Nelson}, Dylan and {Banados}, Eduardo and {Hernquist}, Lars and {Peirani}, S{\'e}bastien and {Vogelsberger}, Mark},
        title = "{Supermassive black holes in cosmological simulations - II: the AGN population and predictions for upcoming X-ray missions}",
      journal = {\mnras},
         year = 2022,
        month = jan,
       volume = {509},
       number = {2},
        pages = {3015-3042},
          doi = {10.1093/mnras/stab3147},
archivePrefix = {arXiv},
       eprint = {2111.01802},
 primaryClass = {astro-ph.GA},
       adsurl = {https://ui.adsabs.harvard.edu/abs/2022MNRAS.509.3015H}
}

@ARTICLE{Habouzit2022b,
       author = {{Habouzit}, M{\'e}lanie and {Onoue}, Masafusa and {Ba{\~n}ados}, Eduardo and {Neeleman}, Marcel and {Angl{\'e}s-Alc{\'a}zar}, Daniel and {Walter}, Fabian and {Pillepich}, Annalisa and {Dav{\'e}}, Romeel and {Jahnke}, Knud and {Dubois}, Yohan},
        title = "{Co-evolution of massive black holes and their host galaxies at high redshift: discrepancies from six cosmological simulations and the key role of JWST}",
      journal = {\mnras},
         year = 2022,
        month = apr,
       volume = {511},
       number = {3},
        pages = {3751-3767},
          doi = {10.1093/mnras/stac225},
archivePrefix = {arXiv},
       eprint = {2201.09892},
 primaryClass = {astro-ph.CO},
       adsurl = {https://ui.adsabs.harvard.edu/abs/2022MNRAS.511.3751H}
}

@ARTICLE{Somerville2015,
       author = {{Somerville}, Rachel S. and {Dav{\'e}}, Romeel},
        title = "{Physical Models of Galaxy Formation in a Cosmological Framework}",
      journal = {\araa},
         year = 2015,
        month = aug,
       volume = {53},
        pages = {51-113},
          doi = {10.1146/annurev-astro-082812-140951},
archivePrefix = {arXiv},
       eprint = {1412.2712},
 primaryClass = {astro-ph.GA},
       adsurl = {https://ui.adsabs.harvard.edu/abs/2015ARA&A..53...51S}
}

@ARTICLE{Fabian2012,
       author = {{Fabian}, A.~C.},
        title = "{Observational Evidence of Active Galactic Nuclei Feedback}",
      journal = {\araa},
         year = 2012,
        month = sep,
       volume = {50},
        pages = {455-489},
          doi = {10.1146/annurev-astro-081811-125521},
archivePrefix = {arXiv},
       eprint = {1204.4114},
 primaryClass = {astro-ph.CO},
       adsurl = {https://ui.adsabs.harvard.edu/abs/2012ARA&A..50..455F}
}

@ARTICLE{Alexander2025,
       author = {{Alexander}, D.~M. and {Hickox}, R.~C. and {Aird}, J. and {Combes}, F. and {Costa}, T. and {Habouzit}, M. and {Harrison}, C.~M. and {Leng}, R.~I. and {Morabito}, L.~K. and {Uckelman}, S.~L. and {Vickers}, P.},
        title = "{What drives the growth of black holes: a decade of progress}",
      journal = {arXiv e-prints},
         year = 2025,
        month = jun,
          eid = {arXiv:2506.19166},
        pages = {arXiv:2506.19166},
          doi = {10.48550/arXiv.2506.19166},
archivePrefix = {arXiv},
       eprint = {2506.19166},
 primaryClass = {astro-ph.GA},
       adsurl = {https://ui.adsabs.harvard.edu/abs/2025arXiv250619166A}
}

@ARTICLE{Booth2010,
       author = {{Booth}, C.~M. and {Schaye}, Joop},
        title = "{Dark matter haloes determine the masses of supermassive black holes}",
      journal = {\mnras},
         year = 2010,
        month = jun,
       volume = {405},
       number = {1},
        pages = {L1-L5},
          doi = {10.1111/j.1745-3933.2010.00832.x},
archivePrefix = {arXiv},
       eprint = {0911.0935},
 primaryClass = {astro-ph.CO},
       adsurl = {https://ui.adsabs.harvard.edu/abs/2010MNRAS.405L...1B}
}

@ARTICLE{Shankar2025,
       author = {{Shankar}, Francesco and {Bernardi}, Mariangela and {Roberts}, Daniel and {Arana-Catania}, Miguel and {Grubenmann}, Tobias and {Habouzit}, Melanie and {Smith}, Amy and {Marsden}, Christopher and {Varadarajan}, Karthik Mahesh and {Alonso Tetilla}, Alba Vega and {Angl{\'e}s-Alc{\'a}zar}, Daniel and {Boco}, Lumen and {Farrah}, Duncan and {Fu}, Hao and {Haniewicz}, Henryk and {Lapi}, Andrea and {Lovell}, Christopher C. and {Menci}, Nicola and {Powell}, Meredith and {Ricci}, Federica},
        title = "{Probing the co-evolution of Supermassive Black Holes and their hosts from scaling relations pairwise residuals: dominance of stellar velocity dispersion and host halo mass}",
      journal = {\mnras},
         year = 2025,
        month = aug,
       volume = {541},
       number = {2},
        pages = {2070-2092},
          doi = {10.1093/mnras/staf747},
archivePrefix = {arXiv},
       eprint = {2505.02920},
 primaryClass = {astro-ph.GA},
       adsurl = {https://ui.adsabs.harvard.edu/abs/2025MNRAS.541.2070S}
}

@ARTICLE{Wang2021,
       author = {{Wang}, Feige and {Yang}, Jinyi and {Fan}, Xiaohui and {Hennawi}, Joseph F. and {Barth}, Aaron J. and {Banados}, Eduardo and {Bian}, Fuyan and {Boutsia}, Konstantina and {Connor}, Thomas and {Davies}, Frederick B. and {Decarli}, Roberto and {Eilers}, Anna-Christina and {Farina}, Emanuele Paolo and {Green}, Richard and {Jiang}, Linhua and {Li}, Jiang-Tao and {Mazzucchelli}, Chiara and {Nanni}, Riccardo and {Schindler}, Jan-Torge and {Venemans}, Bram and {Walter}, Fabian and {Wu}, Xue-Bing and {Yue}, Minghao},
        title = "{A Luminous Quasar at Redshift 7.642}",
      journal = {\apjl},
         year = 2021,
        month = jan,
       volume = {907},
       number = {1},
          eid = {L1},
        pages = {L1},
          doi = {10.3847/2041-8213/abd8c6},
archivePrefix = {arXiv},
       eprint = {2101.03179},
 primaryClass = {astro-ph.GA},
       adsurl = {https://ui.adsabs.harvard.edu/abs/2021ApJ...907L...1W}
}

@ARTICLE{Banados2023,
       author = {{Ba{\~n}ados}, Eduardo and {Schindler}, Jan-Torge and {Venemans}, Bram P. and {Connor}, Thomas and {Decarli}, Roberto and {Farina}, Emanuele Paolo and {Mazzucchelli}, Chiara and {Meyer}, Romain A. and {Stern}, Daniel and {Walter}, Fabian and {Fan}, Xiaohui and {Hennawi}, Joseph F. and {Khusanova}, Yana and {Morrell}, Nidia and {Nanni}, Riccardo and {Noirot}, Ga{\"e}l and {Pensabene}, Antonio and {Rix}, Hans-Walter and {Simon}, Joseph and {Verdoes Kleijn}, Gijs A. and {Xie}, Zhang-Liang and {Yang}, Da-Ming and {Connor}, Andrew},
        title = "{The Pan-STARRS1 z > 5.6 Quasar Survey. II. Discovery of 55 Quasars at 5.6 < z < 6.5}",
      journal = {\apjs},
         year = 2023,
        month = mar,
       volume = {265},
       number = {1},
          eid = {29},
        pages = {29},
          doi = {10.3847/1538-4365/acb3c7},
archivePrefix = {arXiv},
       eprint = {2212.04452},
 primaryClass = {astro-ph.GA},
       adsurl = {https://ui.adsabs.harvard.edu/abs/2023ApJS..265...29B}
}

@ARTICLE{Garcia2022,
       author = {{Garc{\'\i}a-Vergara}, Cristina and {Rybak}, Matus and {Hodge}, Jacqueline and {Hennawi}, Joseph F. and {Decarli}, Roberto and {Gonz{\'a}lez-L{\'o}pez}, Jorge and {Arrigoni-Battaia}, Fabrizio and {Aravena}, Manuel and {Farina}, Emanuele P.},
        title = "{ALMA Reveals a Large Overdensity and Strong Clustering of Galaxies in Quasar Environments at z 4}",
      journal = {\apj},
         year = 2022,
        month = mar,
       volume = {927},
       number = {1},
          eid = {65},
        pages = {65},
          doi = {10.3847/1538-4357/ac469d},
archivePrefix = {arXiv},
       eprint = {2109.09754},
 primaryClass = {astro-ph.GA},
       adsurl = {https://ui.adsabs.harvard.edu/abs/2022ApJ...927...65G}
}

@ARTICLE{Belladitta2022,
       author = {{Belladitta}, S. and {Caccianiga}, A. and {Diana}, A. and {Moretti}, A. and {Severgnini}, P. and {Pedani}, M. and {Cassar{\`a}}, L.~P. and {Spingola}, C. and {Ighina}, L. and {Rossi}, A. and {Della Ceca}, R.},
        title = "{Central engine of the highest redshift blazar}",
      journal = {\aap},
         year = 2022,
        month = apr,
       volume = {660},
          eid = {A74},
        pages = {A74},
          doi = {10.1051/0004-6361/202142335},
archivePrefix = {arXiv},
       eprint = {2201.08863},
 primaryClass = {astro-ph.HE},
       adsurl = {https://ui.adsabs.harvard.edu/abs/2022A&A...660A..74B}
}

@ARTICLE{Yang2021,
       author = {{Yang}, Jinyi and {Wang}, Feige and {Fan}, Xiaohui and {Barth}, Aaron J. and {Hennawi}, Joseph F. and {Nanni}, Riccardo and {Bian}, Fuyan and {Davies}, Frederick B. and {Farina}, Emanuele P. and {Schindler}, Jan-Torge and {Ba{\~n}ados}, Eduardo and {Decarli}, Roberto and {Eilers}, Anna-Christina and {Green}, Richard and {Guo}, Hengxiao and {Jiang}, Linhua and {Li}, Jiang-Tao and {Venemans}, Bram and {Walter}, Fabian and {Wu}, Xue-Bing and {Yue}, Minghao},
        title = "{Probing Early Supermassive Black Hole Growth and Quasar Evolution with Near-infrared Spectroscopy of 37 Reionization-era Quasars at 6.3 < z {\ensuremath{\leq}} 7.64}",
      journal = {\apj},
         year = 2021,
        month = dec,
       volume = {923},
       number = {2},
          eid = {262},
        pages = {262},
          doi = {10.3847/1538-4357/ac2b32},
archivePrefix = {arXiv},
       eprint = {2109.13942},
 primaryClass = {astro-ph.GA},
       adsurl = {https://ui.adsabs.harvard.edu/abs/2021ApJ...923..262Y}
}

@ARTICLE{Yang2020,
       author = {{Yang}, Jinyi and {Wang}, Feige and {Fan}, Xiaohui and {Hennawi}, Joseph F. and {Davies}, Frederick B. and {Yue}, Minghao and {Banados}, Eduardo and {Wu}, Xue-Bing and {Venemans}, Bram and {Barth}, Aaron J. and {Bian}, Fuyan and {Boutsia}, Konstantina and {Decarli}, Roberto and {Farina}, Emanuele Paolo and {Green}, Richard and {Jiang}, Linhua and {Li}, Jiang-Tao and {Mazzucchelli}, Chiara and {Walter}, Fabian},
        title = "{P{\={o}}niu{\={a}}'ena: A Luminous z = 7.5 Quasar Hosting a 1.5 Billion Solar Mass Black Hole}",
      journal = {\apjl},
         year = 2020,
        month = jul,
       volume = {897},
       number = {1},
          eid = {L14},
        pages = {L14},
          doi = {10.3847/2041-8213/ab9c26},
archivePrefix = {arXiv},
       eprint = {2006.13452},
 primaryClass = {astro-ph.GA},
       adsurl = {https://ui.adsabs.harvard.edu/abs/2020ApJ...897L..14Y}
}

@ARTICLE{Farina2022,
       author = {{Farina}, Emanuele Paolo and {Schindler}, Jan-Torge and {Walter}, Fabian and {Ba{\~n}ados}, Eduardo and {Davies}, Frederick B. and {Decarli}, Roberto and {Eilers}, Anna-Christina and {Fan}, Xiaohui and {Hennawi}, Joseph F. and {Mazzucchelli}, Chiara and {Meyer}, Romain A. and {Trakhtenbrot}, Benny and {Volonteri}, Marta and {Wang}, Feige and {Worseck}, G{\'a}bor and {Yang}, Jinyi and {Gutcke}, Thales A. and {Venemans}, Bram P. and {Bosman}, Sarah E.~I. and {Costa}, Tiago and {De Rosa}, Gisella and {Drake}, Alyssa B. and {Onoue}, Masafusa},
        title = "{The X-shooter/ALMA Sample of Quasars in the Epoch of Reionization. II. Black Hole Masses, Eddington Ratios, and the Formation of the First Quasars}",
      journal = {\apj},
         year = 2022,
        month = dec,
       volume = {941},
       number = {2},
          eid = {106},
        pages = {106},
          doi = {10.3847/1538-4357/ac9626},
archivePrefix = {arXiv},
       eprint = {2207.05113},
 primaryClass = {astro-ph.GA},
       adsurl = {https://ui.adsabs.harvard.edu/abs/2022ApJ...941..106F}
}

@ARTICLE{Mazzuccheli2017,
       author = {{Mazzucchelli}, C. and {Ba{\~n}ados}, E. and {Venemans}, B.~P. and {Decarli}, R. and {Farina}, E.~P. and {Walter}, F. and {Eilers}, A. -C. and {Rix}, H. -W. and {Simcoe}, R. and {Stern}, D. and {Fan}, X. and {Schlafly}, E. and {De Rosa}, G. and {Hennawi}, J. and {Chambers}, K.~C. and {Greiner}, J. and {Burgett}, W. and {Draper}, P.~W. and {Kaiser}, N. and {Kudritzki}, R. -P. and {Magnier}, E. and {Metcalfe}, N. and {Waters}, C. and {Wainscoat}, R.~J.},
        title = "{Physical Properties of 15 Quasars at z {\ensuremath{\gtrsim}} 6.5}",
      journal = {\apj},
         year = 2017,
        month = nov,
       volume = {849},
       number = {2},
          eid = {91},
        pages = {91},
          doi = {10.3847/1538-4357/aa9185},
archivePrefix = {arXiv},
       eprint = {1710.01251},
 primaryClass = {astro-ph.GA},
       adsurl = {https://ui.adsabs.harvard.edu/abs/2017ApJ...849...91M}
}

@ARTICLE{Fan2023,
       author = {{Fan}, Xiaohui and {Ba{\~n}ados}, Eduardo and {Simcoe}, Robert A.},
        title = "{Quasars and the Intergalactic Medium at Cosmic Dawn}",
      journal = {\araa},
         year = 2023,
        month = aug,
       volume = {61},
        pages = {373-426},
          doi = {10.1146/annurev-astro-052920-102455},
archivePrefix = {arXiv},
       eprint = {2212.06907},
 primaryClass = {astro-ph.GA},
       adsurl = {https://ui.adsabs.harvard.edu/abs/2023ARA&A..61..373F}
}

@ARTICLE{Belladitta2025,
       author = {{Belladitta}, Silvia and {Ba{\~n}ados}, Eduardo and {Xie}, Zhang-Liang and {Decarli}, Roberto and {Onorato}, Silvia and {Yang}, Jinyi and {Bischetti}, Manuela and {Onoue}, Masafusa and {Loiacono}, Federica and {Mart{\'\i}nez-Ram{\'\i}rez}, Laura N. and {Mazzucchelli}, Chiara and {Davies}, Frederick B. and {Wolf}, Julien and {Schindler}, Jan-Torge and {Fan}, Xiaohui and {Wang}, Feige and {Walter}, Fabian and {Mkrtchyan}, Tatevik and {Stern}, Daniel and {Farina}, Emanuele P. and {Venemans}, Bram P.},
        title = "{Discovery and characterization of 25 new quasars at 4.6 < z < 6.9 from wide-field multiband surveys}",
      journal = {\aap},
         year = 2025,
        month = jul,
       volume = {699},
          eid = {A335},
        pages = {A335},
          doi = {10.1051/0004-6361/202554859},
archivePrefix = {arXiv},
       eprint = {2505.15923},
 primaryClass = {astro-ph.GA},
       adsurl = {https://ui.adsabs.harvard.edu/abs/2025A&A...699A.335B}
}

@ARTICLE{Ubler2024,
       author = {{{\"U}bler}, Hannah and {Maiolino}, Roberto and {P{\'e}rez-Gonz{\'a}lez}, Pablo G. and {D'Eugenio}, Francesco and {Perna}, Michele and {Curti}, Mirko and {Arribas}, Santiago and {Bunker}, Andrew and {Carniani}, Stefano and {Charlot}, St{\'e}phane and {Rodr{\'\i}guez Del Pino}, Bruno and {Baker}, William and {B{\"o}ker}, Torsten and {Cresci}, Giovanni and {Dunlop}, James and {Grogin}, Norman A. and {Jones}, Gareth C. and {Kumari}, Nimisha and {Lamperti}, Isabella and {Laporte}, Nicolas and {Marshall}, Madeline A. and {Mazzolari}, Giovanni and {Parlanti}, Eleonora and {Rawle}, Tim and {Scholtz}, Jan and {Venturi}, Giacomo and {Witstok}, Joris},
        title = "{GA-NIFS: JWST discovers an offset AGN 740 million years after the big bang}",
      journal = {\mnras},
         year = 2024,
        month = jun,
       volume = {531},
       number = {1},
        pages = {355-365},
          doi = {10.1093/mnras/stae943},
archivePrefix = {arXiv},
       eprint = {2312.03589},
 primaryClass = {astro-ph.GA},
       adsurl = {https://ui.adsabs.harvard.edu/abs/2024MNRAS.531..355U}
}

@ARTICLE{Bogdan,
       author = {{Bogd{\'a}n}, {\'A}kos and {Goulding}, Andy D. and {Natarajan}, Priyamvada and {Kov{\'a}cs}, Orsolya E. and {Tremblay}, Grant R. and {Chadayammuri}, Urmila and {Volonteri}, Marta and {Kraft}, Ralph P. and {Forman}, William R. and {Jones}, Christine and {Churazov}, Eugene and {Zhuravleva}, Irina},
        title = "{Evidence for heavy-seed origin of early supermassive black holes from a z {\ensuremath{\approx}} 10 X-ray quasar}",
      journal = {Nature Astronomy},
         year = 2024,
        month = jan,
       volume = {8},
       number = {1},
        pages = {126-133},
          doi = {10.1038/s41550-023-02111-9},
archivePrefix = {arXiv},
       eprint = {2305.15458},
 primaryClass = {astro-ph.GA},
       adsurl = {https://ui.adsabs.harvard.edu/abs/2024NatAs...8..126B}
}

@ARTICLE{Larson2023,
       author = {{Larson}, Rebecca L. and {Finkelstein}, Steven L. and {Kocevski}, Dale D. and {Hutchison}, Taylor A. and {Trump}, Jonathan R. and {Arrabal Haro}, Pablo and {Bromm}, Volker and {Cleri}, Nikko J. and {Dickinson}, Mark and {Fujimoto}, Seiji and {Kartaltepe}, Jeyhan S. and {Koekemoer}, Anton M. and {Papovich}, Casey and {Pirzkal}, Nor and {Tacchella}, Sandro and {Zavala}, Jorge A. and {Bagley}, Micaela and {Behroozi}, Peter and {Champagne}, Jaclyn B. and {Cole}, Justin W. and {Jung}, Intae and {Morales}, Alexa M. and {Yang}, Guang and {Zhang}, Haowen and {Zitrin}, Adi and {Amor{\'\i}n}, Ricardo O. and {Burgarella}, Denis and {Casey}, Caitlin M. and {Ch{\'a}vez Ortiz}, {\'O}scar A. and {Cox}, Isabella G. and {Chworowsky}, Katherine and {Fontana}, Adriano and {Gawiser}, Eric and {Grazian}, Andrea and {Grogin}, Norman A. and {Harish}, Santosh and {Hathi}, Nimish P. and {Hirschmann}, Michaela and {Holwerda}, Benne W. and {Juneau}, St{\'e}phanie and {Leung}, Gene C.~K. and {Lucas}, Ray A. and {McGrath}, Elizabeth J. and {P{\'e}rez-Gonz{\'a}lez}, Pablo G. and {Rigby}, Jane R. and {Seill{\'e}}, Lise-Marie and {Simons}, Raymond C. and {de La Vega}, Alexander and {Weiner}, Benjamin J. and {Wilkins}, Stephen M. and {Yung}, L.~Y. Aaron and {Ceers Team}},
        title = "{A CEERS Discovery of an Accreting Supermassive Black Hole 570 Myr after the Big Bang: Identifying a Progenitor of Massive z > 6 Quasars}",
      journal = {\apjl},
         year = 2023,
        month = aug,
       volume = {953},
       number = {2},
          eid = {L29},
        pages = {L29},
          doi = {10.3847/2041-8213/ace619},
archivePrefix = {arXiv},
       eprint = {2303.08918},
 primaryClass = {astro-ph.GA},
       adsurl = {https://ui.adsabs.harvard.edu/abs/2023ApJ...953L..29L}
}

@ARTICLE{Kokorev2023,
       author = {{Kokorev}, Vasily and {Fujimoto}, Seiji and {Labbe}, Ivo and {Greene}, Jenny E. and {Bezanson}, Rachel and {Dayal}, Pratika and {Nelson}, Erica J. and {Atek}, Hakim and {Brammer}, Gabriel and {Caputi}, Karina I. and {Chemerynska}, Iryna and {Cutler}, Sam E. and {Feldmann}, Robert and {Fudamoto}, Yoshinobu and {Furtak}, Lukas J. and {Goulding}, Andy D. and {de Graaff}, Anna and {Leja}, Joel and {Marchesini}, Danilo and {Miller}, Tim B. and {Nanayakkara}, Themiya and {Oesch}, Pascal A. and {Pan}, Richard and {Price}, Sedona H. and {Setton}, David J. and {Smit}, Renske and {Stefanon}, Mauro and {Wang}, Bingjie and {Weaver}, John R. and {Whitaker}, Katherine E. and {Williams}, Christina C. and {Zitrin}, Adi},
        title = "{UNCOVER: A NIRSpec Identification of a Broad-line AGN at z = 8.50}",
      journal = {\apjl},
         year = 2023,
        month = nov,
       volume = {957},
       number = {1},
          eid = {L7},
        pages = {L7},
          doi = {10.3847/2041-8213/ad037a},
archivePrefix = {arXiv},
       eprint = {2308.11610},
 primaryClass = {astro-ph.GA},
       adsurl = {https://ui.adsabs.harvard.edu/abs/2023ApJ...957L...7K}
}

@ARTICLE{Miaolino2024,
       author = {{Maiolino}, Roberto and {Scholtz}, Jan and {Curtis-Lake}, Emma and {Carniani}, Stefano and {Baker}, William and {de Graaff}, Anna and {Tacchella}, Sandro and {{\"U}bler}, Hannah and {D'Eugenio}, Francesco and {Witstok}, Joris and {Curti}, Mirko and {Arribas}, Santiago and {Bunker}, Andrew J. and {Charlot}, St{\'e}phane and {Chevallard}, Jacopo and {Eisenstein}, Daniel J. and {Egami}, Eiichi and {Ji}, Zhiyuan and {Jones}, Gareth C. and {Lyu}, Jianwei and {Rawle}, Tim and {Robertson}, Brant and {Rujopakarn}, Wiphu and {Perna}, Michele and {Sun}, Fengwu and {Venturi}, Giacomo and {Williams}, Christina C. and {Willott}, Chris},
        title = "{JADES: The diverse population of infant black holes at 4 < z < 11: Merging, tiny, poor, but mighty}",
      journal = {\aap},
         year = 2024,
        month = nov,
       volume = {691},
          eid = {A145},
        pages = {A145},
          doi = {10.1051/0004-6361/202347640},
archivePrefix = {arXiv},
       eprint = {2308.01230},
 primaryClass = {astro-ph.GA},
       adsurl = {https://ui.adsabs.harvard.edu/abs/2024A&A...691A.145M}
}

@ARTICLE{Matsuoka2025,
       author = {{Matsuoka}, Yoshiki and {Onoue}, Masafusa and {Iwasawa}, Kazushi and {Aoki}, Kentaro and {Strauss}, Michael A. and {Silverman}, John D. and {Ding}, Xuheng and {Phillips}, Camryn L. and {Akiyama}, Masayuki and {Arita}, Junya and {Imanishi}, Masatoshi and {Izumi}, Takuma and {Kashikawa}, Nobunari and {Kawaguchi}, Toshihiro and {Kikuta}, Satoshi and {Kohno}, Kotaro and {Lee}, Chien-Hsiu and {Nagao}, Tohru and {Takahashi}, Ayumi and {Toba}, Yoshiki},
        title = "{SHELLQs. Bridging the Gap: JWST Unveils Obscured Quasars in the Most Luminous Galaxies at z > 6}",
      journal = {\apj},
         year = 2025,
        month = jul,
       volume = {988},
       number = {1},
          eid = {57},
        pages = {57},
          doi = {10.3847/1538-4357/addf4e},
archivePrefix = {arXiv},
       eprint = {2505.04825},
 primaryClass = {astro-ph.GA},
       adsurl = {https://ui.adsabs.harvard.edu/abs/2025ApJ...988...57M}
}

@ARTICLE{Banados2026,
       author = {{Banados}, Eduardo},
        title = "{Observations of Early Black Holes Before and After JWST}",
      journal = {arXiv e-prints},
         year = 2026,
        month = mar,
          eid = {arXiv:2603.21976},
        pages = {arXiv:2603.21976},
          doi = {10.48550/arXiv.2603.21976},
archivePrefix = {arXiv},
       eprint = {2603.21976},
 primaryClass = {astro-ph.GA},
       adsurl = {https://ui.adsabs.harvard.edu/abs/2026arXiv260321976B}
}

@ARTICLE{Wang2026,
       author = {{Wang}, Feige and {Champagne}, Jaclyn B. and {Huang}, Jiamu and {Yang}, Jinyi and {Hennawi}, Joseph F. and {Fan}, Xiaohui and {Zhang}, Haowen and {Costa}, Tiago and {Decarli}, Roberto and {Habouzit}, Melanie and {Sun}, Fengwu and {Banados}, Eduardo and {Jin}, Xiangyu and {Kakiichi}, Koki and {Meyer}, Romain A. and {Wu}, Yunjing and {Belladitta}, Silvia and {Blecha}, Laura and {Bosman}, Sarah E.~I. and {Cai}, Zheng and {Connor}, Thomas and {Davies}, Frederick B. and {Eilers}, Anna-Christina and {Haiman}, Zoltan and {Jun}, Hyunsung D. and {Li}, Mingyu and {Li}, Zihao and {Liu}, Weizhe and {Lupi}, Alessandro and {Lyu}, Jianwei and {Mazzucchelli}, Chiara and {Onoue}, Masafusa and {Pudoka}, Maria and {Rojas-Ruiz}, Sofia and {Schindler}, Jan-Torge and {Shen}, Yue and {Tee}, Wei Leong and {Trakhtenbrot}, Benny and {Trebitsch}, Maxime and {Vestergaard}, Marianne and {Volonteri}, Marta and {Walter}, Fabian and {Zhang}, Huanian and {Zou}, Siwei},
        title = "{ASPIRE: The Environments and Dark Matter Halos of Luminous Quasars in the Epoch of Reionization}",
      journal = {arXiv e-prints},
         year = 2026,
        month = feb,
          eid = {arXiv:2602.04979},
        pages = {arXiv:2602.04979},
          doi = {10.48550/arXiv.2602.04979},
archivePrefix = {arXiv},
       eprint = {2602.04979},
 primaryClass = {astro-ph.GA},
       adsurl = {https://ui.adsabs.harvard.edu/abs/2026arXiv260204979W}
}

@ARTICLE{Huang2026,
       author = {{Huang}, Jiamu and {Hennawi}, Joseph and {Pizzati}, Elia and {Wang}, Feige and {Yang}, Jinyi and {Champagne}, Jaclyn B. and {Fan}, Xiaohui and {Ba{\~n}ados}, Eduardo and {Jin}, Xiangyu and {Kakiichi}, Koki and {Meyer}, Romain A. and {Sun}, Fengwu and {Wu}, Yunjing and {Zhang}, Haowen and {Mazzucchelli}, Chiara and {Eilers}, Anna-Christina and {Pudoka}, Maria and {Zhang}, Huanian and {Schindler}, Jan-Torge and {Schaller}, Matthieu and {Schaye}, Joop and {Snyder}, Ben and {Kang}, Yi and {Onorato}, Silvia},
        title = "{Clustering of z\raisebox{-0.5ex}\textasciitilde6.6 Quasars and [O III] Emitters Constrains Host Halo Masses and Duty Cycles in 25 ASPIRE Fields}",
      journal = {arXiv e-prints},
         year = 2026,
        month = feb,
          eid = {arXiv:2602.04974},
        pages = {arXiv:2602.04974},
          doi = {10.48550/arXiv.2602.04974},
archivePrefix = {arXiv},
       eprint = {2602.04974},
 primaryClass = {astro-ph.GA},
       adsurl = {https://ui.adsabs.harvard.edu/abs/2026arXiv260204974H}
}

@ARTICLE{JahnkeMaccio2011,
       author = {{Jahnke}, Knud and {Macci{\`o}}, Andrea V.},
        title = "{The Non-causal Origin of the Black-hole-galaxy Scaling Relations}",
      journal = {\apj},
         year = 2011,
        month = jun,
       volume = {734},
       number = {2},
          eid = {92},
        pages = {92},
          doi = {10.1088/0004-637X/734/2/92},
archivePrefix = {arXiv},
       eprint = {1006.0482},
 primaryClass = {astro-ph.CO},
       adsurl = {https://ui.adsabs.harvard.edu/abs/2011ApJ...734...92J}
}

@ARTICLE{BogdanGoulding2015,
       author = {{Bogd{\'a}n}, {\'A}kos and {Goulding}, Andy D.},
        title = "{Connecting Dark Matter Halos with the Galaxy Center and the Supermassive Black Hole}",
      journal = {\apj},
         year = 2015,
        month = feb,
       volume = {800},
       number = {2},
          eid = {124},
        pages = {124},
          doi = {10.1088/0004-637X/800/2/124},
archivePrefix = {arXiv},
       eprint = {1502.05043},
 primaryClass = {astro-ph.GA},
       adsurl = {https://ui.adsabs.harvard.edu/abs/2015ApJ...800..124B}
}

@ARTICLE{Donnari2019,
       author = {{Donnari}, Martina and {Pillepich}, Annalisa and {Nelson}, Dylan and {Vogelsberger}, Mark and {Genel}, Shy and {Weinberger}, Rainer and {Marinacci}, Federico and {Springel}, Volker and {Hernquist}, Lars},
        title = "{The star formation activity of IllustrisTNG galaxies: main sequence, UVJ diagram, quenched fractions, and systematics}",
      journal = {\mnras},
         year = 2019,
        month = jun,
       volume = {485},
       number = {4},
        pages = {4817-4840},
          doi = {10.1093/mnras/stz712},
archivePrefix = {arXiv},
       eprint = {1812.07584},
 primaryClass = {astro-ph.GA},
       adsurl = {https://ui.adsabs.harvard.edu/abs/2019MNRAS.485.4817D}
}

@ARTICLE{Pizzati2024,
       author = {{Pizzati}, Elia and {Hennawi}, Joseph F. and {Schaye}, Joop and {Schaller}, Matthieu},
        title = "{Revisiting the extreme clustering of z {\ensuremath{\approx}} 4 quasars with large volume cosmological simulations}",
      journal = {\mnras},
         year = 2024,
        month = mar,
       volume = {528},
       number = {3},
        pages = {4466-4489},
          doi = {10.1093/mnras/stae329},
archivePrefix = {arXiv},
       eprint = {2311.17181},
 primaryClass = {astro-ph.GA},
       adsurl = {https://ui.adsabs.harvard.edu/abs/2024MNRAS.528.4466P}
}

@ARTICLE{White2008,
       author = {{White}, Martin and {Martini}, Paul and {Cohn}, J.~D.},
        title = "{Constraints on the correlation between QSO luminosity and host halo mass from high-redshift quasar clustering}",
      journal = {\mnras},
         year = 2008,
        month = nov,
       volume = {390},
       number = {3},
        pages = {1179-1184},
          doi = {10.1111/j.1365-2966.2008.13817.x},
archivePrefix = {arXiv},
       eprint = {0711.4109},
 primaryClass = {astro-ph},
       adsurl = {https://ui.adsabs.harvard.edu/abs/2008MNRAS.390.1179W}
}

@ARTICLE{Shankar2010,
       author = {{Shankar}, Francesco and {Crocce}, Martin and {Miralda-Escud{\'e}}, Jordi and {Fosalba}, Pablo and {Weinberg}, David H.},
        title = "{On the Radiative Efficiencies, Eddington Ratios, and Duty Cycles of Luminous High-redshift Quasars}",
      journal = {\apj},
         year = 2010,
        month = jul,
       volume = {718},
       number = {1},
        pages = {231-250},
          doi = {10.1088/0004-637X/718/1/231},
archivePrefix = {arXiv},
       eprint = {0810.4919},
 primaryClass = {astro-ph},
       adsurl = {https://ui.adsabs.harvard.edu/abs/2010ApJ...718..231S}
}

@ARTICLE{Wyithe2009,
       author = {{Wyithe}, J. Stuart B. and {Loeb}, Abraham},
        title = "{Evidence for merger-driven activity in the clustering of high-redshift quasars}",
      journal = {\mnras},
         year = 2009,
        month = may,
       volume = {395},
       number = {3},
        pages = {1607-1619},
          doi = {10.1111/j.1365-2966.2009.14647.x},
archivePrefix = {arXiv},
       eprint = {0810.3455},
 primaryClass = {astro-ph},
       adsurl = {https://ui.adsabs.harvard.edu/abs/2009MNRAS.395.1607W}
}

@ARTICLE{Zhang2023,
       author = {{Zhang}, Haowen and {Behroozi}, Peter and {Volonteri}, Marta and {Silk}, Joseph and {Fan}, Xiaohui and {Hopkins}, Philip F. and {Yang}, Jinyi and {Aird}, James},
        title = "{TRINITY I: self-consistently modelling the dark matter halo-galaxy-supermassive black hole connection from z = 0-10}",
      journal = {\mnras},
         year = 2023,
        month = jan,
       volume = {518},
       number = {2},
        pages = {2123-2163},
          doi = {10.1093/mnras/stac2633},
archivePrefix = {arXiv},
       eprint = {2105.10474},
 primaryClass = {astro-ph.GA},
       adsurl = {https://ui.adsabs.harvard.edu/abs/2023MNRAS.518.2123Z}
}

@ARTICLE{Shen2020,
       author = {{Shen}, Xuejian and {Hopkins}, Philip F. and {Faucher-Gigu{\`e}re}, Claude-Andr{\'e} and {Alexander}, D.~M. and {Richards}, Gordon T. and {Ross}, Nicholas P. and {Hickox}, R.~C.},
        title = "{The bolometric quasar luminosity function at z = 0-7}",
      journal = {\mnras},
         year = 2020,
        month = jan,
       volume = {495},
       number = {3},
        pages = {3252-3275},
          doi = {10.1093/mnras/staa1381},
archivePrefix = {arXiv},
       eprint = {2001.02696},
 primaryClass = {astro-ph.GA},
       adsurl = {https://ui.adsabs.harvard.edu/abs/2020MNRAS.495.3252S}
}

@ARTICLE{Tenneti2018,
       author = {{Tenneti}, Ananth and {Di Matteo}, Tiziana and {Croft}, Rupert and {Garcia}, ThomasJae and {Feng}, Yu},
        title = "{The descendants of the first quasars in the BlueTides simulation}",
      journal = {\mnras},
         year = 2018,
        month = feb,
       volume = {474},
       number = {1},
        pages = {597-603},
          doi = {10.1093/mnras/stx2788},
archivePrefix = {arXiv},
       eprint = {1708.03373},
 primaryClass = {astro-ph.GA},
       adsurl = {https://ui.adsabs.harvard.edu/abs/2018MNRAS.474..597T}
}

@ARTICLE{Feng2016,
       author = {{Feng}, Yu and {Di-Matteo}, Tiziana and {Croft}, Rupert A. and {Bird}, Simeon and {Battaglia}, Nicholas and {Wilkins}, Stephen},
        title = "{The BlueTides simulation: first galaxies and reionization}",
      journal = {\mnras},
         year = 2016,
        month = jan,
       volume = {455},
       number = {3},
        pages = {2778-2791},
          doi = {10.1093/mnras/stv2484},
archivePrefix = {arXiv},
       eprint = {1504.06619},
 primaryClass = {astro-ph.CO},
       adsurl = {https://ui.adsabs.harvard.edu/abs/2016MNRAS.455.2778F}
}

@ARTICLE{Fanidakis2013,
       author = {{Fanidakis}, N. and {Macci{\`o}}, A.~V. and {Baugh}, C.~M. and {Lacey}, C.~G. and {Frenk}, C.~S.},
        title = "{The most luminous quasars do not live in the most massive dark matter haloes at any redshift}",
      journal = {\mnras},
         year = 2013,
        month = nov,
       volume = {436},
       number = {1},
        pages = {315-326},
          doi = {10.1093/mnras/stt1567},
archivePrefix = {arXiv},
       eprint = {1305.2199},
 primaryClass = {astro-ph.CO},
       adsurl = {https://ui.adsabs.harvard.edu/abs/2013MNRAS.436..315F}
}

@ARTICLE{Costa2024,
       author = {{Costa}, Tiago},
        title = "{The host dark matter haloes of the first quasars}",
      journal = {\mnras},
         year = 2024,
        month = jun,
       volume = {531},
       number = {1},
        pages = {930-944},
          doi = {10.1093/mnras/stae1157},
archivePrefix = {arXiv},
       eprint = {2308.12987},
 primaryClass = {astro-ph.GA},
       adsurl = {https://ui.adsabs.harvard.edu/abs/2024MNRAS.531..930C}
}

@ARTICLE{flamingo_quasar_clustering,
       author = {{Ding}, Boyi and {Pizzati}, Elia and {Schaye}, Joop and {Hennawi}, Joseph F. and {McDonald}, William and {Schaller}, Matthieu},
        title = "{The luminosity function and clustering of bright quasars in the FLAMINGO cosmological simulations}",
      journal = {arXiv e-prints},
         year = 2025,
        month = oct,
          eid = {arXiv:2510.24283},
        pages = {arXiv:2510.24283},
archivePrefix = {arXiv},
       eprint = {2510.24283},
 primaryClass = {astro-ph.GA},
       adsurl = {https://ui.adsabs.harvard.edu/abs/2025arXiv251024283D}
}

@ARTICLE{Habouzit2019,
       author = {{Habouzit}, M{\'e}lanie and {Genel}, Shy and {Somerville}, Rachel S. and {Kocevski}, Dale and {Hirschmann}, Michaela and {Dekel}, Avishai and {Choi}, Ena and {Nelson}, Dylan and {Pillepich}, Annalisa and {Torrey}, Paul and {Hernquist}, Lars and {Vogelsberger}, Mark and {Weinberger}, Rainer and {Springel}, Volker},
        title = "{Linking galaxy structural properties and star formation activity to black hole activity with IllustrisTNG}",
      journal = {\mnras},
         year = 2019,
        month = apr,
       volume = {484},
       number = {4},
        pages = {4413-4443},
          doi = {10.1093/mnras/stz102},
archivePrefix = {arXiv},
       eprint = {1809.05588},
 primaryClass = {astro-ph.GA},
       adsurl = {https://ui.adsabs.harvard.edu/abs/2019MNRAS.484.4413H}
}

@ARTICLE{Richards2006,
       author = {{Richards}, Gordon T. and {Strauss}, Michael A. and {Fan}, Xiaohui and {Hall}, Patrick B. and {Jester}, Sebastian and {Schneider}, Donald P. and {Vanden Berk}, Daniel E. and {Stoughton}, Chris and {Anderson}, Scott F. and {Brunner}, Robert J. and {Gray}, Jim and {Gunn}, James E. and {Ivezi{\'c}}, {\v{Z}}eljko and {Kirkland}, Margaret K. and {Knapp}, G.~R. and {Loveday}, Jon and {Meiksin}, Avery and {Pope}, Adrian and {Szalay}, Alexander S. and {Thakar}, Anirudda R. and {Yanny}, Brian and {York}, Donald G. and {Barentine}, J.~C. and {Brewington}, Howard J. and {Brinkmann}, J. and {Fukugita}, Masataka and {Harvanek}, Michael and {Kent}, Stephen M. and {Kleinman}, S.~J. and {Krzesi{\'n}ski}, Jurek and {Long}, Daniel C. and {Lupton}, Robert H. and {Nash}, Thomas and {Neilsen}, Jr., Eric H. and {Nitta}, Atsuko and {Schlegel}, David J. and {Snedden}, Stephanie A.},
        title = "{The Sloan Digital Sky Survey Quasar Survey: Quasar Luminosity Function from Data Release 3}",
      journal = {\aj},
         year = 2006,
        month = jun,
       volume = {131},
       number = {6},
        pages = {2766-2787},
          doi = {10.1086/503559},
archivePrefix = {arXiv},
       eprint = {astro-ph/0601434},
 primaryClass = {astro-ph},
       adsurl = {https://ui.adsabs.harvard.edu/abs/2006AJ....131.2766R}
}

@ARTICLE{Lusso2015,
       author = {{Lusso}, E. and {Worseck}, G. and {Hennawi}, J.~F. and {Prochaska}, J.~X. and {Vignali}, C. and {Stern}, J. and {O'Meara}, J.~M.},
        title = "{The first ultraviolet quasar-stacked spectrum at z ≃ 2.4 from WFC3}",
      journal = {\mnras},
         year = 2015,
        month = jun,
       volume = {449},
       number = {4},
        pages = {4204-4220},
          doi = {10.1093/mnras/stv516},
archivePrefix = {arXiv},
       eprint = {1503.02075},
 primaryClass = {astro-ph.GA},
       adsurl = {https://ui.adsabs.harvard.edu/abs/2015MNRAS.449.4204L}
}

@ARTICLE{COLOSSUS,
       author = {{Diemer}, Benedikt},
        title = "{COLOSSUS: A Python Toolkit for Cosmology, Large-scale Structure, and Dark Matter Halos}",
      journal = {\apjs},
         year = 2018,
        month = dec,
       volume = {239},
       number = {2},
          eid = {35},
        pages = {35},
          doi = {10.3847/1538-4365/aaee8c},
archivePrefix = {arXiv},
       eprint = {1712.04512},
 primaryClass = {astro-ph.CO},
       adsurl = {https://ui.adsabs.harvard.edu/abs/2018ApJS..239...35D}
}

@Article{Matplotlib,
  Author    = {Hunter, J. D.},
  Title     = {Matplotlib: A 2D graphics environment},
  Journal   = {Computing in Science \& Engineering},
  Volume    = {9},
  Number    = {3},
  Pages     = {90--95},
  publisher = {IEEE COMPUTER SOC},
  doi       = {10.1109/MCSE.2007.55},
  year      = 2007
}

@ARTICLE{Yu2002,
       author = {{Yu}, Qingjuan and {Tremaine}, Scott},
        title = "{Observational constraints on growth of massive black holes}",
      journal = {\mnras},
         year = 2002,
        month = oct,
       volume = {335},
       number = {4},
        pages = {965-976},
          doi = {10.1046/j.1365-8711.2002.05532.x},
archivePrefix = {arXiv},
       eprint = {astro-ph/0203082},
 primaryClass = {astro-ph},
       adsurl = {https://ui.adsabs.harvard.edu/abs/2002MNRAS.335..965Y}
}

@ARTICLE{Knud2025,
       author = {{Jahnke}, Knud},
        title = "{The Soltan argument at redshift 6: UV-luminous quasars contribute less than 10\% to early black hole mass growth}",
      journal = {The Open Journal of Astrophysics},
         year = 2025,
        month = jan,
       volume = {8},
          eid = {9},
        pages = {9},
          doi = {10.33232/001c.129063},
archivePrefix = {arXiv},
       eprint = {2411.03184},
 primaryClass = {astro-ph.GA},
       adsurl = {https://ui.adsabs.harvard.edu/abs/2025OJAp....8E...9J}
}

@ARTICLE{Davies2019,
       author = {{Davies}, Frederick B. and {Hennawi}, Joseph F. and {Eilers}, Anna-Christina},
        title = "{Evidence for Low Radiative Efficiency or Highly Obscured Growth of z > 7 Quasars}",
      journal = {\apjl},
         year = 2019,
        month = oct,
       volume = {884},
       number = {1},
          eid = {L19},
        pages = {L19},
          doi = {10.3847/2041-8213/ab42e3},
archivePrefix = {arXiv},
       eprint = {1906.10130},
 primaryClass = {astro-ph.GA},
       adsurl = {https://ui.adsabs.harvard.edu/abs/2019ApJ...884L..19D}
}

@ARTICLE{Shen2026,
       author = {{Shen}, Xuejian and {Zier}, Oliver and {Smith}, Aaron and {Liu}, Rongrong and {Kannan}, Rahul and {Bulichi}, Teodora-Elena and {Koehler}, Sonja M. and {Springel}, Volker and {Vogelsberger}, Mark and {Hernquist}, Lars and {Naidu}, Rohan P. and {de Graaff}, Anna and {Pizzati}, Elia and {Alexander}, David M. and {Ho}, Luis C. and {Kokorev}, Vasily and {Leung}, Gene and {Eilers}, Anna-Christina and {Hickox}, Ryan C.},
        title = "{The Lumina Project: The Demographics of Active Galactic Nuclei from Quasars to Little Red Dots at $z\geq 3$}",
      journal = {arXiv e-prints},
         year = 2026,
        month = may,
          eid = {arXiv:2605.24112},
        pages = {arXiv:2605.24112},
          doi = {10.48550/arXiv.2605.24112},
archivePrefix = {arXiv},
       eprint = {2605.24112},
 primaryClass = {astro-ph.GA},
       adsurl = {https://ui.adsabs.harvard.edu/abs/2026arXiv260524112S}
}

@ARTICLE{Zier2026,
       author = {{Zier}, Oliver and {Smith}, Aaron and {Shen}, Xuejian and {Liu}, Rongrong and {Kannan}, Rahul and {Koehler}, Sonja M. and {Springel}, Volker and {Pakmor}, R{\"u}diger and {Vogelsberger}, Mark and {Bulichi}, Teodora-Elena and {Hernquist}, Lars},
        title = "{Introducing the Lumina project: large-volume radiation-hydrodynamic simulations of the epochs of hydrogen and helium reionization}",
      journal = {arXiv e-prints},
         year = 2026,
        month = may,
          eid = {arXiv:2605.15310},
        pages = {arXiv:2605.15310},
          doi = {10.48550/arXiv.2605.15310},
archivePrefix = {arXiv},
       eprint = {2605.15310},
 primaryClass = {astro-ph.CO},
       adsurl = {https://ui.adsabs.harvard.edu/abs/2026arXiv260515310Z}
}

@ARTICLE{Tenneti2019,
       author = {{Tenneti}, Ananth and {Wilkins}, Stephen M. and {Di Matteo}, Tiziana and {Croft}, Rupert A.~C. and {Feng}, Yu},
        title = "{A tiny host galaxy for the first giant black hole: z = 7.5 quasar in BlueTides}",
      journal = {\mnras},
         year = 2019,
        month = feb,
       volume = {483},
       number = {1},
        pages = {1388-1399},
          doi = {10.1093/mnras/sty3161},
archivePrefix = {arXiv},
       eprint = {1806.00185},
 primaryClass = {astro-ph.GA},
       adsurl = {https://ui.adsabs.harvard.edu/abs/2019MNRAS.483.1388T}
}

@ARTICLE{Marshall2021,
       author = {{Marshall}, Madeline A. and {Wyithe}, J. Stuart B. and {Windhorst}, Rogier A. and {Di Matteo}, Tiziana and {Ni}, Yueying and {Wilkins}, Stephen and {Croft}, Rupert A.~C. and {Mechtley}, Mira},
        title = "{Observing the host galaxies of high-redshift quasars with JWST: predictions from the BLUETIDES simulation}",
      journal = {\mnras},
         year = 2021,
        month = sep,
       volume = {506},
       number = {1},
        pages = {1209-1228},
          doi = {10.1093/mnras/stab1763},
archivePrefix = {arXiv},
       eprint = {2101.01219},
 primaryClass = {astro-ph.GA},
       adsurl = {https://ui.adsabs.harvard.edu/abs/2021MNRAS.506.1209M}
}

\appendix

\section{The populations of SMBHs at high redshift in \TNG and \flamingo}
\label{app:SMBHpopulations}

\begin{figure*}
\centering
\includegraphics[width=0.3\linewidth]{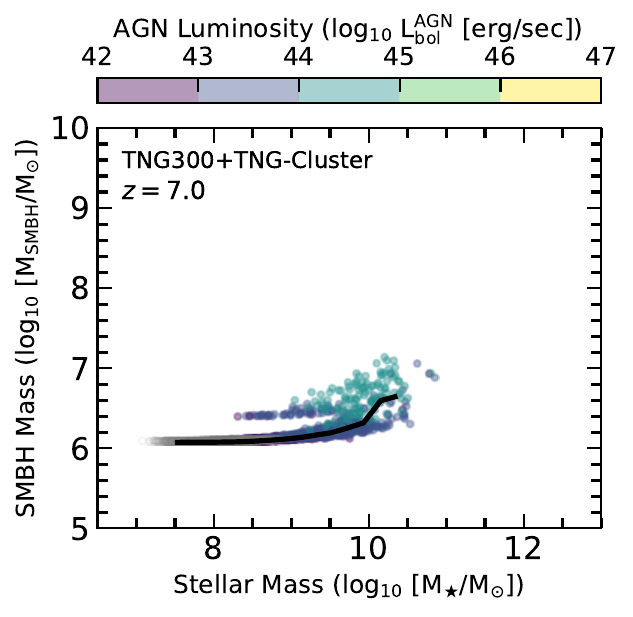}
\includegraphics[width=0.3\linewidth]{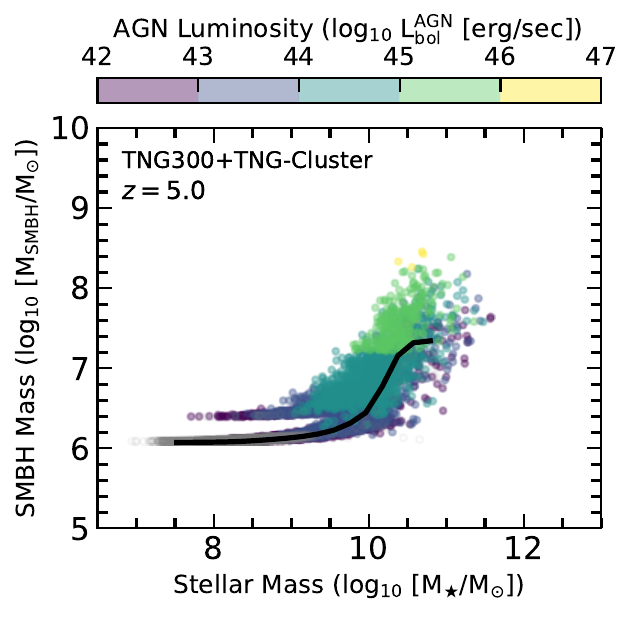}
\includegraphics[width=0.3\linewidth]{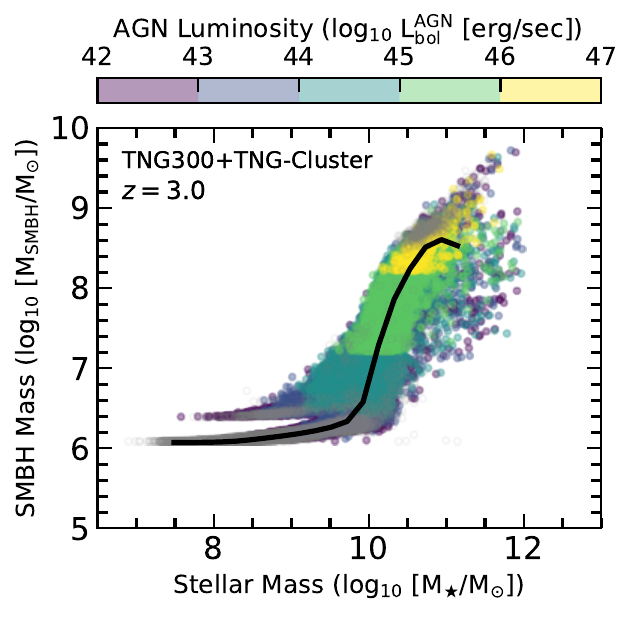}
\includegraphics[width=0.3\linewidth]{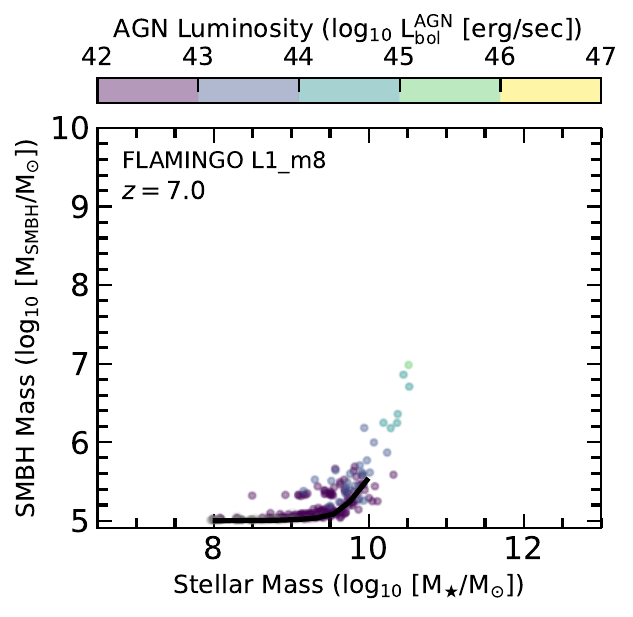}
\includegraphics[width=0.3\linewidth]{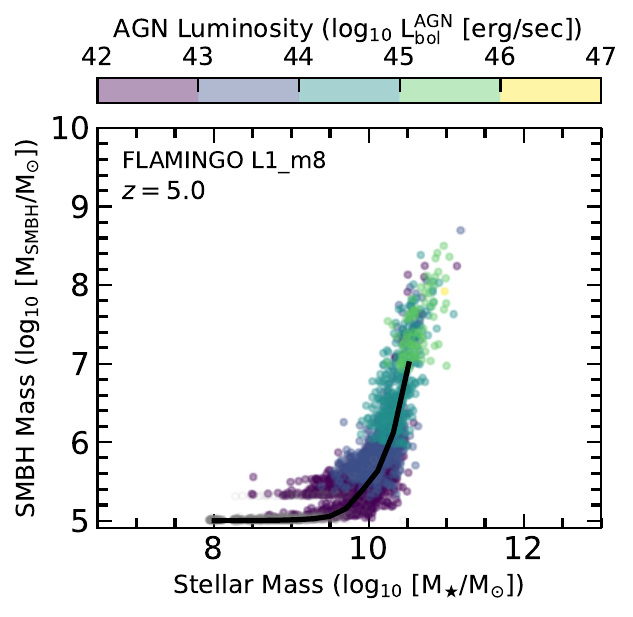}
\includegraphics[width=0.3\linewidth]{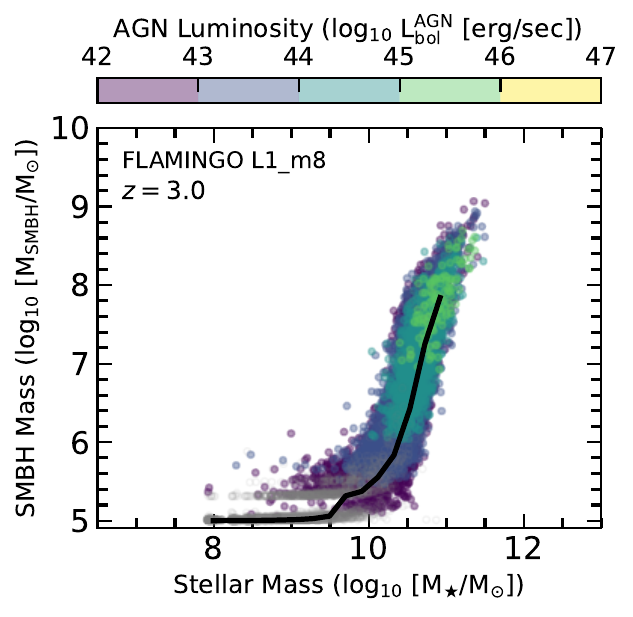}
\includegraphics[width=0.3\linewidth]{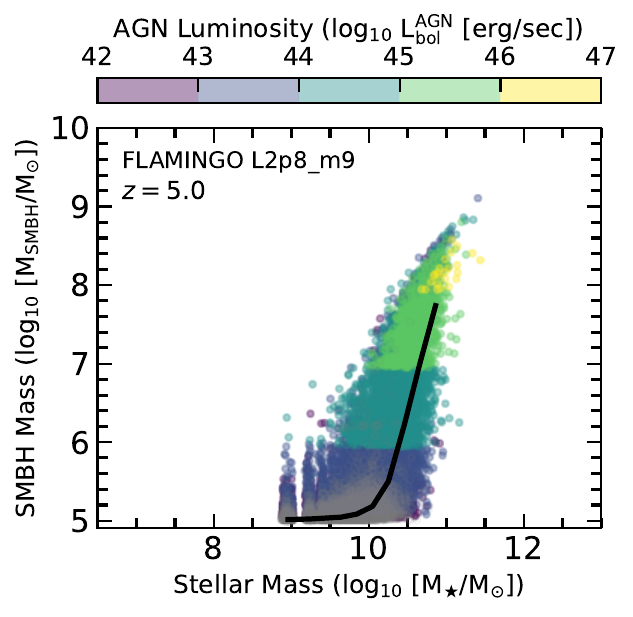}
\includegraphics[width=0.3\linewidth]{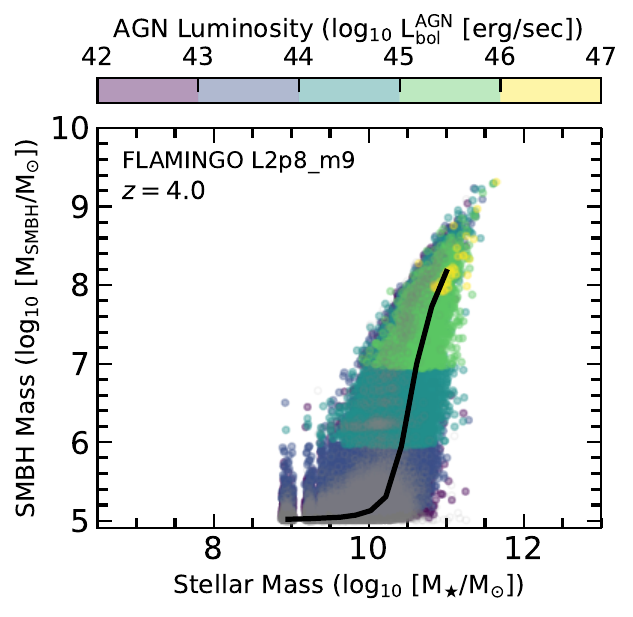}
\includegraphics[width=0.3\linewidth]{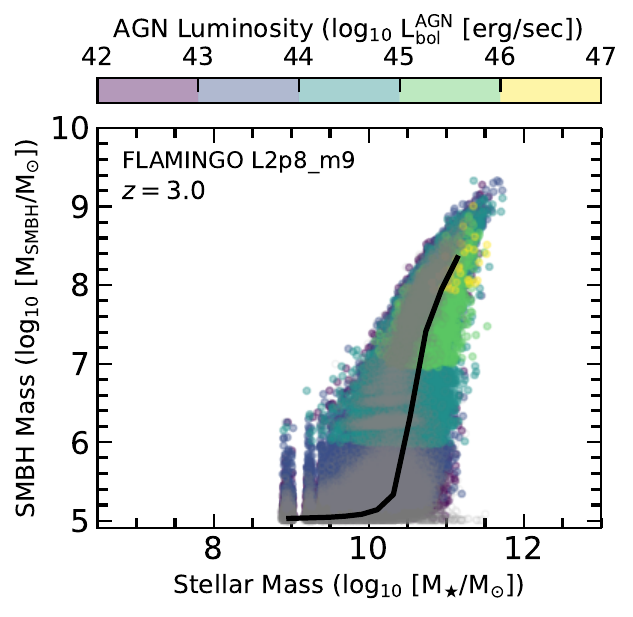}

\caption{\textbf{SMBH mass versus galaxy stellar mass according to the \TNG and \flamingo simulations.} We present the populations of simulated SMBHs i.e. AGN with a scatter plot of all (central) galaxies at $z=7, 5$ and 3 (left to right column) in the \TNG (top row) and in the \flamingo \flone (middle row) simulations, as well as those from the \flamingo \fltwo simulation (bottom row) at $z=5, 4$ and 3. The AGN considered throughout the paper, i.e. those with $\Lbol \ge 10^{42}$ erg~s$^{-1}$, are color coded by their AGN bolometric luminosities whereas the gray markers represent the rest of the population available (and simulated) within each run. At Cosmic Noon or shortly before ($z=3$), these simulations allow us to study large numbers of objects, including SMBHs as massive as $10^{9-10}$\msun. Interestingly, the different underlying models return different relationships among SMBH mass, galaxy stellar mass and AGN bolometric luminosity and, for example, in \TNG a population of very low-luminosity AGN appears at $z<5$ at the highest SMBH and galaxy mass end. Importantly, the $\Lbol \ge 10^{42}$ erg~s$^{-1}$ cut allows us to focus on simulated SMBHs that are not as affected by modeling choices such as the SMBH seeding mass (see almost horizontal tracks at about $10^6$ and $10^5$\msun).}
\label{fig:app_SMBHs_1}
\end{figure*}

\begin{figure*}
\centering

\includegraphics[width=0.3\linewidth]{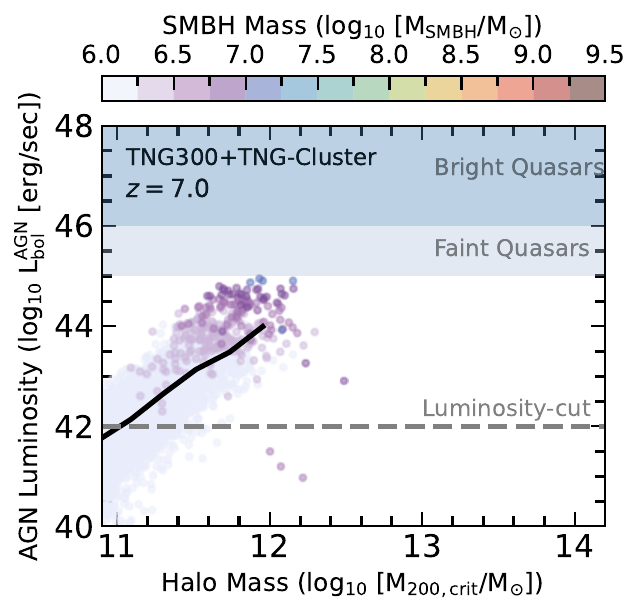}
\includegraphics[width=0.3\linewidth]{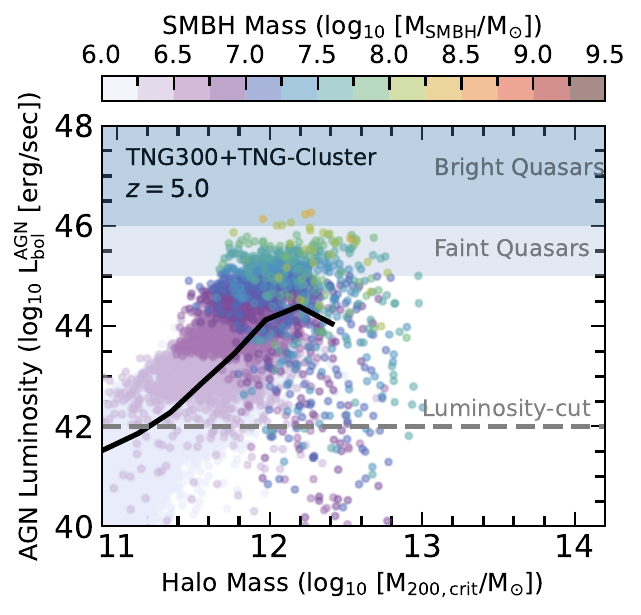}
\includegraphics[width=0.3\linewidth]{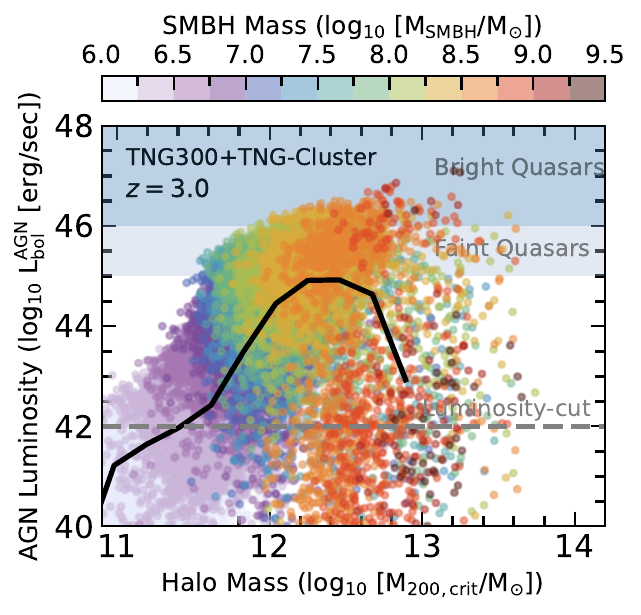}  
\includegraphics[width=0.3\linewidth]{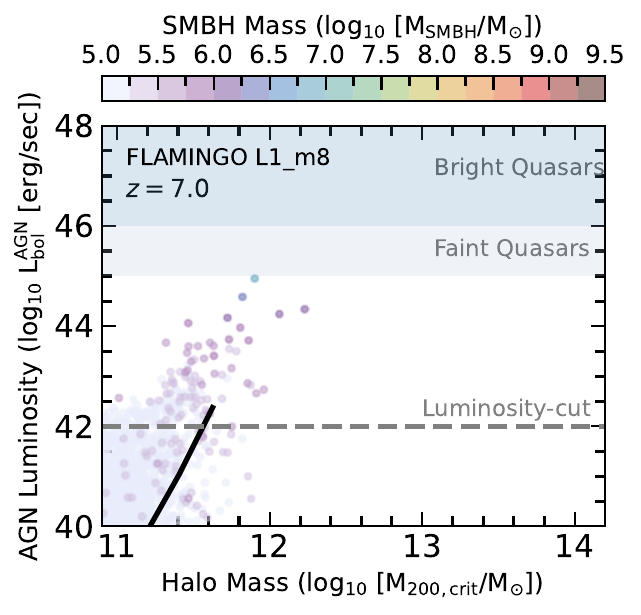}
\includegraphics[width=0.3\linewidth]{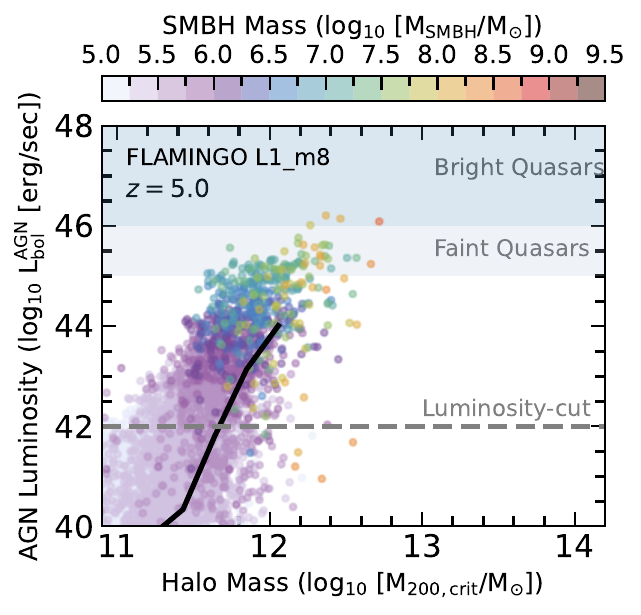}
\includegraphics[width=0.3\linewidth]{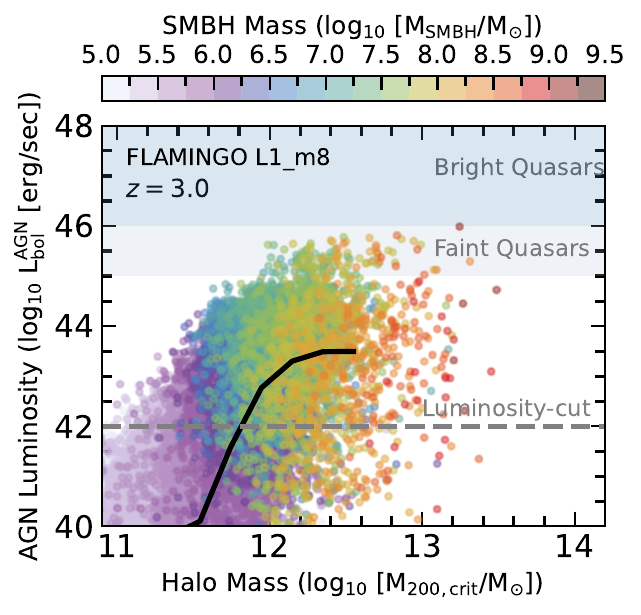}  
 
\caption{\textbf{SMBH masses in the plane of AGN bolometric luminosity and host halo mass for the \TNG (top) and the \flamingo \flone (bottom) simulations.} We showcase the simulated populations at $z=7, 5,$ and 3, from left to right. Each marker represents individual simulated objects, color coded by their SMBH mass. To contextualize the results of the main body of the paper, here we include all simulated AGN, i.e. also those below the luminosity cut of $10^{42}$ erg~s$^{-1}$ adopted throughout (horizontal dashed line), and the solid curves denote the median AGN bolometric luminosity in bins of host halo mass accounting for {\it all} simulated and stored systems -- to be compared with the median curves of the top panels of Figs.~\ref{fig:lum_vs_halo_tng} and \ref{fig:lum_vs_halo_fl}. More massive haloes host on average more massive SMBHs and more luminous AGN, however the latter only up to host halo masses of about $10^{12}$\msun. In both simulation suites, towards lower redshifts, the scatter is also driven by the emergence of low-accreting but massive SMBHs.}
   \label{fig:app_SMBHs_2}
\end{figure*}

Throughout our analysis we restrict to study simulated AGNs with minimum bolometric luminosity of $10^{42}$ erg~s$^{-1}$ (Section~\ref{sec:methods_selection}). However, naturally, in the considered simulations a larger majority of SMBHs actually exhibit lower luminosities, even at the high redshift considered. Moreover, while  their host halo masses are clearly quantified in the main body of this paper, the so-selected AGNs span a very large range of SMBH and host galaxy stellar masses. It is therefore useful to contextualize the main results of the paper within the broad populations of simulated haloes, SMBHs and galaxies. 

In Fig.~\ref{fig:app_SMBHs_1}, we show the relation between SMBH and galaxy stellar mass of all available AGNs residing in central galaxies in the \TNG (top row), \flamingo \flone (middle row), and \flamingo \fltwo simulations (bottom row), at three different example redshifts: $z=7, 5, 3$ for the former two and $z=5,4,$ and 3 for the \fltwo box, from left to right. Individual markers indicate individual objects and the color coding indicates the corresponding AGN bolometric luminosities: colors are reserved for the selected cases ($\Lbol \ge 10^{42}$ erg~s$^{-1}$) whereas gray scatter markers indicate the SMBHs that are simulated (and available in the catalogs) but have not been selected. 

A relatively tight {\it Magorrian} relation naturally emerges in the \TNG and \flamingo simulations already at high redshifts -- see e.g. \cite{Habouzit2021} for similar plots for the smaller-volume simulations of Table~\ref{Tab:simulations}. Black solid curves denote running medians and, as expected, more massive galaxies host more massive SMBHs. However, peculiar features are also manifest.

The horizontal tracks at SMBH masses of about $10^6$ and $10^5$\msun are a manifestation of the mass with which SMBHs are seeded in the \illustrisTNG and \flamingo models, respectively, with the second higher tracks representing SMBHs that have moved up from their seed mass by merging with a just-seeded and not-yet-grown SMBH.  

The adopted AGN luminosity threshold (gray vs. colored markers) enable us to (largely) avoid the regimes that are most directly affected by the seeding prescriptions of the simulations. Within these limits, the AGN populations studied in the paper encompass up to four orders of magnitude in SMBH mass in \flamingo \fltwo and e.g. more than 3 orders of magnitude in SMBH mass in \TNG towards low redshift ($z\sim3$). Analogously, the AGN populations studied in the paper reside in galaxies with stellar mass ranging mostly across $10^{9-12}$\msun at $z\lesssim 3-4$ in \TNG, but with steeper relations, and hence narrow galaxy mass ranges, in the \flamingo simulations. 

In any case, our analyses are based on simulated systems that are for the largest fraction well resolved. For example, in \TNG, galaxies with stellar mass of $10^9$\msun are made of more than 100 stellar particles each (see Table~\ref{Tab:simulations}). \flamingo galaxies are made of fewer resolution elements. However, throughout the paper, we focus on haloes with total mass exceeding $\ge 10^{11}$\msun, to avoid the model-dependent minimum mass for a halo to be seeded with a SMBH (see Table~\ref{Tab:simulations}), which in turn is largely dominated by galaxies with mass $\gtrsim 10^{10}$ at $z\geq3$, thus ensuring a few tens of stellar particles per galaxy even in the lower-resolution \flamingo \fltwo.  

Looking at the AGN bolometric luminosities, because halo, SMBH, and galaxy masses all correlate, we clearly see, as broadcast in the main body of the paper, that more massive systems typically host more luminous AGN, but only up to a certain limit. 

Particularly in \TNG, the high-mass end of the SMBH populations at $z\lesssim5$ can be characterized by very low AGN luminosities. This is particularly clear at $z \sim 3$ (third panel in the top row), where a non-negligible fraction of low-luminosity AGN reside in the most massive galaxies of the simulations. In fact, a number of them exhibit much lower SMBH masses than those in similarly-massive galaxies, {\it falling off} from the main scaling relation; a few, however, and even below the $\Lbol \ge 10^{42}$ erg~s$^{-1}$ threshold, actually are among the most massive SMBHs (probably this massive thanks to merging, see e.g. \citealt{Truong2021}).  All this is a manifestation of the effectiveness of the \illustrisTNG low-accretion kinetic mode feedback at vacating gas from the central regions of galaxies, initiating the quenching of star formation in massive galaxies \citep{Weinberger-feedback, Davies2020, Zinger2020, Kurinchi-Vendhan2024} but also limiting the supply of gas available for accretion onto the SMBHs \citep[see e.g.][for an extensive discussion, albeit in the low-redshift regime]{Kurinchi-Vendhan2025} -- this involving larger fractions of massive galaxies as time proceeds, i.e. towards lower redshifts.   

Also in the \flamingo simulations, at $z\lesssim5$, some of the most massive galaxies (or SMBHs) can host as low-luminosity AGNs as $10^{42-43}$ erg~s$^{-1}$, suggesting a similar but not identical high-mass AGN quenching as in the \illustrisTNG model. In fact, low-luminosity AGNs (even below the adopted threshold) seem to be present (without dominating) the entire SMBH/galaxy mass range, particularly visible in the \flamingo \fltwo run (see gray markers and discussions in \citealt{flamingo_quasar_clustering}). 

The intriguing presence of low-luminosity AGNs at the high end of the galaxy or halo mass range of the \TNG and \flamingo simulations is even more clear in Fig.~\ref{fig:app_SMBHs_2}, which shows the plane of AGN bolometric luminosity vs. host halo mass (studied throughout the main body of the paper) but here color coded by SMBH mass. As expected, more massive haloes host more massive SMBHs, but again only up to a certain mass, and the relationship between AGN bolometric luminosity and SMBH mass is complex, changing somehow below $z\lesssim5$. Towards lower redshifts, both simulation suites predict that high-mass haloes can host both highly- and low-luminosity AGN.

\section{Scatter at fixed AGN bolometric luminosity in smaller-volume simulations} 
\label{app:allsims}

\begin{figure}
\includegraphics[width=1.\linewidth]{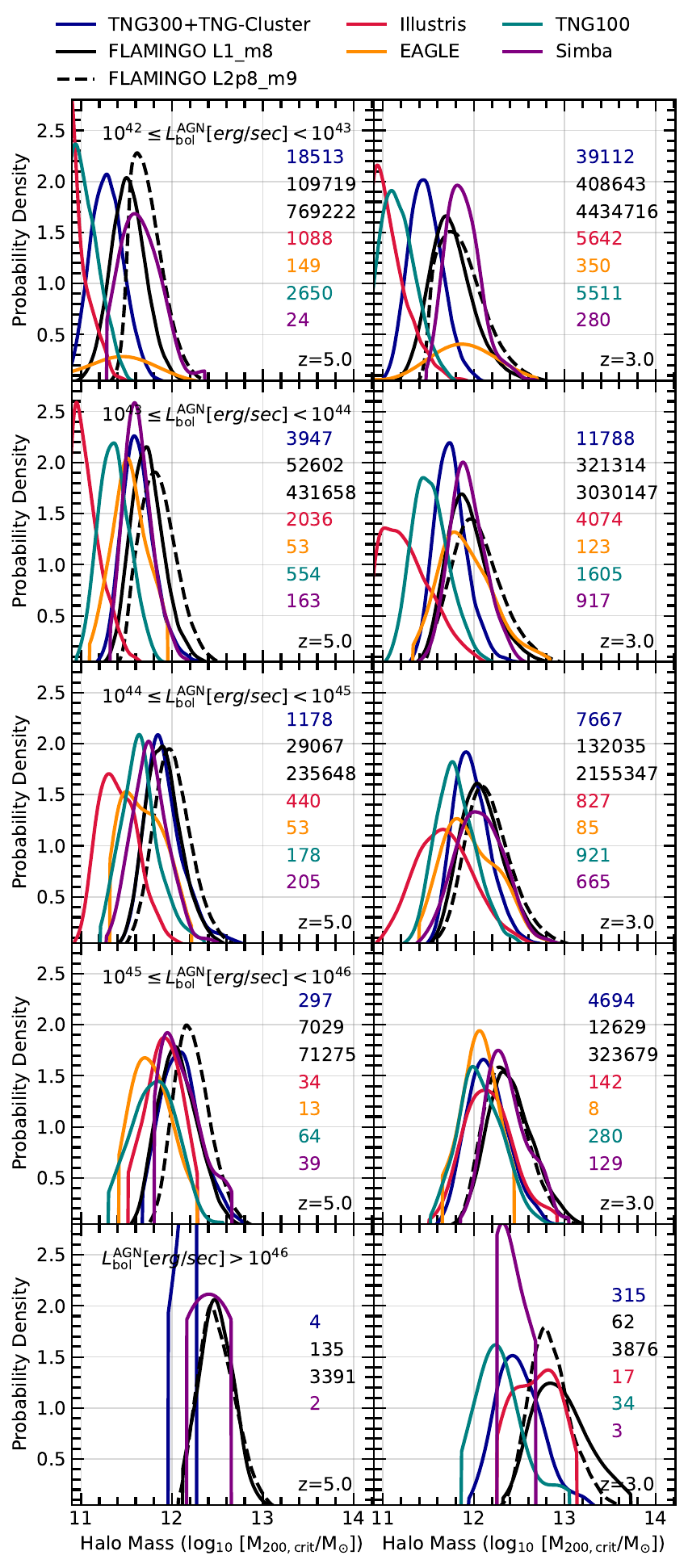}
\caption{\textbf{Conditional probability of getting a certain host halo mass given a certain AGN bolometric luminosity, according to all simulations studied in this paper.} This figure provides analog normalized KDE probability density functions as the bottom rows of Figs.~\ref{fig:lum_vs_halo_tng} and \ref{fig:lum_vs_halo_fl}, with similar annotations and setups. Overall, all simulations show a general increase in the typical halo mass with luminosity and a mild variation with decreasing redshift. The scatter in the relation is also similar across simulations.}
\label{fig:app_pdf_Mhalo_compare_sims}
\end{figure}

For completeness, we provide below the conditional probabilities of obtaining a certain host halo mass given a certain AGN bolometric luminosity (Fig.~\ref{fig:app_pdf_Mhalo_compare_sims}), according to all simulations studied in this paper. We visualize only $z=3$ and $5$, with the latter representing the highest redshift that still provides a statistically-meaningful sample across the studied simulations. The two \flamingo runs exhibit similar distributions and levels of scatter across both redshift and luminosity. The scatter in host halo mass at fixed AGN bolometric luminosity is comparable across simulations and reads about 1 dex, but with some exceptions: \eagle and \illustris at lower AGN luminosities return relatively wider scatters. While the most probable host halo mass depends upon the simulation model, the results generally agree within the overlapping scatter, except for \illustris at lower AGN luminosities. A detailed analysis of the halo mass distributions of AGN populations in these smaller volume simulations is left for future work.

\section{The evolutionary history of the brightest and most massive AGNs at z=3 in \TNG}
\label{App:tracks}
\begin{figure}
\centering
\includegraphics[width=0.9\columnwidth]{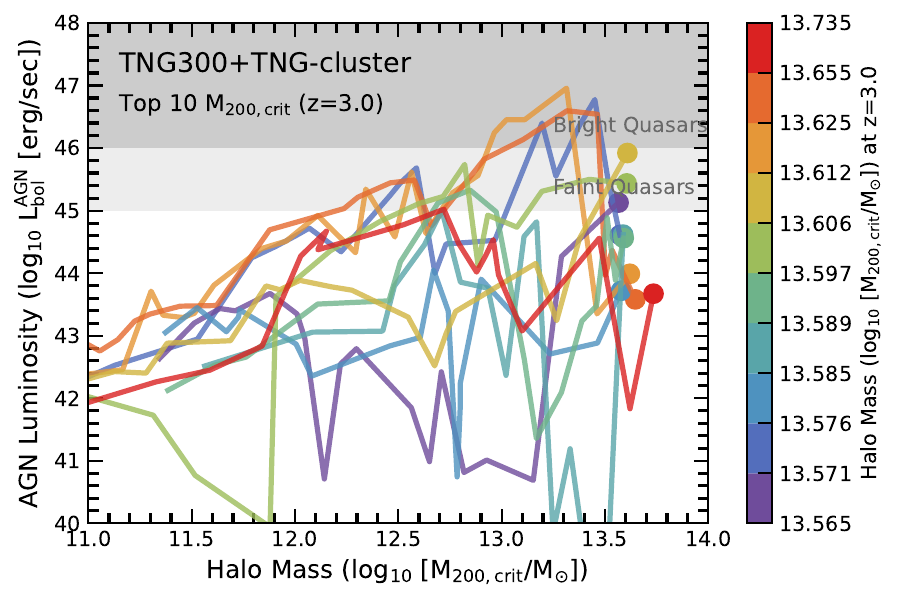}
\includegraphics[width=0.9\columnwidth]{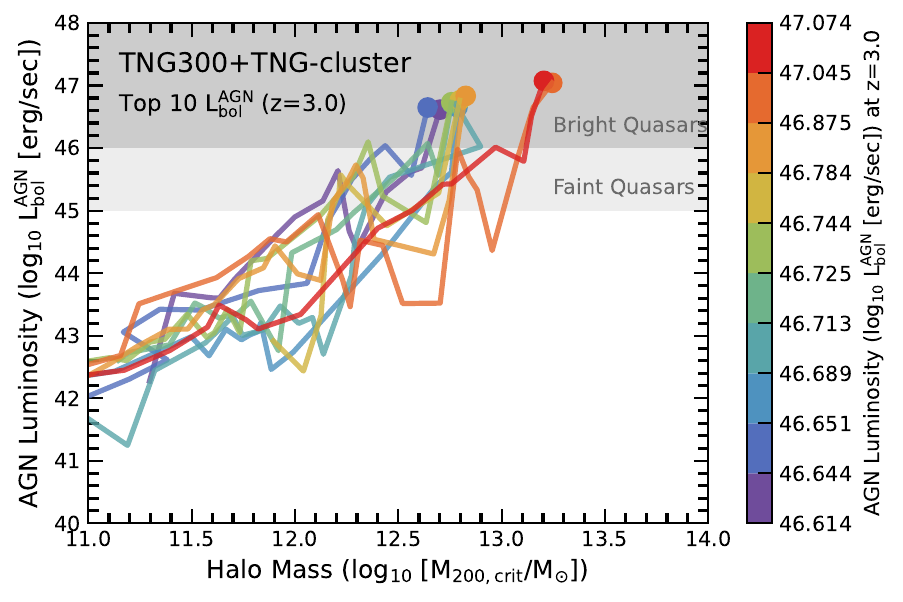}
\caption{{\bf Time evolution of example systems from \TNG}. We show the evolutionary tracks on the plane of AGN bolometric luminosity vs. host halo mass of twenty AGNs: those that reside in the 10 most massive haloes (top) and that are the 10 most luminous at $z=3$ (filled circles), as per the red circles and squares markers, respectively, of Fig~\ref{fig:lum_vs_halo_tng}. The objects are hence selected at $z=3$ and followed back in time along their main progenitor branch, shown with a time cadence of about hundred million years. Both panels show that AGN luminosity typically peaks, at least according to the \illustrisTNG model, when the haloes reach masses of around $10^{12-13.5}$\msun and that low-luminosity AGNs in the most massive haloes at $z=3$ have been more luminous in the past, i.e. prior to the quenching effects of their own feedback.}
\label{Fig:tracks}
\end{figure}

Fig.~\ref{Fig:tracks} shows the evolutionary tracks, sampled at intervals of $100-150$ Myr, for AGNs residing in the 10 most massive haloes (top panel) and the 10 most luminous AGNs (bottom panel) of \TNG at $z=3$. These are the systems highlighted by the red circles and squares, respectively, in Fig.~\ref{fig:lum_vs_halo_tng}.

Both panels indicate that, at least in the \illustrisTNG model, AGNs initially follow the power-law relations discussed in Sections~\ref{sec:results_plane_relation} and \ref{sec:results_plane_zfits} and quantified in Table~\ref{Tab:fits}, until SMBH feedback becomes effective, causing the systems to deviate from this early-growth trajectory.
The precise shape of the evolutionary tracks may depend on many factors and certainly it seems to depend on when feedback starts to be effective, which may in turn depend on how fast or slow the host haloes grow in mass. The later the SMBH quenching kicks in, the better are the chances of AGNs to keep moving along the tracks of ``higher AGN luminosities in more massive host haloes'', possibly reaching quasar-like levels of luminosity.

From our analysis of \TNG, we know that, even neglecting the time variability captured by the model, all AGNs in the most massive haloes at any given epoch, at least at $z < 5-6$, i.e. also those with very low AGN bolometric luminosity at that time -- have been more luminous in the past. However, based on Fig.~\ref{Fig:tracks}, not all massive haloes at e.g. $z=3$ have necessarily experienced a {\it quasar}-like phase in their past, i.e. have not necessarily reached quasar-like bolometric luminosities. In fact, here we explicitly sample their evolutions only every about one hundred million years, the time cadence of the snapshot data. Even if we instead plotted the evolution of SMBHs at the cadence of time resolution of the simulations, establishing conclusively whether all most massive haloes at any given time have undergone a quasar-luminosity phase is not feasible, given that the simulations cannot capture short-lived episodes of quasar activity (as discussed in Section~\ref{sec:disc}).

\section{Summary of observationally inferred halo masses of quasars}
\label{App:obs}

Finally, we summarize in Table~\ref{Tab:observations} the empirically-inferred host halo masses of quasars at the redshifts of interest in this paper. The estimates span a range of observational techniques and assumptions, reflecting the current uncertainties in constraining quasar host halo masses. These measurements provide an observational benchmark against which we compare the host halo masses of quasars predicted by our simulations in Fig.~\ref{fig:quasarmasses}.

\begin{table*}
\setlength{\tabcolsep}{4pt}  
\renewcommand{\arraystretch}{1.1}
\caption{Summary of the empirically-inferred host halo masses of quasars at $z \gtrsim 3$ based on various observational estimates. The columns read: 1-3) reference, redshift range and inferred host halo masses$^{\rm a}$, respectively. 4-5) absolute magnitude of the observed quasars and their corresponding bolometric luminosities; 6) number of quasars for which the measurement was performed; 7) method adopted to infer host halo masses. The first section of the table summarizes all measurements based on quasar-quasar clustering (labeled clustering) and quasar-galaxy clustering (labeled cross-correlation); the second part reports measurements using methods other than clustering.}
\begin{tabularx}{\textwidth}{l c c c c c X}
\hline
Reference & 
Redshift &
Host halo mass$^{\rm a}$ & 
Magnitude & 
Luminosity$^{\rm d}$  & 
$\#$ of quasars & 
Method \\
&&log$_{10}[\Mtwohcrit$/\msun]& (absolute) & $L_{\rm bol} [10^{46}$ erg~s$^{-1}]$  & &  \\
\hline
\cite{Shen2007} & $2.9-3.5$ & $\gtrsim 12.7$ & $[-26, \infty)^{\rm b,d}$ & $\gtrsim 9.8^{\rm d}$ & 2,650 & Clustering \\

\cite{Shen2007}& $3.5-5.3$ & $\gtrsim 13.0$ & $[-26.9, \infty)^{\rm b,d}$ & $\gtrsim 22.4^{\rm d}$ & 1,776 & Clustering \\

\cite{Sarah2015}& $2.2-2.8$ & $12.44^{+0.04}_{-0.05}$ & $[-23.8, -28.7]^{\rm b}$ & $1.4-115$ & 55,826 & Clustering \\

\cite{Sarah2015}& $2.6-3.4$ & $12.00^{+0.04}_{-0.04}$ & $[-24.4, -29.3]$$^{\rm b}$ & $2.3-200$ & 24,724 & Clustering \\

\cite{Timlin2018}& $2.9-5.1$& $12.7-13.4$ & $[-23.8, -27.5]^{\rm b}$ & $1.4-38$ & 1,378 & Clustering \\

\cite{Arita2023}& $5.8-6.6$ & $12.85^{+0.42}_{-0.69}$ & $[-21.5, -27]^{\rm c}$ & $0.2-27$ & 107 & Clustering \\

\cite{Mascarell2025}& $3.0-4.0 $&$\gtrsim 12.82^{+0.1}_{-0.12}$& $[-25, -29]^{\rm c}$ &$4.3-165$&35,466& Clustering\\

\cite{Meng2026} &$5.0-5.7$&$12.2^{+0.2}_{-0.7}$&$[-22.5, -28.5]^{\rm c}$&$0.4-104.6$&2429& Clustering\\

\cite{Meng2026} &$5.7-6.3$&$11.9^{+0.3}_{-0.7}$&$[-22.5, -28.5]^{\rm c}$&$0.4-104.6$&1923&Clustering \\

&&&&&&\\

\cite{He2018} & $3.1-4.6$ & $11.63-12.45$ &  $[-22.3, -24.7]^{\rm c}$ & $0.4-3.3$ & 901 & Cross-correlation \\

\cite{Eilers}& $6.0-7.0$ & $12.43^{+0.13}_{-0.15}$ & $[-26.6,-29.0]^{\rm c}$ & $18.5 - 165$ & 4 & Cross-correlation \\


\cite{Schindler2026}& $7.0-7.5$&$\gtrsim 11.6^{+0.6}_{-0.6}$&$-25.77, -26.62^{\rm c}$ & $9, 19$ &2& Cross-correlation \\

\cite{Wang2026} &$6.5-6.8$& $12.27^{+0.21}_{-0.26}$ & $[-25,-27.5]^{\rm c}$ & $4.3-42$& 25& Cross-correlation \\

\cite{Huang2026}&$6.5-6.8$&$12.1^{+0.3}_{-0.4}$&$[-25.3,-27.4]^{\rm c}$&$5.4-37.7$&25&Cross-correlation \\

&&&&&&\\
\cite{Chen2022} & $5.7-6.6 $& $12.5^{+0.4}_{-0.7}$ & $[-26.6, -29.0]$ & $18.5-165$ & 10 & Density field around quasars \\ 

\cite{Pizzati2024}$^{\rm e}$ & $\sim 4$& $13.0-13.5$ & -- & -- & -- & HoD based on \cite{Shen2007} \\

\cite{Pizzati2024}$^{\rm e}$ & $\sim 2.5$& $12.5-13.0$ & -- &--& -- & HoD based on \cite{Sarah2015} \\

\cite{Pizzati2024b}$^{\rm e}$ & $\sim 6.0$& $12.1-12.8$ & -- &--& -- & HoD based on \cite{Eilers} \\

\cite{Costa2024}$^{\rm f}$ &6& $\gtrsim 12.7$& -- & -- & -- & simulated velocity distribution + observations  \\

\cite{Qinyue2025} &$\sim 6$& $12.5-12.8$ & -- & -- & 2 & gas kinematics  \\

\hline
\end{tabularx}
\begin{tablenotes}[para,flushleft]
\footnotesize
\item[a] All values for host halo masses have been converted to the same cosmology. Namely, we have corrected for the different underlying choices of cosmological parameters (especially $\sigma_8$) by converting the halo mass to a peak height in one cosmology and then using that peak height to infer halo mass using the cosmology in simulations, by using the COLOSSUS cosmology package \citep{COLOSSUS}. On the other hand, clustering measurements typically provide host halo masses in terms of $M_\mr{200,mean}$, defined as the mass enclosed in a radius that corresponds to 200 times the mean density of the Universe at that redshift, rather than $\Mtwohcrit$: the difference between ${M_\mr{200,mean}}$ and $\Mtwohcrit$ at our redshifts of interest is negligible, given that then the mean and critical densities of the Universe in these cosmic epochs are similar.\\ 
\item[b,c] Absolute magnitudes marked with $^{\rm b}$ are in \texttt{SDSS} $i$-band ($M_i$) K-corrected to $z=2$; those marked with $^{\rm c}$ are ultraviolet absolute magnitudes at 1450 \AA ($M_{1450}$).\\
\item[d] To connect to the AGN bolometric luminosities quoted throughout the paper and derived from the simulations outputs, we convert the observed \textit{i}-band or UV magnitudes using the bolometric corrections from \cite{Shen2020}. All values in this column of the Table must be intended as approximate, as ball park estimates. In particular, we first correct the $M_i(z=2)$ observed magnitudes into $M_i(z=0)$ using the \cite{Richards2006} K-correction: $M_i(z=0)=M_i(z=2)+2(1-\alpha)~{\rm log}(1+z)$. The $M_i(z=0)$ (which correspond to $\sim 7480$~\AA) is then converted to the B-band ($4400$~\AA) magnitude using: $M_{4400}=M_{i} + 2.5 \alpha~\mr{log (4400/7480)}=M_{i}+0.35$, where $\alpha=-0.61$ \citep{Lusso2015}. The $M_{4400}$ is then converted to $L_{4400}$  using, $L_{\nu}=10^{-0.4(M_{1450-51.6})}$ in the AB magnitude system. Finally the B-band luminosities thus obtained are converted to bolometric luminosities using bolometric-corrections from Table 2 of \cite{Shen2020}. We adopt the same \citep{Shen2020} bolometric corrections to estimate $L_{1450}$ from the quoted $M_{1450}$, and then deduce a bolometric luminosity.\\
\item[d] We report here the magnitude limits (and corresponding minimum values of AGN bolometric luminosity) for the lower limit only of the redshift range, for brevity. \\  
\item[e] The \cite{Pizzati2024,Pizzati2024b} inferences are based on a conditional luminosity-based halo occupation model (HoD) that constrains host halo masses from combined measurements of clustering and luminosity function of quasars at that redshift.\\
\item[f] Host halo masses from \cite{Costa2024} are based on forward modeling zoom-in simulations to compare with observed velocity distribution of gas around quasars.

\end{tablenotes}
\label{Tab:observations}
\end{table*}

\bsp	
\label{lastpage}
\end{document}